\documentclass[12pt]{article}
\usepackage[english]{babel}
\usepackage[pdftex]{color}
\usepackage{amsmath}
\usepackage{graphicx}
\usepackage{hyperref}
\usepackage{amsmath}
\usepackage{amsfonts}
\usepackage{amssymb}
\renewcommand{\theequation}{\thesection.\arabic{equation}}

\begin{document}

\title{
{\bf{} General Lagrangian Formulation for Higher Spin Fields with
Arbitrary Index Symmetry. I. Bosonic fields}}

\author{\sc I.L. Buchbinder${}^{a,b}$\thanks{joseph@tspu.edu.ru
\hspace{0.5cm} ${}^{\dagger}$reshet@ispms.tsc.ru},
 A. Reshetnyak$^{c\ddagger}$ \\[0.5cm]
\it ${}^a$Departamento de Fisica, UFJF, Juiz de Fora, MG, Brazil\\
 and\\
  \it ${}^b$Department of Theoretical Physics,\\
\it Tomsk State Pedagogical University,\\ \it Tomsk 634041,
Russia\footnote{Permanent
address}\\[0.3cm]
\it ${}^{c}$ Laboratory of Non-Linear Media Physics, Institute of
\\ \it Strength Physics and Materials Science, 634021 Tomsk, Russia}
\date{}

\maketitle 

\begin{abstract}
We construct a Lagrangian description of irreducible integer
higher-spin representations of the Poincare group with an
arbitrary Young tableaux having  $k$ rows, on a basis of the
universal BRST approach. Starting with a description of bosonic
mixed-symmetry higher-spin fields in a flat space of any dimension
in terms of an auxiliary Fock space associated with special
Poincare module, we realize a conversion of the initial operator
constraint system (constructed with respect to the relations
extracting irreducible Poincare-group representations) into a
first-class constraint system. For this purpose, we find, for the
first time, auxiliary representations of the constraint
subalgebra, to be isomorphic due to Howe duality to $sp(2k)$
algebra, and containing the subsystem of second-class constraints
in terms of new  oscillator variables. We propose a universal
procedure of constructing unconstrained gauge-invariant
Lagrangians with reducible gauge symmetries describing the
dynamics of both massless and massive bosonic fields of any spin.
It is shown that the space of BRST cohomologies with a vanishing
ghost number is determined only by the constraints corresponding
to an irreducible Poincare-group representation. As  examples of
the general procedure, we formulate the method of Lagrangian
construction for bosonic fields subject to arbitrary Young
tableaux having  $3$ rows and derive the gauge-invariant
Lagrangian for new model of massless rank-4  tensor field with
spin $(2,1,1)$ and second-stage reducible gauge symmetries.
\end{abstract}

\noindent {\sl Keywords:} \ Higher spin fields; Gauge theories;
Lagrangian formulation; BRST operator; BRST cohomology; Higher
spin symmetry algebra.

\section{Introduction}

Problems of higher-spin field theory attract a considerable
attention for a long time. Interest to higher spin field theory is
mainly stipulated by hope to construct the new types of Lagrangian
models in classical field theory and may be to formulate on this
ground the new approaches to the unification of the fundamental
interactions. Higher-spin field theory is closely related to
superstring theory, which operates with an infinite tower of bosonic
and fermionic higher-spin fields. From this point of view the higher
spin theory can be treated as an approach to study a structure of
superstring theory from field theory side. The aspects of current
state of higher spin field theory are discussed in the reviews
\cite{reviews}. Some newest tendencies in higher spin field theory
are discussed e.g. in \cite{Heslop}--\cite{Boulanger}.

At present, the Lagrangian dynamics of massless and massive free
totally symmetric higher-spin fields is well enough developed in
Minkowski and AdS spaces \cite{massless Minkowski}, \cite{massless
AdS1}, \cite{massless AdS}, \cite{massive Minkowski},
\cite{massive AdS}. It is known that in higher space-time
dimensions, there appear, besides totally symmetric irreducible
representations of Poincare or (A)dS algebras, the mixed-symmetry
representations determined by more than one spin-like parameters
\cite{Labastida}, \cite{Vasilievmix}, \cite{metsaevmixirrep}  and
the problem of their field-\-theoretic description is not so
well-studied as for totally symmetric representations. Indeed, a
simplest mixed-symmetry fields were considered in
\cite{CurtrightAulakh}, the problems of constrained (with imposing
off-shell traceless and Young symmetry constraints on all the
fields and gauge parameters) gauge-invariant description of
mixed-symmetry fields on flat space-time have been developed in
\cite{Siegel}. The aspects of lagrangian formulations for general
mixed-symmetry fields were discussed in \cite{Bekaert1}, the
unconstrained Lagrangians with higher derivatives for such
massless HS fields in "metric-like" formulation on flat background
were derived on a base of Bianchi identities resolution in
\cite{Franciamix}\footnote{In this paper a Lagrangian formulation
is constructed for massless fields corresponding to a reducible
Poincare group representation with arbitrary index symmetry.
Besides, it is suggested the Lagrangian formulation  which uses
the special projector's operators which are not constructed in
explicit form, although for every concrete case of
Poincare-irreducible HS fields  they can be written down. The
authors are grateful to D. Francia having drawn their attention on
the exact formulation of the paper \cite{Franciamix} results.}
 The details of Lagrangian description of  mixed-symmetry
 HS tensors on flat and (A)dS backgrounds in "frame-like"
 formulation were studied in \cite{AlkalaevVasiliev}.
The  interesting approach of derivation  the Labastida-like
\cite{Labastida} constrained Lagrangians  for arbitrary
mixed-symmetric higher spin fields were recently studied on a
basis of detour complexes from the BRST quantization of worldline
diffeomorphism invariant systems in \cite{WaldronBRST}.
  At present,
 the main result within the problem  of Lagrangian construction for
 arbitrary massless mixed-symmetry HS field on a Minkowski space-time of any dimensions
 was obtained in \cite{SkvortsovLF} with use of unfolded form of
 equations of motion for the field in "frame-like"
 formulation. In turn, the corresponding results for the
 mixed-symmetry bosonic fields  in "frame-like" formulation
with off-shell traceless constraints  in the case of
(anti)-de-Sitter case are recently known for the
   Young tableaux with two rows
 \cite{Zinoviev}.

The present work is devoted to the construction of general
gauge-invariant Lagrangians for both massless and massive
mixed-symmetry tensor fields of rank $s_1 + s_2 + ... + s_k$, with
any integer numbers $s_1 \geq s_2 \geq ... \geq s_k \geq 1$ for $k
\leq [d/2]$ in a $d$-dimensional Minkowski space, the fields being
elements of Poincare-group $ISO(1,d-1)$ irreps with a Young
tableaux having $k$ rows. Our approach is based on the BFV--BRST
construction \cite{BFV},
 which was
initially developed for a Hamiltonian quantization of dynamical
systems subject to first-class constraints and is called usually
the BRST construction. The application of the BRST construction to
free higher-spin field theory consists of three steps. First, the
conditions that determine the representations with a given spin
are regarded as a topological (i.e. without Hamiltonian) gauge
system of first- and second-class operator constraints in an
auxiliary Fock space. Second, the subsystem of the initial
constraints, which contains only second-class constraints, is
converted, with a preservation of the initial algebraic structure,
into a system of first-class constraints alone in an enlarged Fock
space (see \cite{conversion} for the development of conversion
methods), with respect to which one constructs the BRST charge.
Third, the Lagrangian for a higher-spin field is constructed in
terms of the BRST charge in such a way that the corresponding
equations of motion reproduce the initial constraints. We
emphasize that this approach automatically implies a
gauge-invariant Lagrangian description with all appropriate
auxiliary and Stuckelberg fields. The BRST approach to Lagrangian
formulation of higher spin field theories has been developed for
arbitrary massless and massive, bosonic and fermionic fields in
Minkowski and AdS spaces in \cite{BurdikPashnev}--\cite{0902BKRT}.
Inclusion of higher-spin field interactions requires a deformation
of BRST charge, the corresponding consideration is discussed in
\cite{brstinter}.

The various aspects of the flat dynamics of mixed-symmetry gauge
fields has been examined in \cite{Francia},  \cite{Zinoviev_m} for
massless bosonic higher-spin fields with two rows of the Young
tableaux \cite{BurdikPashnev}, and recently also for interacting
bosonic higher spin fields \cite{Metsaev-0}, \cite{Tsulaiaint},
\cite{Manvelyan} and for those of lower spins \cite{interlow},
\cite{Bizdadea} on the basis of the BV cohomological deformation
theory \cite{BarnichHenneaux1}. Lagrangian descriptions of
massless mixed-symmetry fermionic higher-spin fields in Minkowski
spaces have been suggested within a ``frame-like'' approach in
\cite{framefermimix}. To be complete, note that for free totally
symmetric higher-spin fields of integer spins the BRST approach
has been used to derive Lagrangians in the flat space
\cite{Pashnev1}, \cite{0505092} and in the (A)dS space
\cite{symint-ads}. The corresponding programme of a Lagrangian
description of fermionic higher-spin fields has been realized in
the flat space \cite{symferm-flat} and in the (A)dS space
\cite{symferm-ads}.

The paper is organized as follows. In Section~\ref{Symmalgebra},
we formulate a closed algebra of operators (using Howe duality),
based on the constraints in an auxiliary Fock space that
determines an massless irreducible representation of the Poincare
group in $\mathbb{R}^{1,d-1}$ with a generalized spin $\mathbf{s}
= (s_1, ..., s_k)$. In Section~\ref{Vermamodule}, we construct an
auxiliary representation for a rank-$k$ symplectic
$sp(2k)$ subalgebra  of an algebra of the initial constraints
corresponding to the subsystem of second-class constraints  in
terms of new (additional) creation and annihilation operators in
Fock space\footnote{Note that a similar construction for bosonic
HS fields subject to Young tableaux with 2 rows in a flat space
has been presented in \cite{0001195} for massless and in
\cite{BuchKrycRysTak} for massive case.}.
 As a result, we carry out a conversion of the initial system of first- and
second-class constraints into a system of first-class constraints
in the space being the tensor product of the initial and new Fock
spaces. Next, we construct the standard BRST operator for the
converted constraint algebra in Section~\ref{BRSToperator}. The
construction of an action and of a sequence of reducible gauge
transformations describing the propagation of a mixed-symmetry
bosonic field of an arbitrary spin is realized in
Section~\ref{LagrFormulation}. We show that the Lagrangian
description for a theory of a massive integer mixed-symmetry HS
field in a $d$-dimensional Minkowski space is deduced by
dimensional reduction of a massless HS field theory of the same
type in a $(d+1)$-dimensional flat space. Then, we sketch a proof
of the fact that the resulting action reproduces the correct
conditions for a field that determine an irreducible
representation of the Poincare group with a fixed
$\mathbf{s}=(s_1,...,s_k)$  spin.  In Section~\ref{examples}, we
show, that general procedure contains, first, earlier known
algorithm of Lagrangian construction  for bosonic fields subject
to Young tableaux with two rows, and, second, a new one for tensor
fields with three rows in the corresponding Young tableaux. Here
we construct the new unconstrained Lagrangian formulation for the
fourth rank massless tensor field with spin $(2, 1, 1)$ which has
not been obtained before.  In Conclusion, we summarize the results
of this work and outline some open problems. Finally, in
Appendix~\ref{addalgebra} we construct auxiliary representation
for $sp(2k)$ algebra. Appendix~\ref{oscrealsp2kdet} is devoted to
obtaining of a polynomial  representation of the operator algebra
given in Table~\ref{table in} in terms of creation and
annihilation operators. In Appendix~\ref{reductionC} we prove that
the constructed Lagrangian reproduces the correct conditions on
the field defining the irreducible representation of the Poincare
group and in Appendix~\ref{example211} the expressions for the
field and all gauge Fock space vectors are written in powers of
ghost creation operators to be used for Lagrangian construction
for the fourth rank tensor.

As a rule we use the conventions of Refs. \cite{BurdikPashnev,
BuchKrycRysTak, 0001195}.

\section{Derivation of Integer
HS Symmetry Algebra on $\mathbb{R}^{1,d-1}$ }\label{Symmalgebra}

In this section, we consider a massless integer spin irreducible
representation of Poincare group in a  Minkowski space
$\mathbb{R}^{1,d-1}$ which is described by a tensor field
$\Phi_{(\mu^1)_{s_1},(\mu^2)_{s_2},...,(\mu^k)_{s_k}}
\hspace{-0.2em}\equiv \hspace{-0.2em}
\Phi_{\mu^1_1\ldots\mu^1_{s_1},\mu^2_1\ldots\mu^2_{s_2},...,}$
${}_{ \mu^k_1\ldots \mu^k_{s_k}}(x)$
 of rank $\sum_{i\geq 1}^k s_i$ and generalized spin
 $\mathbf{s} = (s_1, s_2,
 ... , s_k)$, ($s_1 \geq s_2\geq ... \geq s_k>0, k
\leq [d/2])$ to be corresponding to a Young tableaux with $k$ rows
of length $s_1, s_2, ..., s_k$, respectively
\begin{equation}\label{Young k}
\Phi_{(\mu^1)_{s_1},(\mu^2)_{s_2},...,(\mu^k)_{s_k}}
\hspace{-0.3em}\longleftrightarrow \hspace{-0.3em}
\begin{array}{|c|c|c c c|c|c|c|c|c| c|}\hline
  \!\mu^1_1 \!&\! \mu^1_2\! & \cdot \ & \cdot \ & \cdot \ & \cdot\  & \cdot\  & \cdot\ &
  \cdot\    &\!\! \mu^1_{s_1}\!\! \\
   \hline
    \! \mu^2_1\! &\! \mu^2_2\! & \cdot\
   & \cdot\ & \cdot  & \cdot &  \cdot & \!\!\mu^2_{s_2}\!\!   \\
  \cline{1-8} \!\!\cdots\!\!   \\
   \cline{1-7}
    \! \mu^k_1\! &\! \mu^k_2\! & \cdot\
   & \cdot\ & \cdot  & \cdot &   \!\!\mu^k_{s_k}\!\!   \\
   \cline{1-7}
\end{array}\ ,
\end{equation}
This field is symmetric with respect to the permutations of each
type of Lorentz indices\footnotemark \footnotetext{Throughout the
paper, we use the mostly minus signature $\eta_{\mu\nu} = diag (+,
-,...,-)$, $\mu, \nu = 0,1,...,d-1$.}
 $\mu^i$,
  and
obeys to the Klein-Gordon (\ref{Eq-0b}), divergentless
(\ref{Eq-1b}), traceless (\ref{Eq-2b}) and mixed-symmetry
equations (\ref{Eq-3b}) [for $i,j=1,...,k;\, l_i,m_i=1,...,s_i$]:
\begin{eqnarray}
\label{Eq-0b} &&
\partial^\mu\partial_\mu\Phi_{(\mu^1)_{s_1},(\mu^2)_{s_2},...,(\mu^k)_{s_k}}
 =0,\\
&&\partial^{\mu^i_{l_i}}\Phi_{
(\mu^1)_{s_1},(\mu^2)_{s_2},...,(\mu^k)_{s_k}} =0,  \label{Eq-1b}
\\
&& \eta^{\mu^i_{l_i}\mu^i_{m_i}}\Phi_{
(\mu^1)_{s_1},(\mu^2)_{s_2},...,(\mu^k)_{s_k}}=
\eta^{\mu^i_{l_i}\mu^j_{m_j}}\Phi_{
(\mu^1)_{s_1},(\mu^2)_{s_2},...,(\mu^k)_{s_k}} =0, \quad
 l_i<m_i,  \label{Eq-2b}\\
&& \Phi_{
(\mu^1)_{s_1},...,\{(\mu^i)_{s_i}\underbrace{,...,\mu^j_{1}...}\mu^j_{l_j}\}...\mu^j_{s_j},...(\mu^k)_{s_k}}=0,\quad
i<j,\ 1\leq l_j\leq s_j, \label{Eq-3b}
\end{eqnarray}
where the  bracket below denote that the indices  in it do not
include in  symmetrization, i.e. the symmetrization concerns only
indices $(\mu^i)_{s_i}, \mu^j_{l_j} $ in
$\{(\mu^i)_{s_i}\underbrace{,...,\mu^j_{1}...}\mu^j_{l_j}\}$.

In order to describe all the irreducible representations
simultaneously, we introduce in an usual manner an auxiliary Fock
space $\mathcal{H}$, generated by $k$ pairs of bosonic (symmetric
basis) creation $a^i_{\mu^i}(x)$ and annihilation
$a^{j+}_{\nu^j}(x)$ operators, $i,j =1,...,k, \mu^i,\nu^j
=0,1...,d-1$:\footnote{there exists another realization of
auxiliary Fock space generated by the fermionic oscillators
(antisymmetric basis) $\hat{a}^m_{\mu^m}(x)$,
$\hat{a}^{\hat{n}+}_{\nu^n}(x)$ with anticommutation relations,
$\{\hat{a}^m_{\mu^m},
\hat{a}_{\nu^n}^{n+}\}=-\eta_{\mu^m\nu^n}\delta^{mn}$,   for $m, n
= 1,..., s_1$, and develop the  procedure below following to the
lines of Ref. \cite{brst1} for totally antisymmetric tensors for
$s_1=s_2=...=s_k=1$.}
\begin{eqnarray}\label{comrels}
[a^i_{\mu^i}, a_{\nu^j}^{j+}]=-\eta_{\mu^i\nu^j}\delta^{ij}\,,
\qquad \delta^{ij} = diag(1,1,\ldots 1)\,,
\end{eqnarray}
 and a set of constraints for an arbitrary string-like vector
$|\Phi\rangle \in \mathcal{H}$,
\begin{eqnarray}
\label{PhysState}  \hspace{-2ex}&& \hspace{-2ex} |\Phi\rangle  =
\sum_{s_1=0}^{\infty}\sum_{s_2=0}^{s_1}\cdots\sum_{s_k=0}^{s_{k-1}}
\Phi_{(\mu^1)_{s_1},(\mu^2)_{s_2},...,(\mu^k)_{s_k}}(x)\,
\prod_{i=1}^k\prod_{l_i=1}^{s_i} a^{+\mu^i_{l_i}}_i|0\rangle,\\
\label{l0} \hspace{-2ex}&& \hspace{-3ex} {{l}}_0|\Phi\rangle = 0 ,
\quad l_0 = \partial^\mu\partial_\mu,\\
\label{lilijt} \hspace{-2ex} && \hspace{-2ex} \bigl({l}^i, l^{ij},
t^{i_1j_1} \bigr)|\Phi\rangle  = \bigl(-i a^i_\mu \partial^\mu,
\textstyle\frac{1}{2}a^{i}_\mu a^{j\mu}, a^{i_1+}_\mu
a^{j_1\mu}\bigr) |\Phi\rangle=0,\ i\leq j;\, i_1 < j_1.
\end{eqnarray}
 The set of $(k(k+1)+1)$
primary constraints (\ref{l0}), (\ref{lilijt}) with $\{o_\alpha\}$
= $\bigl\{{{l}}_0, {l}^i, l^{ij}, t^{i_1j_1} \bigr\}$, because of
the property of translational invariance of the vacuum,
$\partial_\mu |0\rangle = 0$, are equivalent to Eqs.
(\ref{Eq-0b})--(\ref{Eq-3b}) for all spins. In turn, if we impose
in addition to the Eqs. (\ref{l0}), (\ref{lilijt}) the additional
constraints with number particles operators, $g_0^i$,
\begin{eqnarray}\label{g0iphys}
g_0^i|\Phi\rangle =(s_i+\frac{d}{2}) |\Phi\rangle, \qquad
 g_0^i = -\frac{1}{2}\{a^{i+}_\mu,  a^{\mu{}i}\},
\end{eqnarray}
where the sign $\{\ ,\ \}$ is  the anticommutator,  then these
combined conditions are equivalent to Eqs.
(\ref{Eq-0b})--(\ref{Eq-3b}) for the field
$\Phi_{(\mu^1)_{s_1},(\mu^2)_{s_2},...,(\mu^k)_{s_k}}(x)$ with
given spin $\mathbf{s} = (s_1, s_2,  ... , s_k)$.

 We refer to the vector (\ref{PhysState}) as the basic vector\footnote{We
 may consider a set of
 all finite string-like vectors for finite upper limits for $s_1$
 which  different choice of a spin $\mathbf{s}$ as the vector space  of
 polynomials $P_k^d(a^{+})$
 in degree $a^{+\mu^i}_i$. The Lorentz algebra on $P_k^d(a^{+})$
 is realized by means of action on it
 the Lorentz transformations, $M^{\mu\nu}=  \sum_{i\geq 1}^k a^{+[\mu}_i a^{
 \nu]i}$ with a standard rule
 $A^{[\mu}B^{\nu]}\equiv A^{\mu}B^{\nu}-A^{\nu}B^{\mu}$, therefore endowing $P_k^d(a^{+})$
 by the structure of Lorentz-module.}.

The procedure of Lagrangian formulation implies  the property of
BFV-BRST operator $Q$, $Q = C^\alpha o_\alpha + more$,  to be
Hermitian, that is equivalent to the requirements: $\{o_\alpha\}^+
= \{o_\alpha\}$ and closedness for $\{o_\alpha\}$ with respect to
the commutator multiplication $[\ ,\ ]$. It is evident, that the
set of $\{o_\alpha\}$ violates above conditions. To provide them
we consider in standard manner  an scalar product on
$\mathcal{H}$,
\begin{eqnarray}
\label{sproduct} \langle{\Psi}|\Phi\rangle & =  & \int
d^dx\sum_{s_1=0}^{\infty}\sum_{s_2=0}^{s_1}\cdots\sum_{s_k=0}^{s_{k-1}}
         \sum_{p_1=0}^{\infty}\sum_{p_2=0}^{p_1}\cdots\sum_{p_k=0}^{p_{k-1}}
\langle 0|\prod_{j=1}^k\prod_{m_j=1}^{p_j}
a^{\nu^j_{m_j}}_j\Psi^*_{(\nu^1)_{p_1},(\nu^2)_{p_2},...,(\nu^k)_{p_k}}(x)\nonumber\\
&& \times
\Phi_{(\mu^1)_{s_1},(\mu^2)_{s_2},...,(\mu^k)_{s_k}}(x)\,
\prod_{i=1}^k\prod_{l_i=1}^{s_i} a^{+\mu^i_{l_i}}_i|0\rangle .
\end{eqnarray}
As the result, the set of $\{o_\alpha\}$ extended by means of the
operators,
\begin{eqnarray} \label{lilijt+} \hspace{-2ex} && \hspace{-2ex} \bigl({l}^{i+},\
l^{ij+},\ t^{i_1j_1+} \bigr)  = \bigl(-i a^{i+}_\mu
\partial^\mu,\ \textstyle\frac{1}{2}a^{i+}_\mu a^{j\mu+},\
a^{i_1}_\mu a^{j_1\mu+}\bigr) ,\ i\leq j;\ i_1 < j_1,
\end{eqnarray}
is Hermitian, with taken into account of self-conjugated
operators, $(l_0^+,\ {g_0^i}^+) = (l_0,\ {g_0^i})$. It is rather
simple exercise  to see the second requirement is fulfilled as
well if the number particles operators $g_0^i$ will be included
into set of all constraints $o_I$ having therefore the structure,
\begin{eqnarray}
\{o_I\} = \{o_\alpha, o_\alpha^+;\ g_0^i\}\equiv \{o_a, o_a^+ ;\
l_0,\ l^i,\ l^{i+};\ g_0^i\}. \label{inconstraints}
\end{eqnarray}
Together the set $\{o_a, o_a^+\}$ in the Eq.
(\ref{inconstraints}), for $\{o_a\} = \{l^{ij}, t^{i_1j_1}\}$ and
the one $\{o_A\}= \{l_0,\ l^i,\ l^{i+}\}$, may be considered from
the Hamiltonian analysis of the dynamical systems
 as the operatorial respective $2k^2$ second-class and $(2k+1)$ first-class constraints subsystems among
$\{o_I\}$ for topological gauge system (i.e. with zero Hamiltonian
) because of,
\begin{eqnarray}
[o_a,\; o_b^+] = f^c_{ab} o_c +\Delta_{ab}(g_0^i),\ [o_A,\;o_B] =
f^C_{AB}o_C, \  [o_a,\; o_B] = f^C_{aB}o_C .
\label{inconstraintsd}
\end{eqnarray}
Here $f^c_{ab}, f^C_{AB}, f^C_{aB}$ are the antisymmetric with
respect to permutations of lower indices constant quantities and
quantities $\Delta_{ab}(g_0^i)$ form the non-degenerate $k^2\times
k^2$ matrix $\|\Delta_{ab}\|$ in the Fock space  $\mathcal{H}$ on
the surface $\Sigma \subset \mathcal{H}$:
$\|\Delta_{ab}\|_{|\Sigma} \ne 0 $, which is determined by the
equations, $(o_a, l_0,\ l^i)|\Phi\rangle = 0$. The set of $o_I$
contains the operators $g_0^i$ are not being by the constraints in
$\mathcal{H}$.

Explicitly, operators $o_I$ satisfy to the Lie-algebra commutation
relations,
\begin{equation}\label{geninalg}
    [o_I,\ o_J]= f^K_{IJ}o_K, \  f^K_{IJ}= - f^K_{JI},
\end{equation}
where the structure constants $f^K_{IJ}$ are used in the
Eq.(\ref{inconstraintsd}), included the constants
$f^{[g_0^i]}_{ab}: f^{[g_0^i]}_{ab}g_0^i \equiv
\Delta_{ab}^{[g_0^i]}(g_0^i)$ there and determined from the
multiplication table~\ref{table in}. \hspace{-1ex}{\begin{table}
{{\footnotesize
\begin{center}
\begin{tabular}{||c||c|c|c|c|c|c|c||c||}\hline\hline
$\hspace{-0.2em}[\; \downarrow, \rightarrow
]\hspace{-0.5em}$\hspace{-0.7em}&
 $t^{i_1j_1}$ & $t^+_{i_1j_1}$ &
$l_0$ & $l^i$ &$l^{i{}+}$ & $l^{i_1j_1}$ &$l^{i_1j_1{}+}$ &
$g^i_0$ \\
\hline\hline $t^{i_2j_2}$
    & $A^{i_2j_2, i_1j_1}$ & $B^{i_2j_2}{}_{i_1j_1}$
   & $0$&\hspace{-0.3em}
    $\hspace{-0.2em}l^{j_2}\delta^{i_2i}$\hspace{-0.5em} &
    \hspace{-0.3em}
    $-l^{i_2+}\delta^{j_2 i}$\hspace{-0.3em}
    &\hspace{-0.7em} $\hspace{-0.7em}l^{\{j_1j_2}\delta^{i_1\}i_2}
    \hspace{-0.9em}$ \hspace{-1.2em}& \hspace{-1.2em}$
    -l^{i_2\{i_1+}\delta^{j_1\}j_2}\hspace{-0.9em}$\hspace{-1.2em}& $F^{i_2j_2,i}$ \\
\hline $t^+_{i_2j_2}$
    & $-B^{i_1j_1}{}_{i_2j_2}$ & $A^+_{i_1j_1, i_2j_2}$
&$0$   & \hspace{-0.3em}
    $\hspace{-0.2em} l_{i_2}\delta^{i}_{j_2}$\hspace{-0.5em} &
    \hspace{-0.3em}
    $-l^+_{j_2}\delta^{i}_{i_2}$\hspace{-0.3em}
    & $l_{i_2}{}^{\{j_1}\delta^{i_1\}}_{j_2}$ & $-l_{j_2}{}^{\{j_1+}
    \delta^{i_1\}}_{i_2}$ & $-F_{i_2j_2}{}^{i+}$\\
\hline $l_0$
    & $0$ & $0$
& $0$   &
    $0$\hspace{-0.5em} & \hspace{-0.3em}
    $0$\hspace{-0.3em}
    & $0$ & $0$ & $0$ \\
\hline $l^j$
   & \hspace{-0.5em}$- l^{j_1}\delta^{i_1j}$ \hspace{-0.5em} &
   \hspace{-0.5em}$
   -l_{i_1}\delta_{j_1}^{j}$ \hspace{-0.9em}  & \hspace{-0.3em}$0$ \hspace{-0.3em} & $0$&
   \hspace{-0.3em}
   $l_0\delta^{ji}$\hspace{-0.3em}
    & $0$ & \hspace{-0.5em}$- \textstyle\frac{1}{2}l^{\{i_1+}\delta^{j_1\}j}$
    \hspace{-0.9em}&$l^j\delta^{ij}$  \\
\hline $l^{j+}$ & \hspace{-0.5em}$l^{i_1+}
   \delta^{j_1j}$\hspace{-0.7em} & \hspace{-0.7em}
   $l_{j_1}^+\delta_{i_1}^{j}$ \hspace{-1.0em} &
   $0$&\hspace{-0.3em}
      \hspace{-0.3em}
   $-l_0\delta^{ji}$\hspace{-0.3em}
    \hspace{-0.3em}
   &\hspace{-0.5em} $0$\hspace{-0.5em}
    &\hspace{-0.7em} $ \textstyle\frac{1}{2}l^{\{i_1}\delta^{j_1\}j}
    $\hspace{-0.7em} & $0$ &$-l^{j+}\delta^{ij}$  \\
\hline $l^{i_2j_2}$
    & \hspace{-0.3em}$\hspace{-0.4em}-l^{j_1\{j_2}\delta^{i_2\}i_1}\hspace{-0.5em}$
    \hspace{-0.5em} &\hspace{-0.5em} $\hspace{-0.4em}
    -l_{i_1}{}^{\{i_2+}\delta^{j_2\}}_{j_1}\hspace{-0.3em}$\hspace{-0.3em}
   & $0$&\hspace{-0.3em}
    $0$\hspace{-0.5em} & \hspace{-0.3em}
    $ \hspace{-0.7em}-\textstyle\frac{1}{2}l^{\{i_2}\delta^{j_2\}i}
    \hspace{-0.5em}$\hspace{-0.3em}
    & $0$ & \hspace{-0.7em}$\hspace{-0.3em}
L^{i_2j_2,i_1j_1}\hspace{-0.3em}$\hspace{-0.7em}& $\hspace{-0.7em}  l^{i\{i_2}\delta^{j_2\}i}\hspace{-0.7em}$\hspace{-0.7em} \\
\hline $l^{i_2j_2+}$
    & $ l^{i_1 \{i_2+}\delta^{j_2\}j_1}$ & $ l_{j_1}{}^{\{j_2+}
    \delta^{i_2\}}_{i_1}$
   & $0$&\hspace{-0.3em}
    $\hspace{-0.2em} \textstyle\frac{1}{2}l^{\{i_2+}\delta^{ij_2\}}$\hspace{-0.5em} & \hspace{-0.3em}
    $0$\hspace{-0.3em}
    & $-L^{i_1j_1,i_2j_2}$ & $0$ &$\hspace{-0.5em}  -l^{i\{i_2+}\delta^{j_2\}i}\hspace{-0.3em}$\hspace{-0.2em} \\
\hline\hline $g^j_0$
    & $-F^{i_1j_1,j}$ & $F_{i_1j_1}{}^{j+}$
   &$0$& \hspace{-0.3em}
    $\hspace{-0.2em}-l^i\delta^{ij}$\hspace{-0.5em} & \hspace{-0.3em}
    $l^{i+}\delta^{ij}$\hspace{-0.3em}
    & \hspace{-0.7em}$\hspace{-0.7em}  -l^{j\{i_1}\delta^{j_1\}j}\hspace{-0.7em}$\hspace{-0.7em} & $ l^{j\{i_1+}\delta^{j_1\}j}$&$0$ \\
   \hline\hline
\end{tabular}
\end{center}}} \vspace{-2ex}\caption{HS symmetry  algebra  $\mathcal{A}(Y(k),
\mathbb{R}^{1,d-1})$.\label{table in} }\end{table}

First note that,  in the table~\ref{table in}, the operators
$t^{i_2j_2}, t^+_{i_2j_2}$ satisfy by the definition the
properties
\begin{equation} \label{thetasymb} (t^{i_2j_2}, t^+_{i_2j_2}) \equiv
(t^{i_2j_2},t^+_{i_2j_2})\theta^{j_2i_2}, \ \theta^{j_2i_2} =
\left\{\begin{array}{cc}
               1, &  j_2>i_2\\
               0, & j_2 \leq i_2 \\
             \end{array}\right.
\end{equation}
with Heaviside $\theta$-symbol\footnote{there are no summation
with respect to the indices $i_2, j_2$ in the
Eqs.(\ref{thetasymb}), the figure brackets for the indices $i_1$,
$i_2$ in the quantity $A^{\{i_1}B^{i_2\}i_3}\theta^{i_3i_2\}}$
mean the symmetrization
 $A^{\{i_1}B^{i_2\}i_3}\theta^{i_3i_2\}}$ =
$A^{i_1}B^{i_2i_3}\theta^{i_3i_2}+
A^{i_2}B^{i_1i_3}\theta^{i_3i_1}$ as well as these indices are
raising and lowering by means of Euclidian metric tensors
$\delta^{ij}$, $\delta_{ij}$, $\delta^{i}_{j}$} $\theta^{ji}$.
Second,  the products $B^{i_2j_2}_{i_1j_1}, A^{i_2j_2, i_1j_1},
F^{i_1j_1,i}, L^{i_2j_2,i_1j_1}$ are determined by the explicit
relations,
\begin{eqnarray}
  {}B^{i_2j_2}{}_{i_1j_1} &=&
  (g_0^{i_2}-g_0^{j_2})\delta^{i_2}_{i_1}\delta^{j_2}_{j_1} +
  (t_{j_1}{}^{j_2}\theta^{j_2}{}_{j_1} + t^{j_2}{}^+_{j_1}\theta_{
  j_1}{}^{j_2})\delta^{i_2}_{i_1}
  -(t^+_{i_1}{}^{i_2}\theta^{i_2}{}_{i_1} + t^{i_2}{}_{i_1}\theta_{i_1}{}^{
  i_2})
  \delta^{j_2}_{j_1}
\,,\label{Bijkl}
\\
   A^{i_2j_2, i_1j_1} &=&  t^{i_1j_2}\delta^{i_2j_1}-
  t^{i_2j_1}\delta^{i_1j_2}  ,   \label{Aijkl}\\
   {} F^{i_2j_2,i} &=&
   t^{i_2j_2}(\delta^{j_2i}-\delta^{i_2i}),\label{Fijk} \\
  L^{i_2j_2,i_1j_1} &=&   \textstyle\frac{1}{4}\Bigl\{\delta^{i_2i_1}
\delta^{j_2j_1}\Bigl[2g_0^{i_2}\delta^{i_2j_2} + g_0^{i_2} +
g_0^{j_2}\Bigr]  - \delta^{j_2\{i_1}\Bigl[t^{j_1\}i_2}\theta^{i_2j_1\}} +t^{i_2j_1\}+}\theta^{j_1\}i_2}\Bigr] \nonumber \\
&& - \delta^{i_2\{i_1}\Bigl[t^{j_1\}j_2}\theta^{j_2j_1\}}
+t^{j_2j_1\}+}\theta^{j_1\}j_2}\Bigr] \Bigr\}
 \,.\label{Lklij}
\end{eqnarray}
They obeys  the obvious additional properties of antisymmetry and
Hermitian conjugation,
\begin{align}
  & A^{i_2j_2,  i_1j_1} = -A^{ i_1j_1, i_2j_2} &&
  {} A^+_{i_1j_1,  i_2j_2}=(A_{i_1j_1,  i_2j_2})^+ = t^+_{i_2j_1}\delta_{j_2i_1}
   -   t^+_{i_1j_2}\delta_{i_2j_1},\\
  & ({L^{i_2j_2,i_1j_1}})^+ =  L^{i_1j_1, i_2j_2} && {F^{i_2j_2,i}}^+
  =(F^{i_2j_2,i})^+= t^{i_2j_2+}(\delta^{j_2i}-\delta^{i_2i})
\end{align}
\vspace{-3,5ex}
\begin{equation}\label{Bijk+}
   {B^{i_2j_2}{}_{i_1j_1}}^+ = (g_0^{i_2}-g_0^{j_2})\delta^{i_2}_{i_1}
   \delta^{j_2}_{j_1} +
  (t^+_{j_1}{}^{j_2}\theta^{j_2}{}_{j_1} + t^{j_2}{}_{j_1}
  \theta_{j_1}{}^{j_2})\delta^{i_2}_{i_1}
  -(t_{i_1}{}^{i_2}\theta^{i_2}{}_{i_1} +
  t^{i_2+}{}_{i_1}\theta_{i_1}{}^{i_2})\delta^{j_2}_{j_1}.
\end{equation}

We call the algebra of these operators the \emph{integer
higher-spin symmetry algebra in Minkowski space with a Young
tableaux having $k$ rows}\footnote{one should not identify the
term "\emph{higher-spin symmetry algebra}" using here for free HS
formulation starting from our paper \cite{0505092} with the
algebraic structure known as "\emph{higher-spin algebra}" (see,
for instance Ref.\cite{Vasiliev_inter}) arising to describe the HS
interactions} and denote it as $\mathcal{A}(Y(k),
\mathbb{R}^{1,d-1})$.

From the table~\ref{table in} it is obvious that D'alambertian
$l_0$ being by the Casimir element of the Poincare algebra
$iso(1,d-1)$  belongs to the center of algebra $\mathcal{A}(Y(k),
\mathbb{R}^{1,d-1})$ as well.

Now, we are in position to describe shortly the structure of the
Lorentz-module $P^d_k(a^+)$ of all finite string-like vectors of
the form given by the Eq. (\ref{PhysState})  (see footnote~4)  on
a base of Howe duality \cite{Howe1}. The Howe dual algebra to
$so(1,d-1)$ is $sp(2k)$  if $k=\left[\frac{d}{2}\right]$ with the
following basis elements \cite{Howe1} for arbitrary $i,j =
1,...,k$,
\begin{equation}\label{basissp2n}
    \hat{l}_{ij} = a_{i{}\mu}^+a^{\mu+}_j,\qquad  \hat{t}_{i}{}^j = \frac{1}{2}\{a_{i{}\mu}^+,\;a^{j{}\mu }\},\qquad \hat{l}^{ij} =
    a_{{}\mu}^ia^{j\mu},
\end{equation}
which is distinguished from the elements of $\mathcal{A}(Y(k),
\mathbb{R}^{1,d-1})$ by the sign "hat". Their non-vanishing
commutator's relations have the form
\begin{align}
 & [\hat{t}_{i_1}{}^{j_1},\; \hat{t}_{i_2}{}^{j_2}] \ =\  \hat{t}_{i_1}{}^{j_2}
  \delta_{i_2}^{j_1} - \hat{t}_{i_2}{}^{j_1}
  \delta_{i_1}^{j_2},&&  [\hat{l}^{i_2{}j_2},\; \hat{l}_{i_1{}j_1}]\ =\ \delta^{\{i_2}_{\{i_1}\hat{t}_{j_1\}}{}^{j_2\}},  \nonumber \\
 & [\hat{t}_{i_1}{}^{j_1},\; \hat{l}_{i_2j_2}]  =  \hat{l}_{i_1\{j_2}\delta_{i_2\}}^{j_1} ,
 &&  [\hat{t}_{i_1}{}^{j_1},\; \hat{l}^{i_2j_2}]\ =\
 -\hat{l}^{j_1\{j_2}\delta^{i_2\}}_{i_1} .
  \label{comrelsp}
\end{align}
It is easy to see that  elements $l^{ij}, l^{ij+}, t^{i_1j_1},
t^+_{i_1j_1}, g_0^i$  from  HS symmetry algebra $\mathcal{A}(Y(k),
\mathbb{R}^{1,d-1})$ are derived from the basis elements of
$sp(2k)$ by the rules,
\begin{equation}\label{sp2nhssa}
    l_{ij}^+ = \frac{1}{2}\hat{l}_{ij}, \quad {l}^{ij} =
    \frac{1}{2}\hat{l}^{ij},\qquad
     {t}_{i}{}^j = \hat{t}_{i}{}^j\theta^{ji},\qquad  {{t}^{j}{}_{i}}^+{} = \hat{t}_{i}{}^j\theta^{ij},\quad
     g_0^i=-
     \hat{t}_{i}{}^i.
\end{equation}
The rest elements $\{l^i, l^{i+}, l_0\}$ of the algebra
$\mathcal{A}(Y(k), \mathbb{R}^{1,d-1})$  forms the subalgebra
which describes the isometries of Minkowski space $R^{1,d-1}$. It
may be realized as direct sum of $k$-dimensional commutative
algebra $T^k = \{l_i\}$ and its dual $T^{k*}=\{l^{i+}\}$,
\begin{equation}\label{TkTk}
    \{l^i, l^{i+}, l_0\} = (T^k \oplus T^{k*}\oplus [T^k,
    T^{k*}]),\quad [T^k,
    T^{k*}] \sim l_0,
\end{equation}
so that integer HS symmetry  algebra $\mathcal{A}(Y(k),
\mathbb{R}^{1,d-1})$ represents the semidirect sum of the
symplectic algebra $sp(2k)$ [as an algebra of internal derivations
of $(T^k \oplus T^{k*})]$ with $(T^k \oplus T^{k*}\oplus [T^k,
    T^{k*}])$\footnote{The construction of algebra  $\mathcal{A}(Y(k), \mathbb{R}^{1,d-1})$ in the
    Eq. (\ref{identalg}) is similar to the realization of the Poincare algebra
    $iso(1,d-1)$ by means of Lorentz algebra and Abelian subalgebra $T(1,d-1)$ of
    space-time translations which looks as follows, $iso(1,d-1) = T(1,d-1)+ \hspace{-1em} \supset  so(1,d-1)$.},
\begin{equation}\label{identalg}
    \mathcal{A}(Y(k), \mathbb{R}^{1,d-1}) = \left(T^k \oplus T^{k*}\oplus [T^k,
    T^{k*}]\right) + \hspace{-1em} \supset  sp(2k).
\end{equation}
Note, the elements $g_0^i$, form a basis in the Cartan subalgebra
whereas $l^{ij}, {t}_{i}{}^j$ are the basis of low-triangular
subalgebra in $sp(2k)$.

Having the identification (\ref{identalg}) we may conclude, the
integer spin finite-dimensional irreducible representations
 of the Lorentz algebra $so(1,d-1)$ subject to Young
tableaux $YT(k)$ realized on the tensor fields (\ref{Young k}) are
equivalently extracted by the annihilation of all elements from
$so(1,d-1)$-module $P^d_k(a^+)$ by the low-triangular subalgebra
of $sp(2k)$ along with the weight conditions given by the Eqs.
(\ref{g0iphys}) with respect to  its Cartan subalgebra which look
as follows,
\begin{equation}\label{Cartcond}
   l^{ij}|\Phi\rangle =0, \qquad {t}_{i}{}^j|\Phi\rangle=0,\qquad \hat{t}_i{}^i |\Phi\rangle \equiv - g_0^i|\Phi\rangle = -
    \left(s_i+\textstyle\frac{d}{2}\right)|\Phi\rangle.
\end{equation}
The integer spin finite-dimensional irreducible representations of
the Poincare algebra $iso(1,d-1)$ are easily obtain from ones for
Lorentz algebra by adding the conditions from $T^k$ Abelian
subalgebra and one for D'Alambertian,
\begin{equation}\label{addcond}
    {l}_i |\Phi\rangle = 0, \qquad  l_0|\Phi\rangle = 0.
\end{equation}
lifting the  set $P^d_k(a^+)$ to Poincare-module (another
realization for Poincare module from Lorentz module see in
ref.\cite{mg}).

Having constructed the HS symmetry algebra, we  can  not still
construct BRST operator $Q$ with respect to the elements $o_I$
from  $\mathcal{A}(Y(k), \mathbb{R}^{1,d-1})$ because of a
presence of the non-degenerate in the Fock space $\mathcal{H}$
operators $g_0^i$ determining following to the Eqs.
(\ref{inconstraints}) the system of $o_I$ as  one with
second-class constraints system.  Due to general property
\cite{BFV} of BFV- method a such BRST operator $Q$ would not
reproduce the right set of initial constraints (\ref{l0}),
(\ref{lilijt}) in the zero ghost $Q$-cohomology subspace of total
Hilbert space, $\mathcal{H}_{tot}$ ($\mathcal{H} \subset
\mathcal{H}_{tot}$). To resolve the problem, we consider the
procedure of conversion the set of $o_I$ into one of $O_I$ which
would be by first-class constraints only on the subspaces with
except for extended number particles operators $G_0^I$.

\section{Deformed  HS symmetry algebra for YT with $k$ rows}\label{Vermamodule}
\setcounter{equation}{0}

In this section, to solve the problem of conversion a set of $o_I$
 operators  we describe the method of auxiliary representation
construction for the symplectic algebra $sp(2k)$ with second-class
constraints alone, in terms of new creation and annihilation
operators from auxiliary Fock space over appropriate
Heisenberg--Weyl algebra and extend the latter to the case of
massive integer HS fields subject to the same Young tableaux
$Y(s_1,...,s_k)$.

\subsection{Note on an additive conversion for  $sp(2k)$ algebra}\label{addconvers}

We consider a standard variant of additive conversion procedure
developed within BRST approach, see for instance,
\cite{BurdikPashnev},  \cite{BuchKrycRysTak}, \cite{0001195} which
implies the enlarging of $o_I$ to $O_I = o_I + o'_I$, where
additional parts $o'_I$ are given on a new Fock space
$\mathcal{H}'$ being independent on $\mathcal{H}'$ in a sense that
$\mathcal{H}'\bigcap \mathcal{H} = \emptyset$. In this case the
elements $O_I$ are given on tensor product $\mathcal{H}\otimes
\mathcal{H}'$ so that the requirement for $O_I$ to be in
involution, i.e. $[O_I,\ O_J] \sim O_K$,  leads to the series of
algebraic relations,
\begin{align}\label{addrel}
&[o'_I,\ o'_J]= f^K_{IJ}o'_K, && [O_I,\ O_J]= f^K_{IJ}O_K
\end{align}
with the same structure constants $f^K_{IJ}$ as ones for  the
algebra determined by the Eqs.(\ref{geninalg}) for the original
set of $o_I$.

Because of only $sp(2k)$ generators are the second-class
constraints  in $\mathcal{A}(Y(k),\mathbb{R}^{1,d-1})$ to be
converted then instead of all $o'_I$ in (\ref{addrel}) it should
be used only part of them, namely $\{o'_a, {o'}^+_a\}$. Therefore,
one should to get new operator realization of  $sp(2k)$ algebra
$o'_I$. We will solve this problem by means of special procedure
known in the mathematical literature as Verma module construction
\cite{Dixmier} for the latter algebra which results explicitly
derived in the appendix~\ref{addalgebra}.

\subsection{Oscillator realization of the
additional parts to constraints}\label{oscrealsp2k}

Because of the algebra of the additional parts $o'_a, {o'}^+_a$ is
the same as one for original elements, i.e. given by the part of
the table~\ref{table in} which corresponds to the symplectic
algebra $sp(2k)$. One should be mentioned that in the case of the
algebra of mixed-symmetry HS fields
$\mathcal{A}(Y(2),\mathbb{R}^{1,d-1})$ the auxiliary
representation of its converted subalgebra $sp(4)$ (isomorphic to
$so(3,2)$ algebra used in \cite{BurdikPashnev}) of second-class
constraints was considered in \cite{BurdikPashnev}. In the more
general case of HS fields subject to Young tableaux with $k\geq 2$
rows we also use the prescription of the work \cite{Burdik}  to
transform special representation
 for $sp(2k)$ algebra (Verma module) in details derived in the
Appendix~\ref{addalgebra} to the oscillator form to be suitable
then for a BRST operator construction. As a result, we obtain
(after some calculations presented in the most part in the
Appendix~\ref{oscrealsp2kdet}), first for the number particles
operators and for ones $t^{\prime+}_{lm}$, $l^{\prime+}_{ij}$,
\begin{eqnarray}
g_0^{\prime i}& = &  \sum_{l\leq m}
 b_{lm}^+b_{lm}(\delta^{il}+\delta^{im}) + \sum_{r< s}d^+_{rs}d_{rs}(\delta^{is}-
 \delta^{ir}) +h^i
 \,,\label{g'0iF} \\
  t^{\prime+}_{lm}   & = & d^+_{lm} - \sum_{n=1}^{l-1}d_{nl}d^+_{nm}
   - \sum_{n=1}^{k}(1+\delta_{nl})b^+_{nm}b_{ln}\,,
 \label{t'+lmtext}
 \\
   l^{\prime+}_{ij} & = & b_{ij}^+\,.
 \label{l'+ijF}
\end{eqnarray}
In turn, for the elements  $l^{\prime }_{lm}$, separately for
$l=m$ and for $l<m$ which correspond to the primary constraints we
get (for $k_{-1}\equiv 1$)
\begin{eqnarray}
\label{l'lltext} l^{\prime }_{ll} &=&  \frac{1}{4}\sum_{n=1,n\neq
l}^{k}b^+_{nn}{b}^2_{ln} +  \frac{1}{2}\sum_{n=1}^{l-1}\Bigl[
\sum_{n'=n+1}^{k}(1+\delta_{n'l})b^+_{nn'}b_{n'l}\end{eqnarray}
\vspace{-3ex}
 \begin{eqnarray}
  && -
\sum_{p=0}^{l-n-1} \Big(\sum_{k_1=n+1}^{l-1}\ldots \sum_{k_p=n+p}^{l-1}\Big\{
 C^{k_{p}l}(d^+,d)- \sum_{n'=k_{p-1}}^{k_p-1}d^+_{n'k_p}d_{n' l} \Big\}\prod_{j=1}^{p}d_{k_{j-1}k_{j}} \Big) \Bigr]{b}_{nl}
 \nonumber\\
 && + \left(\sum_{n= l}^k b^+_{nl}b_{nl}  - \sum_{s>l}d^+_{ls}d_{ls}+\sum_{r<l}d^+_{rl}
 d_{rl} + h^{l}\right)b_{ll}\nonumber\\
 && - \frac{1}{2}\sum_{n=l+1}^{k}\Bigl[d^+_{ln} -
\sum\limits_{n'=1}^{l-1}d^+_{n'n}d_{n'l} -
\sum_{n'=n+1}^{k}(1+\delta_{n'l})b^+_{n'n}b_{n'l}
 \Bigr]{b}_{ln} ,\nonumber\\  \label{l'lmbose}
 l^{\prime }_{lm}&=&
 -
\frac{1}{4}\sum\limits_{n=1}^{m-1}(1+\delta_{nl}) \Bigl[
-\sum_{n'=n}^{k}(1+\delta_{n'm})
 b^+_{n'n}b_{n'm} \\
 && +
\sum_{p=0}^{m-n-1}\Big(\sum_{k_1=n+1}^{m-1}\ldots \sum_{k_p=n+p}^{m-1}\Big\{
 C^{k_{p}m}(d^+,d)- \sum_{n'=k_{p-1}}^{k_p-1}d^+_{n'k_p}d_{n' m} \Big\}\prod_{j=1}^{p}d_{k_{j-1}k_{j}}\Big)\Bigr]b_{nl}
 \nonumber\\
&& - \frac{1}{4}\textstyle\sum\limits_{n=m+1}^{k} \Bigl[ d^+_{mn}-
\sum\limits_{n'=1}^{m-1} d^+_{n'n}d_{n'm} -
\sum\limits_{n'=l+1}^{k}(1+\delta_{n'm})b^+_{n'n}b_{mn'}  \Bigr]{b}_{ln}\nonumber\\
&& + \frac{1}{4}\Bigl(\sum_{n=m}^kb^+_{ln}b_{ln} + \sum_{n=
l+1}^k(1+\delta_{nm})b^+_{nm} b_{nm}  - \sum_{s>l}d^+_{ls}d_{ls} -
\sum_{s>m}d^+_{ms}d_{ms}\nonumber\\
&& +\sum_{r<l}d^+_{rl}d_{rl} +\sum_{r<m}d^+_{rm}d_{rm} + h^{l}+
h^{m}\Bigr)b_{lm} \nonumber\\
&& -\frac{1}{4}  \sum\limits_{n=1}^{l-1} \Bigl[
\sum_{p=0}^{l-n-1} \Big(\sum_{k_1=n+1}^{l-1}\ldots \sum_{k_p=n+p}^{l-1}\Big\{
 C^{k_{p}l}(d^+,d)- \sum_{n'=k_{p-1}}^{k_p-1}d^+_{n'k_p}d_{n'l} \Big\}\prod_{j=1}^{p}d_{k_{j-1}k_{j}} \Big)\nonumber\\
 && -\sum_{n'=n+1}^{k}(1+\delta_{n'l})b^+_{n'n}b_{n'l}
 \Bigr]{b}_{nm} - \frac{1}{4}\textstyle\sum\limits_{n=l+1}^{k}(1+\delta_{nm})
\Bigl[ d^+_{ln} - \sum\limits_{n'=1}^{l-1}d^+_{n'n}d_{n'l}
\Bigr]{b}_{mn}\nonumber, \quad l<m.
 \end{eqnarray}
 At last, for the "mixed-symmetry" operators $t^{\prime }_{lm}$,
 we have,
 \begin{eqnarray}
t^{\prime }_{lm} &=& \sum_{p=0}^{m-l-1}\bigg[\sum_{k_1=l+1}^{m-1}\ldots \sum_{k_p=l+p}^{m-1}
 \Big\{C^{k_{p}m}(d^+,d)- \sum_{n'=k_{p-1}}^{k_p-1}d^+_{n'k_p}d_{n' m} \Big\}\prod_{j=1}^{p}d_{k_{j-1}k_{j}}\bigg]
 \label{t'lmF}\\
  && -\sum_{n=1}^{k}(1+\delta_{nm})b^+_{nl}
b_{nm}
 \,, \qquad k_0\equiv l. \nonumber
\end{eqnarray}
In the Eqs. (\ref{l'lltext})--(\ref{t'lmF}) the operators
$C^{lm}(d,d^+)$ is determined, for $l<m$, by the rule,
 \begin{eqnarray}
 \label{Clm}
C^{lm}(d^+,d)&\equiv &
\Bigl(h^{l}-h^{m}-\sum_{n=m+1}^{k}\bigl(d^+_{ln}d_{ln}+d^+_{mn}d_{mn}\bigl)+
\sum_{n=l+1}^{m-1}d^+_{nm}d_{nm}-d^+_{lm}d_{lm}\Bigr)d_{lm} \\
 &&  + \sum_{n=m+1}^{k}\Bigl\{d^+_{mn}  - \sum_{n'=l+1}^{m-1} d^+_{n'n} d_{n'm}\Bigr\}d_{ln}.\nonumber
  \end{eqnarray}
In the above expressions  quantities $h^i, i=1,...,k$ are the
arbitrary dimensionless constants whose values will be determined
later in the Sec. \ref{LagrFormulation} from a solution of a
special spectral problem. For constructing the additional parts
(\ref{g'0iF})--(\ref{t'lmF})  we have introduced new Fock space
$\mathcal{H}'$ generated by
 bosonic, $b^{+}_{ij}, d^+_{rs}, b_{ij},
d_{rs}$, $i,j,r,s =1,\ldots, k; i\leq j; r<s$, creation and
annihilation operators whose number is equal to ones of the
second-class constraints $o'_a, o^{\prime +}_a$ with the standard
(only nonvanishing) commutation relations
\begin{equation}\label{commrelations}
 [b_{i_1j_1}, b^+_{i_2j_2}] =
 \delta_{i_1i_2}\delta_{j_1j_2}\,, \   \qquad [d_{r_1s_1}\,,d^+_{r_2s_2}]
 =\delta_{r_1r_2}\delta_{s_1s_2}\,,
\end{equation}
 Note that the additional parts $o^{\prime}_a(B,B^+), o^{\prime+}_a(B,B^+)$ as the
 polynomials in the oscillator variables $(B,B^+)\equiv (b_{ij},
d_{rs}; b^{+}_{ij}, d^+_{rs})$ do not obey the usual properties
\begin{equation}
 \left(l^{\prime }_{lm}\right)^+\neq l^{\prime +}_{lm},\ l\leq m,  \qquad
\left(t^{\prime }_{rs}\right)^+\neq t^{\prime +}_{rs},\ r< s.
\label{hermcong}
\end{equation}
if one should use the standard rules of Hermitian conjugation for
the new creation and annihilation operators,
\begin{equation}
(b_{ij})^+=b_{ij}^+, \qquad (d_{rs})^+=d^+_{rs}.
\end{equation}
To restore the proper Hermitian conjugation properties for the
additional parts, we change the scalar product in the Fock space
$\mathcal{H}'$ as follows:
\begin{eqnarray}
\langle{\Phi}_1|\Phi_2\rangle_{\mathrm{new}} =
\langle{\Phi}_1|K'|\Phi_2\rangle\,, \label{newsprod}
\end{eqnarray}
for any vectors $|\Phi_1\rangle, |\Phi_2\rangle$ with some, yet
unknown, operator $K'$. This operator is determined by the
condition that all the operators of the algebra  must have the
proper Hermitian properties with respect to the new scalar
product:
\begin{align}
& \langle{\Phi}_1|K'E^{- \prime\alpha}|\Phi_2\rangle =
\langle{\Phi}_2|K'E^{\prime\alpha}|\Phi_1\rangle^* ,  &&
\langle{\Phi}_1|K'g_0^{\prime i}|\Phi_2\rangle =
\langle{\Phi}_2|K'g_0^{\prime i}|\Phi_1\rangle^*,
\end{align}
for   $(E^{\prime\alpha};E^{-\prime\alpha}) = (l^{\prime }_{lm},
t^{\prime }_{rs}; l^{\prime +}_{lm}, t^{\prime +}_{rs})$. These
relations permit one to determine the operator $K'$, Hermitian
with respect to the usual scalar product $\langle\, |\, \rangle$,
as follows:
\begin{eqnarray}
\label{explicit K} K'=Z^+Z, \qquad
Z=\sum_{(\vec{n}_{lm},\vec{p}_{rs})=(\vec{0},\vec{0})}^{\infty}
\left|\vec{N}\rangle_V\right.\frac{1}{(\vec{n}_{lm})!(\vec{p}_{rs})!}\langle
0|\prod_{r,s>r}{d}_{rs}^{p_{rs}}\prod_{l,m \geq
l}{b}_{lm}^{n_{lm}},
\end{eqnarray}
where $(\vec{n}_{lm})! =\prod^k_{l,m\geq l}{n}_{lm}!$,
$(\vec{p}_{rs})! = \prod^k_{r,s> r}{p}_{rs}!$ and a vector
$\left|\vec{N}\rangle_V\right.$ is determined  in the
Appendix~\ref{oscrealsp2kdet}. The detailed calculation of the
operator $K'$ is described there as well.

Now, we turn to the case of the massive  bosonic HS fields whose
system of second-class constraints contains additionally to
elements of $sp(2k)$ algebra  the constraints  of isometry
subalgebra of  Minkowski space  $l^i, l^+_i, l_0$.

\subsection{Auxiliary representations of the algebra
$\mathcal{A}(Y(k),\mathbb{R}^{1,d-1})$ for massive  HS
fields}\label{auxtheorem}

To construct an analogous oscillator representations for the HS
symmetry algebra of massive bosonic HS fields with mass $m$, where
the wave equation given by (\ref{Eq-0b}) should be changed on
Klein-Gordon equation corresponding to the constraint $l_0$
($l_0=\partial^\mu\partial_\mu +m^2$) acting on the same
string-vector $|\Phi\rangle$ (\ref{PhysState})
\begin{eqnarray}
\label{Eq-0bm} && (\partial^\mu\partial_\mu+
m^2)\Phi_{(\mu^1)_{s_1},(\mu^2)_{s_2},...,(\mu^k)_{s_k}}
 =0.
\end{eqnarray}
  we may to consider the procedure described above in
section~\ref{oscrealsp2k} and in details realized in the
Appendices~\ref{addalgebra},~\ref{oscrealsp2kdet}  for $sp(2k)$
algebra (see final comments in the Appendix~\ref{addalgebram} for
massive case).
    Instead,
we have used the procedure of the dimensional reduction of the
initial algebra $\mathcal{A}(Y(k),\mathbb{R}^{1,d})$ for massless
HS fields in $(d+1)$-dimensional flat background to one with
dimension $d$, $\mathbb{R}^{1,d-1}$.

To do so let's write down the rules of the dimensional reduction
from  $\mathbb{R}^{1,d}$ flat space-time to $\mathbb{R}^{1,d-1}$,
\begin{align} \label{reduction}
   &\partial^{M} = (\partial^{\mu}, -\imath m)\,, &&a^{M}_i = (a^{\mu}_i, b_i)\,, &&
   a^{M{}+}_i = (a^{\mu{}+}_i, b_i^+)\,,  \\
   &M=0,1,\ldots ,d\,, && \mu=0,1,\ldots ,d-1\,, && \eta^{MN} =
   diag (1,-1,\ldots,-1,-1)\,,\label{reduction1}
\end{align}
Doing so we obtain for the set of the original elements $o_I$ from
the massless HS symmetry algebra
$\mathcal{A}(Y(k),\mathbb{R}^{1,d})$ the ones $\tilde{o}_I$ in
massive HS symmetry algebra $\mathcal{A}(Y(k),\mathbb{R}^{1,d-1})$
 as follows,
\begin{align}
    &\tilde{l}_0 = \partial^{M}\partial_{M}=
l_0+ m^2, && \tilde{g}^i_{0} = - a^+_{Mi}a^{M}_i+\frac{d+1}{2} =
g_0^i + b^+_ib_{i} +\frac{1}{2},\label{l0tilde} \\
\label{litilde}& \tilde{l}_i = -ia^M_i\partial_{M}= {l}_i +mb_i,
&& \tilde{l}_i^+
= -ia^{+M}_i\partial_{M}= {l}^+_i + mb^+_i,\\
& \tilde{l}_{ij} = \frac{1}{2}a^M_ia_{Mj} = {l}_{ij} -
\frac{1}{2}b_ib_{j}, && \tilde{l}^+_{ij} =
\frac{1}{2}a^{M+}_ia^+_{Mj} = {l}^+_{ij} -
\frac{1}{2}b^+_ib^+_{j}, \\
& \tilde{t}_{ij} = a^{M+}_ia_{Mj}\theta^{ji} = {t}^+_{ij} -
b^+_{i}b_j\theta^{ji} , && \tilde{t}^+_{ij} =
a^{M}_ia^+_{Mj}\theta^{ji} = {t}^+_{ij} - b_{i}b^+_j\theta^{ji}.
\label{exprnew}
\end{align}
The generators $(\tilde{l}_0, {l}^+_i, {l}_i), {l}_{ij},
{l}^+_{ij}, {t}_{ij}, {t}^+_{ij}, g_0^i)$  satisfy the same
algebraic relations as in the table~\ref{table in} for massless HS
symmetry algebra with except for the commutators,
\begin{equation}\label{ll+}
    [l_i, l^+_j] = \delta_{ij}(\tilde{l}_0 - m^2).
\end{equation}
 Relations  (\ref{l0tilde}), (\ref{ll+}) indicate
the presence of $2k$ additional second-class constraints, $l_i,
l_i^+$, with corresponding oscillator operators $b_i, b_i^+$,
$[b_i, b_j^+] = \delta_{ij}$, in comparison with the massless
case.

It is interesting to see the elements with tilde in the
Eqs.(\ref{l0tilde})--(\ref{exprnew}) satisfy the algebraic
relations for massless HS symmetry algebra
$\mathcal{A}(Y(k),\mathbb{R}^{1,d-1})$ now without central charge
(i.e. those quantities $\tilde{o}_I$ contains the same
second-class constraints as ${o}_I$ in massless case). Therefore,
the converted constraints $O_I$, $O_I = o_I + o'_I$, in massive
case are given by the relations,
\begin{equation}\label{conv}
O_I = \tilde{o}_I + o'_I, \qquad M^2 = m^2+{m'}^2=0,
\end{equation}
where additional parts $o'_I = o'_I(b_{ij},b^+_{ij}, t_{i_1j_1},
t^+_{i_1j_1})$  are determined by the relations
(\ref{g'0iF})--(\ref{t'lmF}).

Thus, the auxiliary representation (Verma module) for  $sp(2k)$
algebra determines with use of the dimensional reduction procedure
the oscillator realization for the additional parts of massive HS
symmetry algebra $\mathcal{A}'(Y(k),\mathbb{R}^{1,d-1})$
completely.

In the next section, we determine the algebra of the extended
constraints and find the BRST operator corresponding to this
algebra.

\section{BRST-BFV operator}\label{BRSToperator}
According to our method we should find the BRST-BFV operator.
Since the algebra under consideration is a Lie algebra
$\mathcal{A}(Y(k),\mathbb{R}^{1,d-1})$ this operator can be
constructed in the standard way due to prescription
\cite{BFV}\footnote{Application of the BRST approach to HS field
theory in AdS space leads to problem of constructing a BRST-BFV
operator for non-linear (super)algebras, see e.g. \cite{bl,
resh}.}. First, we introduce the set of the ghost fields $C^I =
(\eta_0, \eta^i, \eta^+_i, \eta^{ij}, \eta^+_{ij}, \vartheta_{rs},
\vartheta^+_{rs}, \eta^i_{G})$ of the opposite Grassmann parity to
the elements $O_I = (L_0, L^+_i$, $L_i$, $L_{ij}, L^+_{ij},
T_{ij}, T^+_{ij}, G_0^i)$\footnote{for the massless HS fields the
elements $L_0, L^+_i, L_i$ coincide with $l_0, l^+_i, l_i$,
whereas for the massive case $L_0=\tilde{l}_0$, $L^+_i=
\tilde{l}^+_i+l^{\prime+}_i, L_i=\tilde{l}_i+l^{\prime}_i$ account
of the Eqs. (\ref{l0tilde}), (\ref{litilde})} with the properties
\begin{equation}\label{propgho}
    \eta^{ij}= \eta^{ji} , \eta^+_{ij}= \eta^+_{ji}, \vartheta_{rs}= \vartheta_{rs}\theta^{sr},
\vartheta^+_{rs} = \vartheta^+_{rs}\theta^{sr},
\end{equation}
 and their conjugated ghost
momenta $\mathcal{P}_I$ with the same properties as ones for $C^I$
in (\ref{propgho}) with the only nonvanishing anticommutation
relations
\begin{align}\label{ghosts}
& \{\vartheta_{rs},\lambda^+_{tu}\}= \{{\lambda_{tu}},
\vartheta_{rs}^+\}= \delta_{rt}\delta_{su}, &&
\{\eta_i,{\cal{}P}_j^+\}= \{{\cal{}P}_j, \eta_i^+\}=\delta_{ij}\,, \nonumber \\
& \{\eta_{lm},{\cal{}P}_{ij}^+\}= \{{\cal{}P}_{ij}, \eta_{lm}^+\}
=\delta_{li}\delta_{jm}\,,  &&  \{\eta_0,{\cal{}P}_0\}= \imath,\
\{\eta^i_{\mathcal{G}}, {\cal{}P}^j_{\mathcal{G}}\}
 = \imath\delta^{ij} \,;
\end{align}
They  also possess the  standard  ghost number distribution,
$gh(\mathcal{C}^I)$ = $ - gh(\mathcal{P}_I)$ = $1$, providing the
property  $gh({Q}')$ = $1$, and have the Hermitian conjugation
properties of zero-mode pairs,\footnote{By means of the
redefinition $\left( {\cal{}P}_0, {\cal{}P}^i_{G} \right) \mapsto
\imath \left( {\cal{}P}_0, {\cal{}P}^i_{G} \right)$, the BRST
operator (\ref{Q'k}) and relations (\ref{ghosts}) are written in
the notation of \cite{symferm-flat}, \cite{symferm-ads}.}
\begin{eqnarray}\label{Hermnull}
 \left( \eta_0, \eta^i_{{G}},  {\cal{}P}_0,
{\cal{}P}^i_{G} \right)^+ & = & \left( \eta_0, \eta^i_{G},  -
{\cal{}P}_0, -{\cal{}P}^i_{G}\right).
\end{eqnarray}
The BRST operator for the algebra of $O_I$ given by the
table~\ref{table in} can be found in an exact form, with the use
of the $(\mathcal{C} \mathcal{P})$-ordering of the ghost
coordinate $\mathcal{C}^I$ and momenta $\mathcal{P}_I$ operators,
as follows:
\begin{equation}\label{generalQ'}
    Q'  = {O}_I\mathcal{C}^I + \frac{1}{2}
    \mathcal{C}^I\mathcal{C}^Jf^K_{JI}\mathcal{P}_K
\end{equation}
Finally, we have
\begin{eqnarray}
\label{Q'k} {Q}' \hspace{-0.4em} &=&\hspace{-0.4em} \textstyle
 \frac{1}{2}\eta_0L_0+\eta_i^+L^i
+\sum\limits_{l\leq m}\eta_{lm}^+L^{lm} + \sum\limits_{l<
m}\vartheta^+_{lm}T^{lm} + \frac{1}{2}\eta^i_{{G}}{G}_i
 + \frac{\imath}{2}\sum_l\eta_l^+\eta^l{\cal{}P}_0
 \\\hspace{-0.4em}
&& {}\hspace{-0.4em} -\sum\limits_{i<l<j}
\vartheta^+_{lj}\vartheta^+_{i}{}^l \lambda^{ij}+
\textstyle\frac{\imath}{2}
\sum\limits_{l<m}\vartheta_{lm}^+\vartheta^{lm}({\cal P}_G^m-{\cal
P}_G^l)-
\sum\limits_{l<n<m}\vartheta_{lm}^+\vartheta^{l}{}_n\lambda^{nm}
\nonumber\\&& +
\sum\limits_{n<l<m}\vartheta_{lm}^+\vartheta_{n}{}^m\lambda^{+nl}
- \sum_{n,l<m}(1+\delta_{ln})\vartheta_{lm}^+\eta^{l+}{}_{n}
\mathcal{P}^{mn}+
\sum_{n,l<m}(1+\delta_{mn})\vartheta_{lm}^+\eta^{m}{}_{n}
\mathcal{P}^{+ln}\nonumber
\end{eqnarray}
\vspace{-3ex}
 \begin{eqnarray}
 &&  + \frac{\imath}{8}\sum_{l\leq
m}(1+\delta_{lm})
\eta_{lm}^+\eta^{lm}({\cal{}P}^l_{{G}}+{\cal{}P}^m_{{G}})+\frac{1}{2}\sum_{l\leq
m}(1+\delta_{lm})\eta^l_{{G}}\bigl(
\eta_{lm}^+{\cal{}P}^{lm}-\eta_{lm}{\cal{}P}^{lm+}\bigr)
\nonumber\\
\hspace{-0.4em} && \hspace{-0.4em} +
\textstyle\frac{1}{2}\sum\limits_{l<m, n\leq
m}\eta^+_{nm}\eta^{n}{}_l\lambda^{lm} +
\frac{1}{2}\sum\limits_{l<m}(\eta^m_{{G}}-
\eta^l_{{G}})\bigl(\vartheta^+_{lm}\lambda^{lm}- \vartheta_{lm}\lambda^{lm+}\bigr) \nonumber\\
&& - \bigl[\textstyle\frac{1}{2}\sum\limits_{l\leq
m}(1+\delta_{lm})\eta^m\eta_{lm}^+ +
\sum\limits_{l<m}\vartheta_{lm} \eta^{+m}
+\sum\limits_{m<l}\vartheta^+_{ml} \eta^{+m}  
\bigr]\mathcal{P}^l  \nonumber\\
&& +
\frac{1}{2}\sum_l\eta^l_{{G}}\bigl(\eta_l^+\mathcal{P}^l-\eta_l\mathcal{P}^{l+}\bigr)
+h.c.\nonumber
\end{eqnarray}
The property of the BRST operator to be Hermitian is defined by
the rule
\begin{eqnarray}\label{HermQ}
  Q^{\prime +}K = K Q'\,,
  \end{eqnarray}
and is calculated with respect to the scalar product $\langle \ |\
\rangle$ in $\mathcal{H}_{tot}$ with the measure $d^dx$, which, in
its turn, is constructed as the direct product of the scalar
products in $\mathcal{H}, \mathcal{H}'$ and $\mathcal{H}_{gh}$.
The operator $K$ in (\ref{HermQ}) is the tensor product of the
operator $K'$ in $\mathcal{H}'$ and the unit operators in
$\mathcal{H}$, $\mathcal{H}_{gh}$
\begin{eqnarray} \label{tK}
  K &=&  \hat{1} \otimes K' \otimes \hat{1}_{gh}\,.
\end{eqnarray}
 Thus, we have constructed a Hermitian BRST
operator for the entire algebra $\mathcal{A}_c(Y(k),
\mathbb{R}^{1,d-1})$ of $O_I$. In the next section, this operator
will be used to construct a Lagrangian action for bosonic HS
fields of spin $(s_1,..., s_k)$ in a flat space.

\section{Construction of Lagrangian Actions}\label{LagrFormulation}
\setcounter{equation}{0}

The construction of Lagrangians for bosonic  higher-spin fields in
a $d$-dimensional Minkowski  space can be developed by partially
following the algorithm of \cite{BurdikPashnev},
\cite{BuchKrycRysTak}, which is a particular case of our
construction, corresponding to $s_3 = 0$. As a first step, we
extract the dependence of the BRST operator $Q'$ (\ref{Q'k}) on
the ghosts $\eta^i_{G}, {\cal{}P}^i_{G}$, so as to obtain the BRST
operator $Q$ only for the system of converted first-class
constraints $\{O_I\} \setminus \{G^i_0\}$:

Due to special character of the number particles operators $G_0^i$
let's extract the terms proportional to the ghost variables
$\eta_G^i, \mathcal{P}_G^i$ from BRST operator $Q'$ (\ref{Q'k}),
\begin{eqnarray}
\label{Q'} {Q}' \hspace{-0.4em} &=& Q +
\eta^i_{G}(\sigma^i+h^i)+\mathcal{A}^i \mathcal{P}^i_{G}
\end{eqnarray}
where  the operator $Q$ corresponding only to the converted
first-class constraints determines as
\begin{eqnarray}
\label{Q} {Q} \hspace{-0.4em} &=&\hspace{-0.4em} \textstyle
 \frac{1}{2}\eta_0L_0+\eta_i^+L^i
+\sum\limits_{l\leq m}\eta_{lm}^+L^{lm} + \sum\limits_{l<
m}\vartheta^+_{lm}T^{lm}
 + \frac{\imath}{2}\sum_l\eta_l^+\eta^l{\cal{}P}_0
 \nonumber
\\\hspace{-0.4em}
&& {}\hspace{-0.4em} -\sum\limits_{i<l<j}
\vartheta^+_{lj}\vartheta^+_{i}{}^l \lambda^{ij}-
\sum\limits_{l<n<m}\vartheta_{lm}^+\vartheta^{l}{}_n\lambda^{nm} +
\sum\limits_{n<l<m}\vartheta_{lm}^+\vartheta_{n}{}^m\lambda^{+nl}
\nonumber\\&& -
\sum_{n,l<m}(1+\delta_{ln})\vartheta_{lm}^+\eta^{l+}{}_{n}
\mathcal{P}^{mn}+
\sum_{n,l<m}(1+\delta_{mn})\vartheta_{lm}^+\eta^{m}{}_{n}
\mathcal{P}^{+ln}+ \textstyle\frac{1}{2}\sum\limits_{l<m,n\leq
m}\eta^+_{nm}\eta^{n}{}_l\lambda^{lm}
\nonumber\\
\hspace{-0.4em} && \hspace{-0.4em}  
 - \bigl[\textstyle\frac{1}{2}\sum\limits_{l\leq
m}(1+\delta_{lm})\eta^m\eta_{lm}^+ +
\sum\limits_{l<m}\vartheta_{lm} \eta^{+m}
+\sum\limits_{m<l}\vartheta^+_{ml} \eta^{+m} \bigr]\mathcal{P}^l
+h.c.
\end{eqnarray}
The generalized spin operator $\vec{\sigma} =
(\sigma^1,\sigma^2,..., \sigma^k)$, extended by the ghost
Wick-pair variables, has the form
\begin{eqnarray}
\label{sigmai}
  \sigma^i &=& G_0^i - h^i   - \eta_i \mathcal{P}^+_i +
   \eta_i^+ \mathcal{P}_i + \sum_{
m}(1+\delta_{im})(
\eta_{im}^+{\cal{}P}^{im}-\eta_{im}{\cal{}P}^+_{im})\nonumber\\
   &&  + \sum_{l<i}[\vartheta^+_{li}
\lambda^{li} - \vartheta^{li}\lambda^+_{li}]-
\sum_{i<l}[\vartheta^+_{il} \lambda^{il} -
\vartheta^{il}\lambda^+_{il}]\,,
\end{eqnarray}
commutes with operator $Q$
\begin{eqnarray}
 {} [Q,\sigma^i]&=& 0\,,
\end{eqnarray}
whereas the operatorial quantities  $\mathcal{A}^i$ are uniquely
defined from (\ref{Q'k}) as follows
\begin{eqnarray}
\mathcal{A}^i  =
 - {\imath}
\sum\limits_{l<m}\vartheta_{lm}^+\vartheta^{lm}(\delta^{mi}-\delta^{li})
+ \frac{\imath}{4}\sum_{l\leq m}(1+\delta_{lm})
\eta_{lm}^+\eta^{lm}(\delta^{il}+\delta^{mi}).
\end{eqnarray}
We choose a representation of the Hilbert space given by the
relations
\begin{eqnarray}
(\eta_i, \eta_{ij},  \vartheta_{rs}, \mathcal{P}_0, \mathcal{P}_i,
\mathcal{P}_{ij}, \lambda_{rs},
\mathcal{P}^{i}_G)|0\rangle=0,\qquad |0\rangle\in
\mathcal{H}_{tot},
\end{eqnarray}
and suppose that the field vectors $|\chi \rangle$ as well as the
gauge parameters $|\Lambda \rangle$ do not depend on ghosts
$\eta^{i}_G$ for number particle operators $G_0^i$
\begin{eqnarray}
|\chi \rangle &=& \sum_n \prod_{l}^k ( b_l^+ )^{n_{l}}\prod_{i\le
j, r<s}^k( b_{ij}^+ )^{n_{ij}}( d_{rs}^+ )^{p_{rs}}( \eta_0^+
)^{n_{f 0}}\nonumber
\\
&&{}\times  \prod_{i, j, l\le m, n\le o}( \eta_i^+ )^{n_{f i}} (
\mathcal{P}_j^+ )^{n_{p j}} ( \eta_{lm}^+ )^{n_{f lm}} (
\mathcal{P}_{no}^+ )^{n_{pno}} \prod_{r<s, t<u}(
\vartheta_{rs}^+)^{n_{f rs}} ( \lambda_{tu}^+ )^{n_{\lambda tu}}
\nonumber
\\
&&{}\times |\Phi(a^+_i)^{n_{f 0} (n)_{f i}(n)_{p j}(n)_{f lm}
(n)_{pno}(n)_{f rs}(n)_{\lambda
tu}}_{(n)_{l}(n)_{ij}(p)_{rs}}\rangle \,. \label{chi}
\end{eqnarray}
The brackets $(n)_{f i},(n)_{p j}, (n)_{ij}$ in definition of
(\ref{chi}) means, for instance, for $(n)_{ij}$ the set of indices
$(n_{11},...,n_{1k},..., n_{k1},..., n_{kk})$. The  sum above is
taken over $n_{l}$, $n_{ij}$, $p_{rs}$ and  running from $0$ to
infinity, and over the rest $n$'s from $0$ to $1$\footnote{for the
massless basic HS field $\Phi_{(\mu^1)_{s_1}...(\mu^k)_{s_k}}$
there are no operators $b^+_l$ in the decomposition (\ref{chi}),
i.e. indices $(n)_{l} = (0)_{l}$.}. Let us denote by
$|\chi^k\rangle$ the state (\ref{chi}) with the ghost number $-k$,
i.e. $gh(|\chi^k\rangle)=-k$. Thus, the physical state having the
ghost number zero is $|\chi^0\rangle$, the gauge parameters
$|\Lambda \rangle$ having the ghost number $-1$ is
$|\chi^1\rangle$ and so on. Moreover for vanishing of all
auxiliary creation operators $b^+, d^+$ and ghost variables
$\eta_0, \eta^+_i, \mathcal{P}^+_i,...$ the vector
$|\chi^0\rangle$ must contain only physical string-like vector
$|\Phi\rangle = |\Phi(a^+_i)^{(0)_{f o} (0)_{f i}(0)_{p j}(0)_{f
lm} (0)_{pno}(0)_{f rs}(0)_{\lambda tu}}_{(0)_l
(0)_{ij}(0)_{rs}}\rangle$, so that
\begin{eqnarray}\label{decomptot}
|\chi^0\rangle&=&|\Phi\rangle+  |\Phi_A\rangle ,
\end{eqnarray}
where the vector $|\Phi_A\rangle$ includes only the auxiliary
fields as its components. One can show, using the part of
equations of motion and gauge transformations, that the vector
$|\Phi_A\rangle$ can be completely removed. The details are given
in Appendix~\ref{reductionC}.

Since the vectors (\ref{chi}) do not depend on $\eta^{i}_G$ the
equation for the physical state ${Q}'|\chi^0\rangle=0$ and the
tower of the reducible gauge transformations, $\delta|\chi\rangle$
= $Q'|\chi^1\rangle$, $\delta|\chi^1\rangle = Q'|\chi^2\rangle$,
$\ldots$, $\delta|\chi^{(s-1)}\rangle = Q'|\chi^{(s)}\rangle$,
lead to a sequence of relations:
\begin{align}
\label{Qchi} & Q|\chi\rangle=0, && (\sigma^i+h^i)|\chi\rangle=0,
&& \left(\varepsilon, {gh}\right)(|\chi\rangle)=(0,0),
\\
& \delta|\chi\rangle=Q|\chi^1\rangle, &&
(\sigma^i+h^i)|\chi^1\rangle=0, && \left(\varepsilon,
{gh}\right)(|\chi^1\rangle)=(1,-1), \label{QLambda}
\\
& \delta|\chi^1\rangle=Q|\chi^2\rangle, &&
(\sigma^i+h^i)|\chi^2\rangle=0, && \left(\varepsilon,
{gh}\right)(|\chi^2\rangle)=(0,-2),\\
& \ldots && \ldots && \ldots \nonumber\\
  &
\delta|\chi^{s-1}\rangle=Q|\chi^{s}\rangle, &&
(\sigma^i+h^i)|\chi^{s}\rangle=0, && \left(\varepsilon,
{gh}\right)(|\chi^{s}\rangle)= (s\, mod\, 2 ,-s). \label{QLambdai}
\end{align}
Where $s=k(k+1)-1$ is the stage of reducibility both for massless
and for the massive bosonic HS field, because of the only
non-vanishing vector (independent gauge parameter
$|\chi^{k(k+1)}\rangle$) has lowest negative ghost number when all
the anticommuting ghost momenta $\mathcal{P}_I$  compose
$|\chi^{k(k+1)}\rangle$ without presence of the ghost coordinates
$\mathcal{C}^I$ in it. The middle set of equations in
(\ref{Qchi})--(\ref{QLambdai}) determines the possible values of
the parameters $h^i$ and the eigenvectors of the operators
$\sigma^i$. Solving these equations, we obtain a set of
eigenvectors, $|\chi^0\rangle_{(n)_k}$, $|\chi^1\rangle_{(n)_k}$,
$\ldots$, $|\chi^{s}\rangle_{(n)_k}$, $n_1 \geq n_2 \geq \ldots
n_k \geq 0$, and a set of eigenvalues,
\begin{eqnarray}
\label{hi} -h^i &=& n^i+\frac{d-2-4i}{2} \;, \quad
i=1,..,k\,,\quad n_1,...,n_{k-1} \in \mathbb{Z}, n_k \in
\mathbb{N}_0\,,
\end{eqnarray}
for massless  and
\begin{eqnarray}
\label{him} -h^i_m &=& n^i+\frac{d-1-4i}{2} \;, \quad
i=1,..,k\,,\quad n_1,...,n_{k-1} \in \mathbb{Z}, n_k \in
\mathbb{N}_0\,,
\end{eqnarray}
for massive HS fields. The values of $n_i$ are related to the spin
components $s_i$  of the field, because of the proper vector
$|\chi\rangle_{(s_1,...,s_k)}$ corresponding to $(h_1,...,h_k)$
has the leading term $$|\Phi(a^+_i)^{0_{f o} (0)_{f i}(0)_{p
j}(0)_{f lm} (0)_{pno}(0)_{f rs}(0)_{\lambda tu}}_{(0)_l
(0)_{ij}(0)_{rs}}\rangle,$$ independent of auxiliary and ghost
operators, which corresponds to the field
$\Phi_{(\mu^1)_{s_1},...,(\mu^k)_{s_k}}(x)$ with the initial value
of spin $\mathbf{s} = (s_1,...,s_k)$ in the decomposition
(\ref{chi}) and representation (\ref{decomptot}). Let us denote
the eigenvectors of $\sigma_i$ corresponding to the eigenvalues
$(n^i+\frac{d-2-4i}{2})$ as $|\chi\rangle_{(n)_k}$. Thus we may
write
\begin{eqnarray}
\sigma_i | \chi \rangle_{(n)_k} = \left( n^i+\frac{d-2-4i}{2}
\right) | \chi \rangle_{(n)_k} \label{state} \,.
\end{eqnarray}
for massless HS fields and
\begin{eqnarray}
\sigma_i | \chi \rangle_{(n)_k} = \left( n^i+\frac{d-1-4i}{2}
\right) | \chi \rangle_{(n)_k} \label{statem} \,.
\end{eqnarray}
for massive ones.  Then one can show that in order to construct
Lagrangian for the field corresponding to a definite Young tableau
(\ref{Young k}) the numbers $n_i$ must be equal to the numbers of
the boxes in the $i$-th row of the corresponding Young tableau,
i.e. $n_i=s_i$. Thus the state $|\chi\rangle_{(s)_k}$ contains the
physical field (\ref{PhysState}) and all its auxiliary fields. Let
us fix some values of $n_i=s_i$. Then one should substitute $h_i$
corresponding to the chosen $s_i$ (\ref{hi}) or (\ref{him}) into
(\ref{Q'k}), (\ref{Qchi})--(\ref{QLambdai}). Thus, the equation of
motion (\ref{Qchi}) corresponding to the field with given spin
$(s_1,...,s_k)$ has the form
\begin{eqnarray}
Q_{(n)_k}|\chi^0\rangle_{(n)_k}=0. \label{Q12}
\end{eqnarray}

Since the BRST-BFV operator ${Q}'$ is nilpotent (\ref{Q'k}) at any
values of $h_i$ we have a sequence of reducible gauge
transformations
\begin{eqnarray}
\label{dx0} \delta|\chi^0 \rangle_{(n)_k}
=Q_{(n)_k}|\chi^{1}\rangle_{(n)_k} \,, &\qquad&
\delta|\chi^{1}\rangle_{(n)_k} =Q_{(n)_k}|\chi^{2}\rangle_{(n)_k}
\,,
\\
\ldots &\ldots& \ldots
\nonumber \\
\label{dxs} \delta|\chi^{k(k+1)-1} \rangle_{(n)_k}
=Q_{(n)_k}|\chi^{k(k+1)}\rangle_{(n)_k} \,, &\qquad&
\delta|\chi^{k(k+1)} \rangle_{(n)_k} =0 \,.
\end{eqnarray}
One can show that $Q_{(n)_k}$ is nilpotent when acting  on $|\chi
\rangle_{(n)_k}$
\begin{eqnarray}
Q_{(n)_k}^2|\chi \rangle_{(n)_k}\equiv0.
\end{eqnarray}
Thus we have obtained equation of motion (\ref{Q12}) of arbitrary
integer spin gauge theory subject to $YT(s_1,...,s_k)$ with mixed
symmetry in any space-time dimension and its tower of reducible
gauge transformations (\ref{dx0})--(\ref{dxs}).

We next find a corresponding Lagrangian. Analogously to the
bosonic one and two row cases \cite{0505092},
\cite{BurdikPashnev}, \cite{BuchKrycRysTak} one can show that
Lagrangian action for fixed spin $(n)_k=(s)_k$ is defined up to an
overall factor as follows
\begin{eqnarray}
\mathcal{S}_{(s)_k} = \int d \eta_0 \; {}_{(s)_k}\langle \chi^0
|K_{(s)_k} Q_{(s)_k}| \chi^0 \rangle_{(s)_k} \label{Scl}
\end{eqnarray}
where the standard scalar product for the creation and
annihilation operators is assumed with measure $d^dx$ over
Minkowski space. The vector $| \chi^0 \rangle_{(s)_k}$   and the
operator $K_{(s)_k}$ in (\ref{Scl}) are  respectively the vector
$| \chi \rangle$ (\ref{chi}) subject to spin distribution
relations (\ref{state}) for massless or (\ref{statem}) for massive
HS tensor field $\Phi_{(\mu^1)_{s_1},...,(\mu^k)_{s_k}}(x)$ with
vanishing value of ghost number and operator $K$ (\ref{tK}) where
the following substitution is done
$h_i\to-(n_i+\frac{d-2-4i+\theta(m)}{2})$, with $\theta$-function
$\theta(x)$, to be equal 1  for $x>0$ and vanishing in other case.
The former choice corresponds to a theory of massive HS bosonic
field whereas the latter choice (for $\theta(m)=0$) to a theory of
massless HS bosonic field.

One can prove that the Lagrangian action (\ref{Scl}) indeed
reproduces the basic conditions (\ref{Eq-0b})--(\ref{Eq-3b}) for
massless and (\ref{Eq-0bm}), (\ref{Eq-1b})--(\ref{Eq-3b}) for
massive HS fields. Such a proof is a test of correctness of the
approach under consideration. The details of the proof are given
in Appendix~\ref{reductionC}. Relations (\ref{dx0}), (\ref{dxs})
and (\ref{Scl}) are our final results. These results provide a
complete solution of general problem of Lagrangian construction
for arbitrary massless and massive higher spin bosonic fields with
indices corresponding to arbitrary Young tableaux.  General action
(\ref{Scl}) gives, in principle, a straight recept to obtain the
Lagrangian for any component field from general vector $| \chi^0
\rangle_{(s)_k}$ since the only what we should do, if to use
quantum mechanical terminology, is computations of vacuum
expectation values of products of some number of creation and
annihilation operators.

Indeed, the general expressions (\ref{dx0}), (\ref{dxs}),
(\ref{Scl}) for Lagrangian formulation can be presented in the
component form without presence of operatorial variables for any
concrete initial tensor field
$\Phi_{(\mu^1)_{s_1},...,(\mu^k)_{s_k}}(x)$, analogously to the
particular examples considered for totally-symmetric (for $k=1$)
 massless \cite{0001195} and massive \cite{0505092},
for mixed-symmetry (for $k=2$) massless \cite{BurdikPashnev} and
massive \cite{BuchKrycRysTak} bosonic HS fields on an arbitrary
flat background. To obtain the result, we have to derive component
Lagrangian formulation for a given HS tensor field  with   $k>2$
rows in the corresponding Young tableaux in terms of only initial
and auxiliary tensor fields having a decomposition (\ref{chi}) by
means of a standard calculation of the scalar products with
accurate treatment of the action of all ghost,  creation (initial
$a^{i+}_{\mu_i}$, auxiliary $b^+_i, b^+_{ij}, d^+_{rs}$) and
annihilation ($a^{i}_{\mu_i}$, $b_i, b_{ij}, d_{rs}$)  operators
composing the vectors $| \chi^0 \rangle_{(s)_k}$, $\ldots$,
$|\chi^{k(k+1)}\rangle_{(s)_k}$ and operators $K_{(s)_k}$,
$Q_{(s)_k}$ in the Eqs.(\ref{dx0}), (\ref{dxs}), (\ref{Scl}). This
analysis shows that there exists a possibility to obtain the
component Lagrangians from the general action (\ref{Scl}).
However, it is worth emphasizing that such component Lagrangians
will look complicate enough and cumbersome and we do not see
necessity to write down it here since all general property can be
studied and understood on the base of action (\ref{Scl}).

Construction of the Lagrangians describing propagation of all
massless or of all massive bosonic fields in flat space
simultaneously is analogous to that for the totally-symmetric case
\cite{0505092} and we do not consider it here.

In what follows, we consider some examples of the Lagrangian
formulation procedure.

\section{Examples}\label{examples}
\setcounter{equation}{0}

Here, we shall realize the general prescriptions of our Lagrangian
formulation in the case of mixed-symmetry bosonic fields of lowest
value of rows and spins.

\subsection{Spin-$(s_1,s_2)$ mixed-symmetric field}
Let us consider the mixed-symmetric field with two families of
indices corresponding to spin-$(s_1,s_2)$. In this case we expect
that our result will be reduced to that considered for massless
case in \cite{BurdikPashnev} and for massive in
\cite{BuchKrycRysTak}, where respectively the mixed-symmetric
massless and massive bosonic fields subject to $Y(s_1,s_2)$ were
considered. According to our procedure we have
$(n_1,n_2)=(s_1,s_2)$, $n_i=0$, for $i=3,\ldots , k$. One can show
that if given $n_i = 0$ then in (\ref{PhysState}) and (\ref{chi})
all the components related with the  rows $i \geq 3 $ in the Young
tableaux must be vanish, i.e.
\begin{eqnarray}
n_{l} &=& n_{1j}=n_{2m}= p_{1s}=p_{2t}=n_{f i}=n_{p j}= n_{f
1j}=n_{f 2m}=n_{p 1j}=n_{p 2o}\nonumber \\
&=& n_{f 1s}=n_{f 2t}=n_{\lambda 1s}=n_{\lambda 2t} =n_{s_m}
=0,\texttt{ for }l, j, m, s, t, i, o
>2.
\end{eqnarray}
Thus the state vector is reduced to
\begin{eqnarray}
&& |\chi \rangle \ =\ \sum_n \prod_{l}^2 ( b_l^+
)^{n_{l}}\prod_{i\le j}^2( b_{ij}^+ )^{n_{ij}}( d_{12}^+
)^{p_{12}}( \eta_0^+ )^{n_{f 0}}\prod_{i, j, l\le m, n\le o}^2(
\eta_i^+ )^{n_{f i}} ( \mathcal{P}_j^+ )^{n_{p j}} ( \eta_{lm}^+
)^{n_{f lm}} ( \mathcal{P}_{no}^+ )^{n_{pno}} \nonumber
\\
&&{}\qquad\quad \times ( \vartheta_{12}^+)^{n_{f 12}} (
\lambda_{12}^+ )^{n_{\lambda 12}}|\Phi(a^+_1,a^+_2)^{n_{f 0}
(n)_{f i}(n)_{p j}(n)_{f lm} (n)_{pno} n_{f 12}n_{\lambda
12}}_{(n)_{l}(n)_{ij}p_{12}}\rangle\,, \label{x2}\\
&& {}|\Phi(a^+_1,a^+_2)^{n_{f 0} \ldots n_{\lambda
12}}_{(n)_{l}(n)_{ij} p_{12}}\rangle \ = \
\sum_{p_1=0}^{\infty}\sum_{p_2=0}^{p_1}
\Phi_{(\mu^1)_{p_1},(\mu^2)_{p_2}, (0)_{s_3}...,(0)_{s_k}}(x)\,
\prod_{l_1=1}^{p_1}a^{+\mu^1_{l_1}}_1\prod_{l_2=1}^{p_2}
a^{+\mu^2_{l_2}}_2|0\rangle\label{x2cl}
\end{eqnarray}
which corresponds to that in \cite{BuchKrycRysTak} and for
$(n)_{l}=(0)_{l}$ to \cite{BurdikPashnev}. The  operator
$C^{lm}(d^+,d)$ in (\ref{Clm}) now has the only non-vanishing
value for $C^{12}(d^+,d)$,
\begin{eqnarray}
 \label{C122}
C^{12}(d^+,d) &\equiv &
\bigl(h^{1}-h^{2}-d^+_{12}d_{12}\bigr)d_{12},
 \end{eqnarray}
 so that the expression for $sp(4)$ algebra auxiliary representation
  may be
 easily derived  from the Eqs.(\ref{g'0iF})--(\ref{t'lmF}).
 Then one can easily show that equations (\ref{Q12}),
(\ref{dx0}), (\ref{dxs}), (\ref{Scl}) with $|\chi\rangle$ as in
(\ref{x2}), (\ref{x2cl}) reproduce the same relations as those in
\cite{BurdikPashnev} for massless and in \cite{BuchKrycRysTak} for
massive case.

\subsection{Spin-$(s_1,s_2,s_3)$ general mixed-symnmetric field}

Now we consider a new yet unknown Lagrangian formulation for
mixed-symmetric HS field
$\Phi_{(\mu^1)_{s_1},(\mu^2)_{s_2},(\mu^3)_{s_3}}$ with three
group of symmetric indices subject to $Y(s_1,s_2,s_3)$. The values
of spin $(s_1,s_2,s_3)$, for $s_1\geq s_2\geq s_3$, can be
composed from the set of coefficients $(n_l,n_{ij}, p_{rs},
n_{f{}0}$, $n_{f{}i}$, $n_{p{}j},n_{f{}lm}, n_{p{}no}$, $n_{f{}rs},
n_{\lambda{}tu}, p_i)$, for $l,i,j,r,s,l,m,n,o, t,u =1,2,3, \
i\leq j, r<s, l\leq m, n\leq o$, $t<u$, in (\ref{chi}) and
(\ref{PhysState})\footnote{we change the indices $s_i$ given in
(\ref{PhysState}) for the vector in initial Fock space
$\mathcal{H}$ on $p_i$ because of the usage of $s_i$ for the value
of generalized spin of the basic HS field
$\Phi_{(\mu^1)_{s_1},(\mu^2)_{s_2},(\mu^3)_{s_3}}$} to be
restricted for all the vectors $| \chi^l \rangle_{(s)_3}$,
$l=0,\ldots, 12$ in view of the spectral problem solution
(\ref{state}), (\ref{statem}) by the formulae
\begin{eqnarray}\label{nidecompos}
s_i &= & p_i+ \Theta(m)n_{i} +
\sum_{j=1}(1+\delta_{ij})(n_{ij}+n_{f{}ij}+n_{p{}ij} )
+n_{f{}i}+n_{p{}i} \nonumber \\
&{}& +\sum_{r<i} (p_{ri} + n_{f{}ri}+ n_{\lambda{}ri}) -\sum_{r>i}
(p_{ir}+ n_{f{}ir}+ n_{\lambda{}ir} )\,,\  i=1,2,3.
\end{eqnarray}
In addition to  restrictions (\ref{nidecompos}),  being valid
for general case of HS field subject to $Y(s_1,\ldots, s_k)$ as
well, the subset of "ghost" numbers $(n_{f{}0}, n_{f{}i}$,
$n_{p{}j},n_{f{}lm}, n_{p{}no}$, $n_{f{}rs}, n_{\lambda{}tu})$in
(\ref{chi}) and (\ref{PhysState}) for fixed values of $s_i$,
satisfies the following equations for $|\chi^l\rangle_{(s)_3}$, $l
= 0,\ldots , k(k+1)=12$,
\begin{eqnarray}\label{ghnum}
   |\chi^l\rangle_{(s)_3}  &:& n_{f{}0}+
     \sum_{i}\bigl(n_{f{}i}- n_{p{}i}\bigr)+
\sum_{i\leq j}\bigl(n_{f{}ij}-n_{p{}ij} \bigr) +\sum_{r<s} \bigl(
n_{f{}rs}- n_{\lambda{}rs}\bigr) = -l ,
\end{eqnarray}
which follows from ghost number distributions
(\ref{Qchi})--(\ref{QLambdai}). Above $(k+1[k(k+1)])$ relations
(\ref{nidecompos}), (\ref{ghnum}) (for given example equal to
$(3+12)$) express the fact of the general homogeneity of the
vectors $|\chi^l\rangle_{(s)_3}$ with respect to spin and ghost
number distributions.

The corresponding BRST operator $Q$ in (\ref{Q}) for $25$
 constraints $(L_0, L_i, L_{ij}, T_{rs}, L^+_i,
L^+_{ij}, T^+_{rs})$ has the form,
\begin{eqnarray}
\label{Q3} {Q} \hspace{-0.4em} &=&\hspace{-0.4em} \textstyle
 \frac{1}{2}\eta_0L_0+\eta_i^+L^i
+\sum\limits_{l\leq m}\eta_{lm}^+L^{lm} + \sum\limits_{l<
m}\vartheta^+_{lm}T^{lm}
 + \frac{\imath}{2}\sum\limits_l\eta_l^+\eta^l{\cal{}P}_0 -
\vartheta^+_{23}(\vartheta^+_{12}\lambda^{13}-
\vartheta_{13}\lambda^{+}_{12})
\\ \hspace{-0.4em}
&& {}\hspace{-0.4em} - \vartheta_{13}^+\vartheta_{12}\lambda^{23}
- \sum_{n,l<m}(1+\delta_{ln})\vartheta_{lm}^+\eta^{l+}{}_{n}
\mathcal{P}^{mn}+
\sum_{n,l<m}(1+\delta_{mn})\vartheta_{lm}^+\eta^{m}{}_{n}
\mathcal{P}^{+ln}
\nonumber\\
\hspace{-0.4em} && \hspace{-0.4em} +
\textstyle\frac{1}{2}\hspace{-0.2em}\sum\limits_{l<m,n\leq
m}\hspace{-0.4em}\eta^+_{nm}\eta^{n}{}_l\lambda^{lm}
 - \bigl[\textstyle\frac{1}{2}\sum\limits_{l\leq
m}(1+\delta_{lm})\eta^m\eta_{lm}^+ +
\sum\limits_{l<m}\vartheta_{lm} \eta^{+m}
+\sum\limits_{m<l}\vartheta^+_{ml} \eta^{+m} \bigr]\mathcal{P}^l
\hspace{-0.1em}+\hspace{-0.1em}h.c. \nonumber
\end{eqnarray}
and is nilpotent after substitution $h_i \to
-(p_i+\frac{d-2-4i+\theta(m)}{2})$, for $i=1,2,3$, in it,  when
restricted on to Hilbert subspace in $\mathcal{H}_{tot}$ to be
formed by the  vectors $|\chi^l\rangle_{(s)_3}$ (\ref{Qchi}) to be
proper for the spin operator $(\sigma^1,\sigma^2,\sigma^3)$
(\ref{sigmai}).

The explicit form of the additional parts to the constraints
$o'_a$ is determined by the relations
(\ref{g'0iF})--(\ref{t'lmF}), but for $k=3$ rows in YT, so that,
the operators $t^{\prime +}_{lm}$, for $l,m=1,2,3$; $l<m$ in
(\ref{t'+lmtext}) and $C^{12}(d^+,d), C^{13}(d^+,d),
C^{23}(d^+,d)$ in (\ref{Clm}) are written as follows,
 \begin{eqnarray}
  t^{\prime+}_{12}   & = & d^+_{12}
   - \sum\nolimits_{n=1}^{3}(1+\delta_{1n})b^+_{n2}b_{1n}\,,
 \label{t'+12}
 \\
  t^{\prime+}_{13}   & = & d^+_{13}
   - \sum\nolimits_{n=1}^{3}(1+\delta_{1n})b^+_{n3}b_{1n}\,,
 \label{t'+13}
 \\
  t^{\prime+}_{23}   & = & d^+_{23} - d_{12}d^+_{13}
   - \sum\nolimits_{n=1}^{3}(1+\delta_{n2})b^+_{n3}b_{2n}\,,
 \label{t'+23}
 \\
 \label{C12}
C^{12}(d^+,d)&\equiv &
\bigl(h^{1}-h^{2}-d^+_{12}d_{12}-d^+_{13}d_{13}-d^+_{23}d_{23}\bigr)d_{12}
+ d^+_{23} d_{13},\\
  \label{C13}
C^{13}(d^+,d)&\equiv & \bigl(h^{1}-h^{3} -d^+_{13}d_{13} +
d^+_{23}d_{23}\bigr)d_{13}  ,\\
  \label{C23}
C^{23}(d^+,d)&\equiv &
\bigl(h^{2}-h^{3}-d^+_{23}d_{23}\bigr)d_{23}.
  \end{eqnarray}
It  permit to present, first, the expressions for the elements,
 $l^{\prime}_{ll}$ as,
\begin{eqnarray}
\label{l'11text} l^{\prime }_{11} &=&
\frac{1}{4}\sum_{n=2}^{3}\Bigl[b^+_{nn}{b}_{1n} - 2d^+_{1n}+
2\delta_{2n}b^+_{n3}b_{13}
 \Bigr]{b}_{1n} +
\Bigl(\sum_{n= 1}^3
b^+_{1n}b_{1n}  - \sum_{n= 2}^3d^+_{1n}d_{1n} + h^{1}\Bigr)b_{11}, \\
\label{l'22text} l^{\prime }_{22} &=&  \frac{1}{4}\Bigl(
b^+_{11}{b}^2_{12}+ b^+_{33}{b}^2_{23}\Bigr)+ \frac{1}{2}\Bigl[
2b^+_{12}b_{22}+ b^+_{13}b_{23}-
 C^{12}(d^+,d) \Bigr]{b}_{12}
 \\
 && + \left(b^+_{22}b_{22}+b^+_{23}b_{23}  - d^+_{23}d_{23}+ d^+_{12}
 d_{12} + h^{2}\right)b_{22} - \frac{1}{2}\Bigl[d^+_{23} -
d^+_{13}d_{12} \Bigr]{b}_{23} , \nonumber\\
\label{l'33text} l^{\prime }_{33} &=&
\frac{1}{4}\Bigl(b^+_{11}{b}^2_{13}+b^+_{22}{b}^2_{23}\Bigr) +
\frac{1}{2}\Bigl[ d^+_{12}d_{13}
+2b^+_{23}b_{33} - C^{23}(d^+,d)\Bigr]{b}_{23} \\
  && +
\frac{1}{2}\Bigl[b^+_{12}b_{23}+2b^+_{13}b_{33}
-C^{13}(d^+,d)-\big(C^{23}(d^+,d) - d^+_{12}d_{13}\big)d_{12} \Bigr]{b}_{13}
 \nonumber\\
 && + \left( b^+_{33}b_{33}  + d^+_{13}
 d_{13}+d^+_{23}
 d_{23} + h^{3}\right)b_{33}\nonumber ,
  \end{eqnarray}
 second for $l^{\prime}_{lm}$, for $l<m$,
\begin{eqnarray}
  \label{l'12bose}
 l^{\prime }_{12}&=&
  \frac{1}{4}\Bigl(\sum_{n=2}^3\bigl[b^+_{1n}b_{1n} +
(1+\delta_{n2})b^+_{2n} b_{2n}\bigr]   - d^+_{13}d_{13}-
d^+_{23}d_{23} + h^{1}+ h^{2}\Bigr)b_{12}\\
&& - \frac{1}{2} \bigl[ C^{12}(d^+,d)
-\sum_{n=1}^{3}(1+\delta_{n2})
 b^+_{1n}b_{n2}
\bigr]b_{11}- \frac{1}{4} \bigl[
2d^+_{12}{b}_{22}+d^+_{13}{b}_{23} \bigr]
  \nonumber\\
 &&   - \frac{1}{4} \Bigl[ d^+_{23}-
 d^+_{13}d_{12} - \sum\limits_{n=2}^3
(1+\delta_{n2})b^+_{n3}b_{2n} \Bigr]{b}_{13}\nonumber, \\
  \label{l'13bose}
 l^{\prime }_{13}&=&
\frac{1}{4}\Bigl(b^+_{13}b_{13} +
\sum_{n=2}^3(1+\delta_{n3})b^+_{n3} b_{n3}  - d^+_{12}d_{12}+
d^+_{23}d_{23}+ h^{1}+ h^{3}\Bigr)b_{13}  \\
 && +  \frac{1}{4} \Bigl[ \sum_{n=2}^{3}(1+\delta_{n3})
 b^+_{2n}b_{n3} + d^+_{12}d_{13} - C^{23}(d^+,d)\Bigr]b_{12 }  - \frac{1}{4} \Bigl[ d^+_{12}{b}_{23}+2d^+_{13}{b}_{33}
\Bigr] \nonumber \\
&& + \frac{1}{2} \Bigl[ \sum_{n=1}^{3}(1+\delta_{n3})
 b^+_{1n}b_{n3} - C^{13}(d^+,d)-\big(C^{23}(d^+,d) - d^+_{12}d_{13}\big)d_{12}\Bigr]b_{11 }\nonumber,
\end{eqnarray}
\begin{eqnarray}
  \label{l'23bose}
 l^{\prime }_{23}&=&
 \frac{1}{4} \Bigl[ \sum_{n=1}^{3}(1+\delta_{n3})
 b^+_{2n}b_{n3} - C^{13}(d^+,d)- \big(C^{23}(d^+,d) - d^+_{12}d_{13}\big)d_{12}\Bigr]b_{12
 }\\
&& +  \frac{1}{2} \Bigl[ \sum_{n=2}^{3}(1+\delta_{n3})
 b^+_{2n}b_{n3} + d^+_{12}d_{13} - C^{23}(d^+,d)\Bigr]b_{22 }
\nonumber \\
&& + \frac{1}{4}\Bigl(b^+_{23}b_{23} + 2b^+_{33} b_{33}
+d^+_{12}d_{12} +d^+_{13}d_{13} +  h^{2}+
h^{3}\Bigr)b_{23} \nonumber\\
&& -\frac{1}{4} \Bigl[  C^{12}(d^+,d)
-\sum_{n=2}^{3}(1+\delta_{n2})b^+_{1n}b_{2n}
 \Bigr]{b}_{13} - \frac{1}{2}
\Bigl[ d^+_{23} - d^+_{13}d_{12} \Bigr]{b}_{33}\nonumber,
\end{eqnarray}
 and, third, for $t^{\prime}_{lm}$,
 \begin{eqnarray}
t^{\prime }_{12} &=&
 C^{12}(d^+,d)
 -\sum\nolimits_{n=1}^{3}(1+\delta_{n2})b^+_{n1}
b_{n2}
 \,,  \label{t'12F}\\
t^{\prime }_{13} &=&
 C^{13}(d^+,d)+ \big(C^{23}(d^+,d)- d_{12}^+d_{13}\big) d_{12}
  -\sum\nolimits_{n=1}^{3}(1+\delta_{n3})b^+_{n1}
b_{n3}
 \,,\label{t'13F}\\
t^{\prime }_{23} &=& - d^+_{12}d_{13} +
 C^{23}(d^+,d)
  -\sum\nolimits_{n=1}^{3}(1+\delta_{n3})b^+_{n2}
b_{n3}
 \,.\label{t'23F}
\end{eqnarray}
Relations (\ref{t'+12})--(\ref{t'23F}) together with Eqs.
(\ref{g'0iF}), (\ref{l'+ijF}) for the value of $k=3$ compose the
scalar oscillator realization of $sp(6)$ algebra  over
Heisenberg-Weyl algebra $A_{9}$ with $18$ independent operators
$b^+_{ij}, b_{ij}, d^+_{rs}, d_{rs}$, for $i\leq j, r<m$.

At last, the Lagrangian equations of motion (\ref{Q12}), set of
reducible Abelian gauge transformations (\ref{dx0})--(\ref{dxs})
and proper unconstrained action $\mathcal{S}_{(s)_3}$ (\ref{Scl})
have the final respective form for the HS field of spin
$(s_1,s_2,s_3)$,
\begin{eqnarray}
Q_{(s)_3}|\chi^0\rangle_{(s)_3}=0; &\qquad& \label{Q123}
\\
\label{dx03} \delta|\chi^0 \rangle_{(s)_3}
=Q_{(s)_3}|\chi^{1}\rangle_{(s)_3} \,, &\qquad&
\delta|\chi^{1}\rangle_{(s)_3} =Q_{(s)_3}|\chi^{2}\rangle_{(s)_3}
\,,
\\
\ldots &\ldots& \ldots
\nonumber \\
\label{dxs3} \delta|\chi^{11} \rangle_{(s)_3}
=Q_{(s)_3}|\chi^{12}\rangle_{(s)_3} \,, &\qquad&
\delta|\chi^{12} \rangle_{(s)_3} =0\, ;
\end{eqnarray}
\vspace{-3ex}
\begin{eqnarray}
 \mathcal{S}_{(s)_3}
= \int d \eta_0 \; {}_{(s)_3}\langle \chi^0 |K_{(s)_3} Q_{(s)_3}|
\chi^0 \rangle_{(s)_3}, &\qquad& \label{Scl3}
\end{eqnarray}
where operator $K_{(s)_3}$ is determined by the relations
(\ref{explicit K}), (\ref{tK}), (\ref{Ka}) for $k=3$. The
corresponding Lagrangian formulation is at most $11$-th stage
reducible gauge theory for free arbitrary HS bosonic field subject
to $Y(s_1,s_2,s_3)$ Young tableaux on Minkowski
$\mathbb{R}^{1,d-1}$ space. We will use obtained Lagrangian
formulation (\ref{Q123})--(\ref{Scl3}) to find the component
Lagrangian formulation for the field $\Phi_{\mu \nu,\rho,\sigma}$
with spin $\mathbf{s} =(2,1,1)$.

\subsection{Spin-$(2,1,1)$ mixed-symmetric massless field} \label{ex211}
Here, we apply the general prescriptions of our Lagrangian
formulation  for rank-$4$ tensor field,
$\Phi_{\mu\nu,\rho,\sigma}$, to be symmetric in indices $\mu,
\nu$, i.e. $\Phi_{\mu\nu,\rho,\sigma} =
\Phi_{\nu\mu,\rho,\sigma}$, starting from the analysis of tower of
gauge transformations on a base of cohomological resolution
complex.

\subsubsection{Reducible gauge transformations for the gauge
parameters}\label{ggtrex}

In  the case of spin-$(2,1,1)$ field, we have $(h^1,h^2,h^3) =
(\{1-\frac{d}{2}\},\{4-\frac{d}{2}\},\{6-\frac{d}{2}\})$.
Therefore, due to analysis of the system of three spin
(\ref{nidecompos}) and one (\ref{ghnum}) ghost number equations on
all the indices of powers in the decomposition (\ref{chi}) and
(\ref{PhysState}) for the field $|\chi^0\rangle_{(s)_3}$ and each
of the gauge parameters $|\chi^l\rangle_{(s)_3}$,
${(s)_3}=(2,1,1)$ and $l = 1,\ldots , 12$, the gauge theory is the
$L = 4$th stage of reducibility ($L = 5 < L_{3}=11$).

The first lowest and, therefore, independent gauge parameter
$|\chi^5\rangle_{(s)_3}$  is determnined only one component scalar
field $\phi^5(x)$ (\ref{x-5})  (see Appendix~\ref{example211},
where all the explicit expressions for the vectors
$|\chi^l\rangle_{(s)_3}$,  $l = 0,\ldots , 5$ are derived from the
general Eq. {\ref{chi}}).
 For the reducible gauge parameter of the fourth level
$|\chi^{4}\rangle_{(s)_3}$ given by the Eqs. (\ref{x-4}),
(\ref{x-4decomp})--(\ref{x-4decompf}) we have from the  equation
in (\ref{dxs}) or equivalently in (\ref{dx03}), (\ref{dxs3}),
$\delta|\chi^{4}\rangle_{(s)_3} = Q_{(s)_3}
|\chi^{5}\rangle_{(s)_3}$,
 the gauge
transformations  for the components of fourth level gauge
parameter, with omitting coordinates $x, (x \in
\mathbb{R}^{1,d-1})$ in arguments, has the form for component
functions,
\begin{align}\label{4level}
   & \delta\phi^{4}_0 = \square \phi^{5}, & \delta\phi^{4}_1 = -
\phi^{5}, && \delta\phi^{4}_2 = \phi^{5},\\
& \delta\phi^{4}_3 = -\phi^{5}, &  \delta\phi^{4}_4 =2\phi^{5} ,
&&\delta\phi^{4}_{11} =0,\\
& \delta\phi^{4}_5 = - \delta\phi^{4}_6 = \phi^{5}, &
\delta\phi^{4}_{5|\mu\nu} =\frac{1}{2}\eta_{\mu\nu}\phi^{5} ,
&&\delta\phi^{4}_{9|\mu} =0,\\
& \delta\phi^{4}_{10|\mu} = -i\partial_{\mu}\phi^{5}, &
\delta\phi^{4}_{7|\mu} =i\partial_{\mu}\phi^{5} ,
&&\delta\phi^{\prime4}_{7|\mu}=\delta\phi^{4}_{8|\mu}
=0\label{4levelf}.
\end{align}
To derive Eqs.(\ref{4level})--(\ref{4levelf}), we have taken into
account the definition of BRST operator (\ref{Q3}) and structure
of the additional parts for constraints
(\ref{t'+12})--(\ref{t'23F}).

Then,  we  impose the gauge conditions
(\ref{gk(k+1)-1})\footnote{General gauge-fixing procedure
described in the Appendix~\ref{reductionC} must be adapted because
of as it is mentioned in the Appendix~\ref{example211} our example
of the field $\Phi_{\mu\nu,\rho,\sigma}$ has not the form of
general mixed-symmetric field considered in the
Appendix~\ref{reductionC} which should be at least characterized
by Young tableaux $Y(7,7,7)$ for $k=3$ rows} for the first lowest
dependent gauge parameter $|\chi^{4}\rangle_{(s)_3}$. As the
solution of the equation,
$b_{11}^+\mathcal{P}_{11}^+|\chi^{4}\rangle_{(s)_3}=0$, obtained
from general consideration, we have  the vector
$|{\chi}{}^{4}_g\rangle_{(s)_3}$ with vanishing function
$\phi^{4}_{5}$, so that the theory become by the third stage
reducible gauge theory, with vanishing $\phi^{5}$ and the rest
independent component functions in
$|{\chi}{}^{4}_g\rangle_{(s)_3}$ which has the form of the
Eq.(\ref{x-4}) but with $\phi^{4}_{5} =0$.

In turn, the general gauge conditions (\ref{gk(k+1)-2}) applied to
the second lowest dependent  gauge parameter
$|\chi^{3}\rangle_{(s)_3}$ given by the Eq. (\ref{x-3}) and
Eqs.(\ref{x-3decomp})-(\ref{x-3decompf}), has the form
\begin{eqnarray}\label{gk(k+1)-2ex}
b_{11}\mathcal{P}_{11}^+|\chi^{3}\rangle_{(s)_3}=0, &\qquad&
b_{12}\mathcal{P}_{11}^+\mathcal{P}_{12}^+|\chi^{3}\rangle_{(s)_3}=0,
\end{eqnarray}
and lead to  vanishing of the component functions, $\phi^{3}_{p}$,
$\phi^{3}_{05}$, $\phi^{\prime 3}_{11}$, $\phi^{3}_{b|\mu}$, $
\phi^{\prime 3}_{28|\mu}$, $\phi^{\prime\prime3}_{28|\mu}$, for
$p=7,10,11, 13, 45$, $b=28, 29, 36$.

To find the degree of freedom of which component functions in the
independent reduced vector $|{\chi}{}^{4}_g\rangle_{(s)_3}$
correspond to the vanishing of the above third level components
$\phi^{3}_{\cdots}$ we should to consider the explicit form of the
gauge transformations for the components in
$|{\chi}{}^{3}\rangle_{(s)_3}$. They are given by the Eqs.
(\ref{dx03}), (\ref{dxs3}), namely,
$\delta|\chi^{3}\rangle_{(s)_3} = Q_{(s)_3}
|\chi^{4}_g\rangle_{(s)_3}$. We do not make here corresponding
sequence of the component relations which follows from the above
relation, but the restrictions on above $\phi^{3}_{\cdots}$ gauge
parameters are due to used degrees of freedom from the vector
$|\chi^{4}_g\rangle_{(s)_3}$ related respectively to the
components $\phi^{4}_{s}$, $\phi^{\prime 4}_{7|\mu}$,
$\phi^{4}_{t|\mu}$, for $s=2,1,3,6,11, 0, 4$,   $t=8,10, 9, 7$
which we must set to 0 in  (\ref{x-4}) for
$|\chi^{4}_g\rangle_{(s)_3}$. In addition, to the conditions
(\ref{gk(k+1)-2ex}) we may to gauge away the third level
components $\phi^{3}_{7|\mu\nu}$ by means of the rest fourth level
parameter $\phi^{4}_{5|\mu\nu}$ so that all the degrees of freedom
from the vector $|\chi^{4}_g\rangle_{(s)_3}$ are completelu used.
As the result, the theory becomes by the second-stage reducible
theory. The only component fields  from  the reduced third level
vector in $\mathcal{H}_{tot}$ $|\chi^{3}_g\rangle_{(s)_3}$
(\ref{x-3}) survive with except for gauged away
$\phi^{3}_{7|\mu\nu}$, $\phi^{3}_{p}$, $\phi^{3}_{05}$,
$\phi^{3}_{b|\mu}$, $\phi^{\prime 3}_{11}, \phi^{\prime
3}_{28|\mu}$, $\phi^{\prime\prime3}_{28|\mu}$, for $p=7,10,11, 13,
45$, $b=28, 29, 36$.

Then, for the reducible gauge parameter of the second level
$|\chi^{2}\rangle_{(s)_3}$ given by the Eqs.(x-2),
(\ref{x-2decomp})--(\ref{x-2decompf}) we impose the general gauge
conditions (\ref{gk(k+1)-3}) having for our case the form,
\begin{eqnarray}\label{gk(k+1)-3ex}
\Bigl(b_{11}\mathcal{P}_{11}^+,\quad
b_{12}\mathcal{P}_{11}^+\mathcal{P}_{12}^+,\quad
b_{13}\prod_{i}^3\mathcal{P}_{1i}^+\Bigr)|\chi^{2}\rangle_{(s)_3}=0.
\end{eqnarray}
Eqs. (\ref{gk(k+1)-3ex}) lead to vanishing of the component
functions, $\phi^{2}_{p}$, $\phi^{\prime 2}_{r}$,
$\phi^{\prime\prime 2}_{1}, \phi^{\prime\prime\prime 2}_{1}$,
$\phi^{2}_{b|\mu}$, $\phi^{\prime  2}_{c|\mu}, \phi^{\prime\prime
2}_{c|\mu}$, $\phi^{\prime\prime\prime  2}_{d|\mu}$, $\phi^{(iv)
2}_{46|\mu}$, $\phi^{(v)  2}_{46|\mu}$, $\phi^{ 2}_{e|\mu\nu},
\phi^{\prime 2}_{100|\mu\nu}$, $\phi^{ 2}_{100|\mu,\nu}$
 for $p=1,6,7,16,  47, 51, 53, 56, 69$,
$92, 95 - 100 $, $r = 1, 16, 47, 100$, $b=46, 50, 52, 55, 66, 67,
68, 75 $, $c=46, 52, 55$,  $67 $, $d= 46, 55$, $e=99, 100$. In
turn, the $\eta_0$-dependent terms in $|\chi^{2}_0\rangle_{(s)_3}$
have the same structure as ones in $|\chi^{3}\rangle_{(s)_3}$
(\ref{x-3}) and therefore the same components as in
$|\chi^{3}\rangle_{(s)_3}$ should vanish in
$|\chi^{2}_0\rangle_{(s)_3}$, i.e. of the $\phi^{2}_{0v}$,
$\phi^{2}_{0w|\mu}$, $\phi^{\prime2}_{011}$,
$\phi^{\prime2}_{028|\mu}$, $\phi^{\prime\prime2}_{028|\mu}$ for
$v=7,10, 11, 13, 45$, $w=28, 29, 36$ and of
$\phi^{2}_{07|\mu\nu}$.

The last restrictions  are in one-to-one correspondence (due to
homogeneity in $\eta_0$  the BRST operator $Q$ (\ref{Q3})) with
the used degrees of freedom from the vector
$|\chi^{3}_0\rangle_{(s)_3}$ related to the components
 $\phi^{3}_{0s}$, $\phi^{3}_{0t|\mu}$, for $s=2,1,3,4,6,11$,
$t=7,8,9,10$, $\phi^{\prime 3}_{07|\mu}$ and of
$\phi^{3}_{05|\mu\nu}$ which we must set to 0 in
$\eta_0|\chi^{3}_0\rangle_{(s)_3}$ in (\ref{x-3}) so that
$\eta_0$-dependent terms have used totally.

To make this conclusion we must obtain  the explicit form of the
gauge transformations for the component parameters  in
$|{\chi}{}^{2}\rangle_{(s)_3}$. They are generated by the standard
Eqs. (\ref{dx03}), (\ref{dxs3}), namely,
$\delta|\chi^{2}\rangle_{(s)_3} = Q_{(s)_3}
|\chi^{3}_g\rangle_{(s)_3}$. We do not write below corresponding
sequence of the component relations (vanishing gauge
transformations for $|\chi^{2}\rangle_{(s)_3}$), which forms due
to the gauge (\ref{gk(k+1)-3ex}) a system of not only the
algebraic linear more than one hundred equations in all components
(reminding the case of "unfolded" Vasiliev equations). Its
solution is found by the Gauss-like procedure so that all the
degrees of freedom from the vector $|\chi^{3}\rangle_{(s)_3}$ with
except for the vectors $\phi^3_{1|\mu}, \phi^3_{8|\mu} $ and
scalar $\phi^3_{9}$ should be used to gauge away the above
components from $eta_0$-independent part of
$|\chi^{2}\rangle_{(s)_3}$.

Then, from  the residual gauge transformations for the second
level
 parameters  in the vectors $|\phi^{2}_{9} \rangle_{(1,0,0)},
|\phi^{2}_{10} \rangle_{0_3}$ we find
\begin{align}\label{resgt2}
&\delta\phi^{2}_{9|\mu} = \phi^{3}_{8|\mu}, && \delta\phi^{2}_{10}
= \phi^{3}_{9}
\end{align}
and therefore,  the components $\phi^{2}_{9|\mu}$, $\phi^{2}_{10}$
should gauged away by means of respective third level parameters
$\phi^{3}_{8|\mu}, \phi^{3}_{9}$.

As the result,  the final gauge transformations with non-vanishing
third level gauge parameter  $\phi^{3}_{1|\mu}$ have the form
\begin{align}\label{fgt2}
&\delta\phi^{2}_{01|\mu} =  \Box\phi^{3}_{1|\mu} ,&
\delta\phi^{\prime\prime\prime 2}_{1|\mu,\nu}= -i
\partial_\nu\phi^{3}_{1|\mu},
&& \delta\phi^{2}_{16|\mu,\nu}= i\partial_\nu\phi^{3}_{1|\mu} ,
\\
\label{fgt2f}&\delta\phi^{2}_{56|\mu\nu} =  -\textstyle\frac{i}{2}
\partial^{}_{\{\nu}\phi^{3}_{1|\mu\}}\footnotemark,&
\delta\phi^{2}_{2} = i\partial^\mu\phi^{3}_{1|\mu}.
\end{align}\footnotetext{we have used in the Eqs. (\ref{fgt2f}) and will use later
 both for symmetrization and
antisymmetrization of Lorentz indices the conventions,
$A_{\{\mu,\nu\}} = A_{\mu,\nu}+A_{\nu,\mu}$,  $A_{[\mu,\nu]} =
A_{\mu,\nu}- A_{\nu,\mu}$, i.e. $A_{\mu,\nu} =
\frac{1}{2}A_{\{\mu,\nu\}}+\frac{1}{2}A_{[\mu,\nu]}$}All the rest
components in the gauge fixed vector $|\chi^{2}_g\rangle_{(s)_3}$
(obtained from $|\chi^{2}\rangle_{(s)_3}$ (\ref{x-2}) in the
result of the gauge procedure resolution above) do not contain
totally or partially
 the components from the vectors $|\phi^{2}_{\cdots}
\rangle_{(\cdots)}$ in $\mathcal{H}\bigotimes\mathcal{H}'$ whose
components were gauged away.

Next, for the reducible gauge parameter of the first level
$|\chi^{1}\rangle_{(s)_3}$ given by the Eqs.(\ref{x-1}),
(\ref{x-1decomp})--(\ref{301}) the general gauge conditions
(\ref{Apgauge}), for $k=3$ and $l= 11$
  having now the
form,
\begin{eqnarray} \label{Apgaugex-1}
\Bigl(b_{11}\mathcal{P}_{11}^+,\quad
b_{12}\prod_i^2\mathcal{P}_{1i}^+,\quad
b_{13}\prod_{i}^3\mathcal{P}_{1i}^+,\quad
b_{22}\prod_{i}^3\mathcal{P}_{1i}^+\mathcal{P}_{22}^+\Bigr)|\chi^1\rangle_{(s)_3}=0.
\end{eqnarray}
 lead to  vanishing of the
component functions, $\phi^{1}_{p}$, $\phi^{\prime1}_{r}$,
$\phi^{\prime\prime 1}_{t}$, $\phi^{\prime\prime\prime 1}_{55}$,
$\phi^{\prime\prime\prime 1}_{154}$, $\phi^{\prime\prime\prime
1}_{163}$, $\phi^{1}_{b|\mu}$, $\phi^{\prime (\prime\prime)
1}_{c|\mu}$, $\phi^{\prime\prime\prime 1}_{d|\mu}$, $\phi^{(iv)
1}_{d|\mu}$, $\phi^{(v) 1}_{75|\mu}$, $\phi^{(6)
1}_{1|\mu}$--$\phi^{(12) 1}_{1|\mu}$ $\phi^{1}_{e|\mu\nu}$,
$\phi^{\prime 1}_{f|\mu\nu}$, $\phi^{\prime\prime 1}_{h|\mu\nu}$,
$\phi^{\prime\prime\prime 1}_{h|\mu\nu}$, $\phi^{1}_{g|\mu,\nu}$,
$\phi^{\prime 1}_{h|\mu,\nu}$ for $p=13-15, 21, 49$, $55, 59, 60,
62, 69, 70$, $89, 91, 95, 96, 105, 136$, $139, 146, 147, 150, 153,
154, 156$, $159-163$; $r= 15, 49, 55, 69$, $70, 95, 136$, $150$,
$154, 159, 162, 163$; $t = 55, 69, 154, 163$; $b= 1, 54, 66-68$,
$74, 75, 94, 99, 100$, $155, 157, 158$; $c = 1, 68, 74, 75, 100,
157 $; $d= 1, 75$; $e = 154, 161-163$; $f = 154, 162, 163$;
$h=154, 163$; $g=154, 162, 163$.

In turn,  the same  components from the $\eta_0$-dependent part of
$|\chi^{1}\rangle_{(s)_3}$, i.e. in $|\chi^{1}_0\rangle_{(s)_3}$
should be gauged away as ones in the vector
$|\chi^{2}\rangle_{(s)_3}$. Therefore the only vector
$\phi^2_{01|\mu}$ in the vector $|\phi^{2}_{01} \rangle_{(1,0,0)}$
of $\eta_0$- dependent components from $|\chi^{2}\rangle_{(s)_3}$
survives.

Not presenting  here explicitly a whole system  of  linear
equations on the components (which should be vanished) of
$|\chi^{1}\rangle_{(s)_3}$ to be easily derived from the  standard
gauge transformations given by the Eqs. (\ref{dx03}),
(\ref{dxs3}), namely, $\delta|\chi^{1}\rangle_{(s)_3} = Q_{(s)_3}
|\chi^{2}_g\rangle_{(s)_3}$, we list the results of its
Lorentz-covariant resolution.

First, the corresponding gauge transformations for
$\eta_0$-dependent parameters in $|\chi^{1}_0\rangle_{(s)_3}$ has
the same form as ones for $|\chi^{2}_g\rangle_{(s)_3}$ with only
opposite sign $"-"$ due to anticommuting $\eta_0$-multiplier and
with addition of the terms proportional to d'Alamberian "$\Box$",
\begin{align}\label{gtr10}
& \delta\phi^{\prime\prime\prime 1}_{01|\mu,\nu}= i
\partial_\nu\phi^{2}_{01|\mu}+\Box\phi^{\prime\prime\prime 2}_{1|\mu,\nu},
&& \delta\phi^{1}_{016|\mu,\nu}=
-i\partial_\nu\phi^{2}_{01|\mu}+\Box\phi^{2}_{16|\mu,\nu} ,
\\\label{gtr10f}
&\delta\phi^{1}_{056|\mu\nu} =  \textstyle\frac{i}{2}
\partial^{}_{\{\nu}\phi^{2}_{01|\mu\}}+\Box \phi^{2}_{56|\mu\nu},&&
\delta\phi^{1}_{02} =
-i\partial^\mu\phi^{2}_{01|\mu}+\Box\phi^{2}_{2}.\nonumber
\end{align}
Second, from the resolution of this system it follows the
relations
\begin{equation}\label{resol2115}
\phi^{2}_{56|\mu\nu} = -
\frac{1}{2}\phi^{2}_{16|\{\mu,\nu\}}\texttt{ and }
\phi^{\prime\prime\prime2}_{1|\mu,\nu}\ = \ -
\phi^{2}_{16|\mu,\nu}.
\end{equation}
Third, all the component second level parameters from the vector
$|\chi^{2}_g\rangle_{(s)_3}$ must be used (as the solution of this
system) to gauge away above components from
$|\chi^{1}\rangle_{(s)_3}$ except for the rest gauge parameters of
the second level $\phi^{2}_{01|\mu}$,
 $\phi^{2}_{11|\mu}$, $\phi^{2}_{12}$, $\phi^{2}_{13|\mu}$,
$\phi^{\prime 2}_{13|\mu}$, $\phi^{2}_{14}$, $\phi^{2}_{15}$,
$\phi^{2}_{18|\mu}$, $\phi^{2}_{19}$, $\phi^{2}_{48|\mu}$,
$\phi^{\prime2}_{48|\mu}$, $\phi^{2}_{49|\mu}$,
$\phi^{2}_{54|\mu}$, $\phi^{2}_{70|\mu}$, $\phi^{2}_{102|\mu}$ and
gauge dependent $\phi^{\prime\prime\prime 2}_{ 1|\mu,\nu}$,
$\phi^{2}_{ 16|\mu,\nu}, \phi^{2}_{ 56|\mu\nu}$  $\phi^{2}_{2}$.

However, we have the residual gauge transformations on the not
gauge-fixed (due to  (\ref{Apgaugex-1})) components in
$|\chi^{1}\rangle_{(s)_3}$ to be obtained from the same general
relation, $\delta|\chi^{1}\rangle_{(s)_3} = Q_{(s)_3}
|\chi^{2}_g\rangle_{(s)_3}$,
\begin{align}\label{addgt1}
& \delta \phi^{1}_{20}
=\textstyle\frac{1}{2}\phi^{2}_{16|\mu,}{}^\mu
-\textstyle\frac{1}{2}\phi^{2}_{2} -\phi^{2}_{15}, & \delta
\phi^{1}_{12} = \phi^{2}_{15} -\phi^{2}_{12},\\ & \label{addgt2}
\delta \phi^{1}_{8} = -\phi^{2}_{14}, & \delta \phi^{\prime 1}_{8}
= \phi^{2}_{12} + \phi^{2}_{14},\\
\label{addgt3} & \delta \phi^{1}_{17|\mu} =  i
\partial^\nu \phi^{2}_{16|\nu,\mu}-i
\partial_\mu \phi^{2}_{2} -\phi^{2}_{18|\mu},&
\delta \phi^{\prime1}_{17|\mu} = \phi^{2}_{18|\mu},\\
\label{addgt4}&\delta \phi^{1}_{18|\mu} = \phi^{2}_{01|\mu} +i
\partial^\nu \phi^{2}_{16|\mu,\nu}-\phi^{2}_{18|\mu},&
\delta \phi^{1}_{7|\mu} = -\phi^{2}_{01|\mu} +i
\partial^\nu \phi^{\prime\prime\prime 2}_{1|\mu,\nu}-\phi^{2}_{102|\mu},\\
\label{addgt5}&\delta \phi^{1}_{76|\mu} = -\phi^{2}_{01|\mu} +i
\partial_\mu \phi^{2}_{2}+ 2i \partial^\nu \phi^{2}_{56|\mu\nu}, &
\delta \phi^{1}_{19} = \phi^{2}_{19} -\phi^{2}_{14},\\
\label{addgt6} &
 \delta \phi^{\prime1}_{2|\mu} = - \phi^{2}_{11|\mu}+ \phi^{2}_{13|\mu},
&\delta \phi^{1}_{2|\mu} = i \partial^\nu \phi^{\prime\prime\prime
2}_{ 1|\nu,\mu}+i \partial_\mu \phi^{2}_{2}-\phi^{2}_{13|\mu},\\
\label{addgt7} & \delta \phi^{\prime\prime1}_{2|\mu} = -
\phi^{\prime 2}_{13|\mu}, & \delta
\phi^{\prime\prime\prime1}_{2|\mu} =  \phi^{2}_{11|\mu}+
\phi^{\prime 2}_{13|\mu},\\
\label{addgt8}&\delta \phi^{1}_{72|\mu} =i\partial_\mu
\phi^{2}_{15}
+\phi^{2}_{49|\mu}-\phi^{2}_{48|\mu}+\textstyle\frac{1}{2} \phi^{
2}_{13|\mu}, &  \delta \phi^{1}_{71|\mu} = i\partial_\mu
\phi^{2}_{14} - \phi^{2}_{48|\mu}+\phi^{\prime 2}_{48|\mu},\\
\label{addgt9} & \delta \phi^{\prime1}_{71|\mu} = - \phi^{\prime
2}_{48|\mu}, & \delta \phi^{\prime 1}_{72|\mu} =
-\phi^{2}_{49|\mu}-\phi^{\prime2}_{48|\mu}+\textstyle\frac{1}{2}
\phi^{\prime 2}_{13|\mu},\\
\label{addgt10} & \delta \phi^{1}_{52|\mu} =
\phi^{2}_{54|\mu}+\phi^{2}_{48|\mu}, & \delta
\phi^{\prime1}_{52|\mu} =
-\phi^{2}_{54|\mu}+\phi^{\prime2}_{48|\mu},\\
\label{addgt11}& \delta \phi^{1}_{97|\mu} =
-\phi^{2}_{70|\mu}+\phi^{2}_{48|\mu}-\phi^{2}_{13|\mu}, & \delta
\phi^{\prime1}_{97|\mu} =
\phi^{2}_{70|\mu}+\phi^{\prime2}_{48|\mu}-\phi^{\prime2}_{13|\mu},\\
\label{addgt12}& \delta \phi^{1}_{73|\mu} =
\phi^{2}_{102|\mu}+2\phi^{2}_{49|\mu}. &
\end{align}
Now, we may remove the fields $\phi^{1}_{8},  \phi^{\prime
1}_{8}$, $ \phi^{1}_{12}, \phi^{1}_{19}$,
$\phi^{\prime1}_{17|\mu},
\phi^{\prime1}_{2|\mu}-\phi^{\prime\prime\prime1}_{2|\mu}$,
$\phi^{1}_{71|\mu}, \phi^{1}_{73|\mu}, \phi^{\prime1}_{71|\mu} $
and $ \phi^{\prime1}_{m|\mu} $ for $m=72, 52, 97$  by means of $
\phi^{2}_{n},$ $n=14, 12, 15,  19$,  $\phi^{2}_{18|\mu},
\phi^{2}_{13|\mu}, \phi^{\prime2}_{13|\mu}, \phi^{2}_{11|\mu}$,
$\phi^{2}_{48|\mu}, \phi^{\prime2}_{48|\mu}, \phi^{2}_{102|\mu} $
and $\phi^{2}_{l|\mu} $ for $l= 49, 54, 70$.

For the surviving in (\ref{addgt1})--(\ref{addgt12}) components
from the vectors $|\phi^{1}_{d} \rangle_{(s)_3}$, for $d= 2, 11,
17, 20$, $76,$ $77$, we have the gauge transformations
\begin{align}
\label{addgt13}    &\delta \phi^{1}_{2|\mu} = i
\partial_\mu\phi^2_2 + i
\partial^\nu\phi^{\prime\prime\prime2}_{1|\nu,\mu},
    &&
\delta \phi^{1}_{7|\mu} = -\phi^{2}_{01|\mu} +i
\partial^\nu \phi^{\prime\prime\prime 2}_{1|\mu,\nu},\\
& \delta \phi^{1}_{17|\mu} =  i \partial^\nu
\phi^{2}_{16|\nu,\mu}-i \partial_\mu \phi^{2}_{2}, \label{addgt14}
&&\delta \phi^{1}_{20} =
\textstyle\frac{1}{2}\phi^{2}_{16|\mu,}{}^\mu
-\textstyle\frac{1}{2}\phi^{2}_{2} ,\\
    &\delta \phi^{1}_{76|\mu} = -\phi^{2}_{01|\mu} +i \partial_\mu \phi^{2}_{2}+ 2i \partial^\nu \phi^{2}_{56|\mu\nu}
, \label{addgt15} &&\delta \phi^{ 1}_{11} =
\textstyle\frac{1}{2}\phi^{\prime\prime\prime2}_{1|\mu,}{}^\mu
+\textstyle\frac{1}{2}\phi^{2}_{2},\\
\label{addgt16} &\delta
\phi^{1}_{77} =  \phi^{2}_{56|\mu}{}^\mu +\phi^{2}_{2}.
\end{align}
From the relations (\ref{addgt13}), (\ref{addgt15})   we may gauge
away the first level parameters $\phi^{ 1}_{7|\mu}$ and
 $\phi^{ 1}_{11} $   expressing the rest
free second level parameters $\phi^{ 2}_{2}, \phi^{ 2}_{01|\mu}$
in terms of independent tensor
$\phi^{\prime\prime\prime2}_{1|\mu,\nu}$ as,
\begin{equation}\label{express1}
\phi^{ 2}_{2} = - \phi^{\prime\prime\prime2}_{1|\mu,}{}^\mu,
\qquad \phi^{ 2}_{01|\mu} = i
\partial^\nu \phi^{\prime\prime\prime 2}_{1|\mu,\nu}.
\end{equation}
Summarizing, the final gauge transformations for the first level
gauge parameters reads as follows,
\begin{align}\label{gtr1f1}
 & \delta
\phi^{(vi) 1}_{1|\mu,\nu,\rho} = i
\partial_{\nu}\phi^{
\prime\prime\prime2}_{1|\mu,\rho} - i
\partial_{\rho}\phi^{\prime\prime\prime
2}_{1|\mu,\nu} ,  && \delta \phi^{1}_{2|\mu} = - i \partial_\mu
 \phi^{\prime\prime\prime2}_{1|\nu,}{}^\nu + i
\partial^\nu\phi^{\prime\prime\prime2}_{1|\nu,\mu},\\
\label{gtr1f2}& \delta \phi^{1}_{75|\mu\nu,\rho} = -
\frac{i}{2}\partial_{\{\nu} \phi^{\prime\prime\prime
2}_{1|\mu\},\rho} + \frac{i}{2}\partial_{\rho}
\phi^{\prime\prime\prime 2}_{1|\{\mu,\nu\}} , && \delta \phi^{
1}_{100|\mu\nu,\rho} =
\textstyle\frac{i}{2}\partial_{\{\nu}\phi^{\prime\prime\prime2}_{1|\mu\},\rho}-
\frac{i}{2}\partial_{\rho}\phi^{\prime\prime\prime2}_{1|\{\mu,\nu\}},\\
\label{gtr1f3}& \delta \phi^{1}_{17|\mu} =  i \partial_\mu
 \phi^{\prime\prime\prime2}_{1|\nu,}{}^\nu - i
\partial^\nu\phi^{\prime\prime\prime2}_{1|\nu,\mu},&& \delta \phi^{1}_{76|\mu} = i \partial^\nu \phi^{\prime\prime\prime2}_{1|\nu,\mu}- i
\partial_\mu\phi^{\prime\prime\prime2}_{1|\nu,}{}^\nu ,
\end{align}
and for $\eta_0$-dependent terms from (\ref{gtr10}) due to Eqs.
(\ref{resol2115}), (\ref{express1}) as
\begin{align}\label{gtr10f1}
& \delta\phi^{\prime\prime\prime 1}_{01|\mu,\nu}= -
\partial_\nu
\partial^\rho \phi^{\prime\prime\prime 2}_{1|\mu,\rho}+
\Box\phi^{\prime\prime\prime 2}_{1|\mu,\nu}, &&
\delta\phi^{1}_{016|\mu,\nu}=
\partial_\nu
\partial^\rho \phi^{\prime\prime\prime 2}_{1|\mu,\rho}-
\Box\phi^{\prime\prime\prime2}_{1|\mu,\nu} ,
\\
\label{gtr10f2}&\delta\phi^{1}_{056|\mu\nu} =
-\textstyle\frac{1}{2}
\partial^{}_{\{\nu}
\partial^\rho \phi^{\prime\prime\prime 2}_{1|\mu\},\rho} +
\textstyle\frac{1}{2}\Box
\phi^{\prime\prime\prime2}_{1|\{\mu,\nu\}},&& \delta\phi^{1}_{02}
= \partial^\mu
\partial^\nu \phi^{\prime\prime\prime 2}_{1|\mu,\nu}-
\Box  \phi^{\prime\prime\prime2}_{1|\mu,}{}^\mu.
\end{align}
Note,  the only 2-rank tensor $\phi^{\prime\prime\prime
2}_{1|\mu,\nu}$ left in the second level vector
$|\chi^{2}_g\rangle_{(s)_3}$, and, the structure of first level
vector $|\chi^{1}\rangle_{(s)_3}$ (\ref{x-1}) is simplified to the
reduced vector $|\chi^{1}_g\rangle_{(s)_3}$ having the contents as
in (\ref{x-1}) but without the componets which have been gauged
due to general gauge (\ref{Apgaugex-1}) and after resolution of
the equations (\ref{addgt1})--(\ref{addgt16}).

 Let us turn to the gauge transformation for the fields.

\subsubsection{Gauge transformations for the fields}\label{gtrex}

For the  dependent field vector $|\chi^{}\rangle_{(s)_3}$ the
conditions (\ref{nidecompos}), (\ref{ghnum}) allow one, first, to
extract the dependence on the ghost variables as it was given in
the Appendix~\ref{ex211} by the Eqs.(\ref{x-0}), whose componets
are determined by the relations
(\ref{x-3decomp})--(\ref{x-3decompf}),
(\ref{x-2decomp})--(\ref{x-2decompf}),
(\ref{x-1decomp})--(\ref{301}) but for zeroth level vector and by
the Eqs.(\ref{basdecomp})--(\ref{xdecompf}).  The general gauge
conditions (\ref{G1}) applied for $k=3$ and $l= 11$   have the
form for $|\chi\rangle_{(s)_3}$,
\begin{eqnarray}
&& \Bigl(b_{11}\mathcal{P}_{11}^+,\
b_{12}\prod_i^2\mathcal{P}_{1i}^+,\
b_{13}\prod_{i}^3\mathcal{P}_{1i}^+,\
b_{22}\prod_{i}^3\mathcal{P}_{1i}^+\mathcal{P}_{22}^+, \
b_{23}\prod_{i}^3\mathcal{P}_{1i}^+\mathcal{P}_{22}^+\mathcal{P}_{23}^+,
 \Bigr)|\chi\rangle_{(s)_3} =0 , \label{Apgaugex-g}\\
&&  \Bigl(d_{12}\prod_{i\leq
j}^3\mathcal{P}_{ij}^+\lambda_{12}^+,\ d_{13}\prod_{i\leq
j}^3\mathcal{P}_{ij}^+\lambda_{12}^+\lambda_{13}^+, \
d_{23}\prod_{i\leq
j}^3\mathcal{P}_{ij}^+\lambda_{12}^+\lambda_{13}^+\lambda_{23}^+\Bigr)|\chi\rangle_{(s)_3}=0.\nonumber
\end{eqnarray}
 Eqs. (\ref{Apgaugex-g}) lead, first, to  vanishing of all the
component functions, from  the vector $|\Psi\rangle_{2,1,1}$
(\ref{basdecomp}) with except for initial tensor
$\Phi_{\mu\nu,\rho,\sigma}$. Second, it leads to vanishing of the
fields containing auxiliary oscilators $b^+_{ij}, d^+_{rs}$ in the
vectors $|\phi_{1}\rangle_{\cdots}$ --
$|\phi_{7}\rangle_{\cdots}$, $|\phi_{9}\rangle_{\cdots}$ --
$|\phi_{12}\rangle_{\cdots}$, $|\phi_{n}\rangle_{\cdots}$ for
$n=14, 16, 18, 20$, $22, 49, 50$, $53, 59, 61, 63$, $66-68, 70,
71, 73$, $74, 76 - 80, 82-84$, $92, 96, 98, 101-103, 105, 145,
154$, $159, 162 -170, 178-189$. As we have already discussed above
the peculiarities of the gauge-fixing of $\eta_0$-dependent part
of $|\chi^{1}_0\rangle_{(s)_3}$,   the same components from
$|\chi^{}_0\rangle_{(s)_3}$ should be gauged away as ones in the
$\eta_0$-independent part of the vector
$|\chi^{1}\rangle_{(s)_3}$.

Following to the procedure developing in the previous subsection
we wiil not write down explicitly the whole system  of  linear
equations on the components (which should be vanished) from
$|\chi^{}\rangle_{(s)_3}$ to be easily derived from the  standard
gauge transformations given by the first relations in the Eqs.
(\ref{dx03}), i.e. by $\delta|\chi^{}\rangle_{(s)_3} = Q_{(s)_3}
|\chi^{1}_g\rangle_{(s)_3}$.  Let us only demonstrate the results
of its Lorentz-covariant resolution.

First,  the only tensors $\phi^{(vi) }_{01|\mu,\nu,\rho},
\phi^{}_{02|\mu}, \phi^{}_{075|\mu\nu,\rho}$, $\phi^{
}_{0100|\mu\nu,\rho}, \phi^{}_{017|\mu}, \phi^{}_{076|\mu}$
in the
 $\eta_0$- dependent
components from $|\chi^{}\rangle_{(s)_3}$ survives. Their
 gauge transformations  have the analogous form as
ones for $|\chi^{1}_g\rangle_{(s)_3}$ with only opposite sign
$"-"$ due to anticommuting of $\eta_0$-multiplier and adding of
the terms proportional to d'Alamberian "$\Box$",
\begin{align}\label{gtr00}
 & \delta
\phi^{(vi) }_{01|\mu,\nu,\rho} = -i
\partial_{\nu}\phi^{
\prime\prime\prime1}_{01|\mu,\rho} + i
\partial_{\rho}\phi^{\prime\prime\prime
1}_{01|\mu,\nu} + \Box \phi^{(vi) 1}_{1|\mu,\nu,\rho} , \\
 \label{gtr001}&
\delta \phi^{}_{02|\mu} =  i
\partial_\mu
 \phi^{\prime\prime\prime1}_{01|\nu,}{}^\nu - i
\partial^\nu\phi^{\prime\prime\prime1}_{01|\nu,\mu}+\Box \phi^{1}_{2|\mu},\\
\label{gtr002}& \delta \phi^{}_{075|\mu\nu,\rho} =
\frac{i}{2}\partial_{\{\nu} \phi^{\prime\prime\prime
1}_{01|\mu\},\rho} - \frac{i}{2}\partial_{\rho}
\phi^{\prime\prime\prime 1}_{01|\{\mu,\nu\}}+\Box
\phi^{1}_{75|\mu\nu,\rho} ,\\
\label{gtr003} & \delta \phi^{ }_{0100|\mu\nu,\rho} =
-\textstyle\frac{i}{2}\partial_{\{\nu}\phi^{\prime\prime\prime1}_{01|\mu\},\rho}+
\frac{i}{2}\partial_{\rho}\phi^{\prime\prime\prime1}_{01|\{\mu,\nu\}}
+\Box \phi^{ 1}_{100|\mu\nu,\rho},\\
\label{gtr004} & \delta \phi^{}_{017|\mu} =  -i \partial_\mu
 \phi^{\prime\prime\prime1}_{01|\nu,}{}^\nu + i
\partial^\nu\phi^{\prime\prime\prime1}_{01|\nu,\mu}+\Box \phi^{1}_{17|\mu},\\
\label{gtr005}& \delta \phi^{}_{076|\mu} = -i \partial^\nu
\phi^{\prime\prime\prime1}_{01|\nu,\mu}+ i
\partial_\mu\phi^{\prime\prime\prime1}_{01|\nu,}{}^\nu +\Box
\phi^{1}_{76|\mu}.
\end{align}
Besides, the relations like (\ref{resol2115}), (\ref{express1})
hold
\begin{equation}\label{resol1115}
\phi^{1}_{056|\mu\nu} = -
\frac{1}{2}\phi^{1}_{016|\{\mu,\nu\}}\texttt{ and }
\phi^{\prime\prime\prime1}_{01|\mu,\nu}\ = \ -
\phi^{1}_{016|\mu,\nu}, \texttt{ and }\phi^{ 1}_{02} = -
\phi^{\prime\prime\prime1}_{01|\mu,}{}^\mu.
\end{equation}
Second, the resolution of the system implies the expression of the
tensors $\phi^{1}_{75|\mu\nu,\rho}$, $\phi^{1}_{100|\mu\nu,\rho}$,
$\phi^{\prime\prime\prime1}_{01|\mu,\nu}$ and vectors
$\phi^{1}_{17|\mu}$, $\phi^{1}_{2|\mu}$ in terms of one first
level tensor components $\phi^{(vi)1}_{1|\mu,\rho,\nu}$ as
follows,
\begin{eqnarray}\label{symmprop1}
 && \hspace{-3.0em}  \frac{1}{2}
\phi^{(vi)1}_{1|\{\mu,\hat{\rho},\nu\}}=-
\phi^{1}_{100|\mu\nu,\rho}, \quad
\Bigl(\phi^{(vi)1}_{1|\mu,\{\nu,\rho\}} = 0 \Longrightarrow
\phi^{(vi)1}_{1|\mu,[\nu,\rho]} \ne 0\Bigr)\texttt{ and }
\frac{1}{2}\phi^{(vi)1}_{1|\{\mu,\nu\},\rho}= -
\phi^{1}_{75|\mu\nu,\rho},\\
&& \label{symmprop2} \phi^{(vi)1}_{1|\nu,\mu,}{}^\nu=
\phi^{1}_{17|\mu}, \qquad
\phi^{(vi)1}_{1|\nu,}{}^{\nu}{}_{,\mu}=\phi^{1}_{2|\mu}, \qquad i
\partial^\rho \phi^{(vi)1}_{1|\mu,\nu,\rho} =
\phi^{\prime\prime\prime1}_{01|\mu,\nu} ,
\end{eqnarray}
where the sign "$\hat{}$" in
$\phi^{(vi)1}_{1|\{\mu,\hat{\rho},\nu\}}$ means the absence of
symmetry with respect to index $\rho$ and the second relation in
(\ref{symmprop1}) implies that tensor
$\phi^{(vi)1}_{1|\mu,\nu,\rho}$ is by the antisymmetric in the
last indices.

Third, all the component first level parameters from the vector
$|\chi^{1}_g\rangle_{(s)_3}$ should be used (as the solution of
this system) to gauge away following from the gauge
(\ref{Apgaugex-g}) components from $|\chi^{}\rangle_{(s)_3}$
except for the rest gauge parameters
 $\phi^{1}_{18|\mu}$,  and
gauge dependent $\phi^{(vi)1}_{1|\mu,\nu,\rho}$,   $\phi^{1}_{
76|\mu}$.

Fourth,  we have the residual gauge transformations on the not
gauge-fixed (due to  (\ref{Apgaugex-g})) components in
$|\chi^{}\rangle_{(s)_3}$ to be obtained from the same general
relation, $\delta|\chi^{}\rangle_{(s)_3} = Q_{(s)_3}
|\chi^{1}_g\rangle_{(s)_3}$. So, from the gauge transformations
for the fields $\phi^{}_{8|\mu}$, $\phi^{}_{104|\mu}$
\begin{equation}\label{gauge18}
    \delta\phi^{}_{8|\mu}= \frac{1}{2}\phi^{1}_{18|\mu}, \quad \delta\phi^{}_{104|\mu}= -\phi^{1}_{100|\mu\nu,}{}^\nu - \frac{1}{2}\phi^{1}_{76|\mu}-\frac{1}{2}\phi^{1}_{18|\mu}.
\end{equation}
we find, that fields $\phi^{}_{8|\mu}$, $\phi^{}_{104|\mu}$ are
removed by means of the vector $\phi^{1}_{18|\mu}$ and the
relation
\begin{equation}\label{resgauge18}
    \phi^{1}_{76|\mu} = -2\phi^{1}_{100|\mu\nu,}{}^\nu \Longrightarrow \phi^{1}_{76|\mu} = \frac{1}{2}
\phi^{(vi)1}_{1|}{}^\nu{}_{,[\nu,\mu]}.
\end{equation}
Finally, we may write down all the gauge transformations which
contain the only surviving tensors
$\phi^{(vi)1}_{1|\mu,[\nu,\rho]}, \phi^{1}_{2|\mu}$,
$\phi^{1}_{17|\mu}$, $\phi^{1}_{100|\mu\nu,\rho}$,
$\phi^{1}_{75|\mu\nu,\rho}$, $\phi^{1}_{76|\mu}$,
$\phi^{\prime\prime\prime1}_{01|\mu,\nu}$ where the latter ones
are expressed due to Eqs. (\ref{symmprop1}), (\ref{symmprop2}),
(\ref{resgauge18})  in terms of the components of single tensor
$\phi^{(vi)1}_{1|\mu,[\nu,\rho]}$ as follows
\begin{align}
\label{gtrf1}    & \delta \Phi_{\mu\nu,\rho,\sigma} = -
\textstyle\frac{i}{4}
\partial_{\{\mu} \phi^{(vi)1}_{1|\nu\},[\rho,\sigma]}+
\frac{{i}}{4}
\partial_{\rho} \phi^{(vi)1}_{1|\{\mu,[\nu\},\sigma]} + \frac{{i}}{4}
\partial_{\sigma} \phi^{(vi)1}_{1|\{\mu,[\hat{\rho}, \nu]\}} ,\\
\label{gtrf2} & \delta \phi^{(v)}_{1| \mu,\nu} =\textstyle
\frac{i}{2}\partial_\mu\phi^{(vi)1}_{1|}{}^\rho{}_{,[\rho,\nu]} +
\frac{i}{2}\partial_\nu\phi^{(vi)1}_{1|}{}^\rho{}_{,[\mu,\rho]} -
\frac{i}{2}
\partial^\rho\phi^{(vi)1}_{1|
\rho,[\mu,\nu]},\\
\label{gtrf3}&
\delta\phi^{\prime\prime\prime}_{76|\mu,\nu}=\textstyle
\frac{i}{2}\partial_\nu\phi^{(vi)1}_{1|}{}^\rho{}_{,[\rho,\mu]}-
\frac{i}{2}\partial_\mu\phi^{(vi)1}_{1|}{}^\rho{}_{,[\rho,\nu]}  +
\frac{i}{2}\partial^\rho\phi^{(vi)1}_{1|\rho,[\mu,\nu]} ,\\
\label{gtrf4}&  \delta\phi^{}_{101|\mu,\nu}=-\Bigl(\textstyle
\frac{i}{2}\partial_\nu\phi^{(vi)1}_{1|}{}^\rho{}_{,[\rho,\mu]}-
\frac{i}{2}\partial_\mu\phi^{(vi)1}_{1|}{}^\rho{}_{,[\rho,\nu]}  +
\frac{i}{2}\partial^\rho\phi^{(vi)1}_{1|\rho,[\mu,\nu]}\Bigr) ,
\end{align}
and for $\eta_0$-dependent terms in field vector
$|\chi\rangle_{(s)_3}$ derived from the Eqs.
(\ref{gtr00})--(\ref{gtr005})
\begin{align}\label{gtr00f}
 & \delta
\phi^{(vi) }_{01|\mu,\nu,\rho} = \textstyle- \frac{1}{2}
\partial_{[\rho}\partial^{\sigma}\phi^{(vi)1}_{1|\mu,[\nu],\sigma]}
+ \frac{1}{2}\Box \phi^{(vi)1}_{1|\mu,[\nu,\rho]} , \\
\label{gtr00f1}& \delta \phi^{}_{02|\mu} =  -
\textstyle\frac{1}{2}\partial_\mu
 \partial^{\rho}\phi^{(vi)1}_{1|}{}^\nu{}_{,[\nu,\rho]} + \frac{1}{2}
\partial^\nu\partial^{\rho}\phi^{(vi)1}_{1|\nu,[\mu,\rho]}
+
\textstyle\frac{1}{2}\Box \phi^{(vi)1}_{1|}{}^\nu{}_{,[\nu,\mu]}, \\
\label{gtr00f2}& \delta \phi^{}_{075|\mu\nu,\rho} =
-\textstyle\frac{1}{4}\partial_{\{\nu}
\partial^{\sigma}\phi^{(vi)1}_{1|\mu\},[\rho,\sigma]}
+ \frac{1}{4}\partial_{\rho}
\partial^{\sigma}\phi^{(vi)1}_{1|\{\mu,[\nu\},\sigma]}
- \frac{1}{4}\Box
\phi^{(vi)1}_{1|\{\mu,[\nu\},\rho]} ,\\
 \label{gtr00f3}& \delta \phi^{
}_{0100|\mu\nu,\rho} = \textstyle\frac{1}{4}\partial_{\{\nu}
\partial^{\sigma}\phi^{(vi)1}_{1|\mu\},[\rho,\sigma]}
- \frac{1}{4}\partial_{\rho}
\partial^{\sigma}\phi^{(vi)1}_{1|\{\mu,[\nu,\sigma]}
+ \frac{1}{4}\Box
\phi^{(vi)1}_{1|\{\mu,[\nu\},\rho]} ,\\
\label{gtr00f4} & \delta \phi^{}_{017|\mu} =
\textstyle\frac{1}{2}\partial_\mu
 \partial^{\rho}\phi^{(vi)1}_{1|}{}^\nu{}_{,[\nu,\rho]} - \frac{1}{2}
\partial^\nu\partial^{\rho}\phi^{(vi)1}_{1|\nu,[\mu,\rho]}-
\textstyle\frac{1}{2}\Box
\phi^{(vi)1}_{1|}{}^\nu{}_{,[\nu,\mu]},\\
\label{gtr00f5} & \delta \phi^{}_{076|\mu} = \textstyle  -
\textstyle\frac{1}{2}\partial_\mu
 \partial^{\rho}\phi^{(vi)1}_{1|}{}^\nu{}_{,[\nu,\rho]} + \frac{1}{2}
\partial^\nu\partial^{\rho}\phi^{(vi)1}_{1|\nu,[\mu,\rho]}+
\textstyle\frac{1}{2}\Box \phi^{(vi)1}_{1|}{}^\nu{}_{,[\nu,\mu]} .
\end{align}
The gauge transformations for only independent (because of the
Eqs. (\ref{resol1115}))  gauge parameters
$\phi^{(vi)1}_{1|\rho,[\mu,\nu]}$ now reduce to the final answer
\begin{equation}\label{gtrfin1}
\delta \phi^{(vi) 1}_{1|\mu,[\nu,\rho]} = 2i
\partial_{[\nu}\phi^{
\prime\prime\prime2}_{1|\mu,\rho]}\texttt{ and } \delta\phi^{
\prime\prime\prime2}_{1|\mu,\rho} = -i\partial_\rho \phi^{
3}_{1|\mu]}.
\end{equation}
The structure of the gauge parameter $|\chi^1_g\rangle_{(s)_3}$
now is completely determined, whereas the field vector
$|\chi\rangle_{(s)_3}$ is reduced to $|\chi_g\rangle_{(s)_3}$
which has the same structure as in (\ref{x-0}) but with
sighnificantly less number of components.  For instance, gauged
vector $|\Psi_g\rangle_{(2,1,1)}$ contains only initial tensor,
$|\Phi\rangle_{(2,1,1)} =
a^{+\mu}_1a^{+\nu}_1a^{+\rho}_2a^{+\sigma}_3|0\rangle
\Phi_{\mu\nu,\rho,\sigma}$  and $|\chi_{0|g}\rangle_{(s)_3}$ has
the same structure as $|\chi^1_{g}\rangle_{(s)_3}$.

Let us turn to the removing of the rest auxiliary fields in the
field vector $|\chi_g\rangle_{(s)_3}$ by means of the resolution
of the part of the equations of motion.

\subsubsection{Gauge-invariant unconstrained
Lagrangian}\label{glex}

Now, we should find the result of the BRST operator $Q$ action on
the gauged field vector
 $|\chi_g\rangle_{(s)_3}$, in
 order to  solve algebraic equations of motion from the last
general relation (\ref{Q123}), i.e.,  $Q|\chi_g\rangle_{(s)_3} =
0$.  We start from the obvious consequences that, all the fields
$|\phi_{n}\rangle_{(s)_3}$, for $n=26, \ldots , 42, 85, \ldots, 90
$ and $n= 106, 108, \ldots , 161, 190$  with $\mathcal{P}_{ij}^+$
multipliers should vanish except for the already gauged scalars
$\phi^{}_{144}-\phi^{}_{146}, \phi^{}_{154}, \phi^{}_{159}$ (as in
case of gauge fixing procedure for $|\chi\rangle_{(s)_3}$).
Second,
 we resolve $\eta_0$-dependent part of the equations as it have already done for
the gauge vector $|\chi^1_g\rangle_{(s)_3}$ and  find
\begin{align}
\label{ident0f} & \Bigl(\phi^{(vi)}_{01|\mu,\{\nu,\rho\}} = 0
\Longrightarrow \phi^{(vi)}_{01|\mu,[\nu,\rho]} \ne 0\Bigr), &
\phi^{}_{076|\mu} = \frac{1}{2}
\phi^{(vi)}_{01|}{}^\nu{}_{,[\nu,\mu]}, &&
\frac{1}{2}\phi^{(vi)}_{01|}{}^\nu{}_{,[\mu,\nu]}=
\phi^{}_{017|\mu},\\& \label{ident1f}  \frac{1}{2}
\phi^{(vi)}_{01|\{\mu,[\hat{\rho},\nu]\}}=
\phi^{}_{0100|\mu[\rho,\nu]},
&\frac{1}{2}\phi^{(vi)}_{01|}{}^{\nu}{}_{,[\nu,\mu]}=\phi^{}_{02|\mu},
, && \frac{1}{2}\phi^{(vi)}_{01|\{\mu,[\nu\},\rho]}= -
\phi^{}_{075|\mu[\nu,\rho]}.
\end{align}
Therefore, the only gauge transformations (\ref{gtr00f}) for the
independent field $\phi^{(vi)}_{01|\mu,[\nu,\rho]}$ remains and
are written as,
\begin{equation}\label{gtr00fin}
  \delta
\phi^{(vi) }_{01|\mu,[\nu,\rho]} = \textstyle-
\partial_{[\rho}\partial^{\sigma}\phi^{(vi)1}_{1|\mu,[\nu],\sigma]}
+ \Box \phi^{(vi)1}_{1|\mu,[\nu,\rho]} .
\end{equation}
 Then, from the system of the $\eta_0$-independent part of the
Lagrangian equations of motion, $Q|\chi_g\rangle_{(s)_3} = 0$,
considered for the linear terms in ghost $\mathcal{C}^I$, then  in
third $\mathcal{P}_K\mathcal{C}^I\mathcal{C}^J$ and finally, in
the fifth
$\mathcal{P}_{K_1}\mathcal{P}_{K_2}\mathcal{C}^I\mathcal{C}^J\mathcal{C}^K$
powers in ghost we get the set of equations, which permits to
completely remove the rest auxiliary (gauge-independent) fields in
$|\chi_g\rangle_{(s)_3}$:
\begin{eqnarray}
&& \phi^{\prime\prime}_{1|\mu\nu}, \phi^{(iv)}_{1|\mu\nu},
\phi^{(6)}_{1|\mu,\nu}, \phi^{}_{2|\mu,\nu},
\phi^{\prime\prime\prime}_{2|\mu,\nu}, \phi^{}_{6|\mu},
\phi^{}_{13}, \phi^{}_{15|\mu}-\phi^{}_{17|\mu}, \phi^{}_{21},
\phi^{}_{23}, \phi^{}_{24|\mu}, \phi^{}_{43|\mu},
\phi^{}_{44|\mu}, \label{lemalg}\\
&& \phi^{}_{45}-\phi^{}_{48}, \phi^{\prime}_{45},
\phi^{}_{49|\mu\nu}-\phi^{}_{51|\mu\nu},
\phi^{\prime}_{51|\mu,\nu}, \phi^{}_{52}, \phi^{}_{53|\mu\nu},
\phi^{}_{54}, \phi^{}_{55|\mu}-\phi^{}_{58|\mu},
\phi^{\prime}_{57|\mu}, \phi^{}_{60|\mu}, \phi^{}_{62},\nonumber\\
&& \phi^{}_{64}, \phi^{}_{66|\mu}, \phi^{}_{69},
\phi^{}_{70|\mu\nu}, \phi^{}_{70|\mu,\nu}, \phi^{}_{72},
\phi^{}_{73|\mu\nu}, \phi^{}_{74|\mu}, \phi^{}_{75|\mu},
\phi^{}_{78|\mu\nu}, \phi^{}_{79|\mu},
\phi^{\prime\prime}_{79|\mu}, \phi^{}_{81|\mu},
\phi^{}_{83|\mu\nu\rho},\nonumber\\
&& \phi^{}_{84|\mu\nu\rho}, \phi^{\prime}_{84|\mu\nu,\rho},
\phi^{\prime\prime}_{84|\mu,\nu\rho}, \phi^{}_{91},
\phi^{}_{92|\mu\nu}, \phi^{}_{93|\mu}, \phi^{}_{94},
\phi^{}_{95|\mu}, \phi^{}_{97}, \phi^{}_{98|\mu\nu}, \phi^{}_{99},
\phi^{}_{100|\mu}, \phi^{}_{102|\mu\nu}, \nonumber\\
&&\phi^{}_{103|\mu}, \phi^{}_{104|\mu}, \phi^{}_{105|\mu\nu\rho},
\phi^{}_{107}, \phi^{}_{162|\mu\nu,\rho},
\phi^{}_{164|\mu\nu\rho},
\phi^{}_{165|\mu\nu}-\phi^{\prime\prime\prime}_{165|\mu,\nu},
\phi^{}_{166|\mu\nu}, \phi^{}_{167|\mu\nu},
\phi^{}_{169|\mu\nu\rho\sigma},
\nonumber\\
&& \phi^{}_{170|\mu\nu\rho\sigma},
\phi^{}_{170|\mu\nu\rho,\sigma}, \phi^{}_{171|\mu},
\phi^{}_{172|\mu\nu }, \phi^{}_{173|\mu}, \phi^{\prime}_{173|\mu},
\phi^{}_{174|\mu}, \phi^{}_{175|\mu\nu\rho}, \phi^{}_{176|\mu},
\phi^{}_{177}, \phi^{}_{178|\mu\nu},\nonumber\\
&&  \phi^{}_{179|\mu\nu\rho}, \phi^{}_{179|\mu\nu,\rho},
\phi^{}_{180|\mu\nu\rho}, \phi^{}_{182|\mu\nu},
\phi^{}_{183|\mu\nu\rho\sigma},
\phi^{}_{184|\mu\nu\rho}-\phi^{\prime\prime}_{184|\mu,\nu\rho},
\phi^{}_{185|\mu\nu\rho}, \phi^{}_{185|\mu\nu,\rho},\nonumber\\
&& \phi^{\prime}_{186|\mu,\nu}, \phi^{}_{189|\mu\nu\rho\sigma},
\phi^{}_{189|\mu\nu\rho,\sigma}\footnotemark \nonumber
\end{eqnarray}\footnotetext{all the rest component fields in
$|\chi\rangle_{(s)_3}$ was gauged away within the procedure
described in the Subsection~\ref{gtrex} except for the fields with
nontrivial gauge transformations in (\ref{gtrf1})--(\ref{gtrf4})}
and to derive the relations
\begin{eqnarray}
 \label{bas23ant}  &&\hspace{-2em}
\Phi_{\mu\nu,\{\rho,\sigma\}}=0 \Longrightarrow
\Phi_{\mu\nu,[\rho,\sigma]}\ne 0, \qquad
        \Phi_{\{\mu\nu,\rho\},\sigma}=0,\qquad
\quad \Phi_{\{\mu\nu,\hat{\sigma},\rho\}} =0,\\
\label{basphi1}&& \hspace{-2em}\Phi^\rho{}_{\rho,\mu,\nu}
  =  \phi^{(v)}_{1|\mu,\nu}, \qquad   \Phi^\rho{}_{\mu,\rho,\nu} \ =\
\frac{{1}}{2}
 \phi^{\prime\prime\prime}_{76|\mu,\nu}, \qquad  \Phi^\rho{}_{\mu,\nu,\rho} =
\frac{{1}}{2}
 \phi^{}_{101|\mu,\nu}\\
&&\label{basphi2}
\hspace{-2em}\Bigl(\phi^{\prime\prime\prime}_{76|\mu,\nu} = -
\phi^{}_{101|\mu,\nu},  \phi^{(v)}_{1|\nu,\mu}= -
\phi^{}_{101|\mu,\nu} ,  \phi^{(v)}_{1|\{\mu,\nu\}} =0\Bigr)
\Longrightarrow
\phi^{(v)}_{1|[\mu,\nu]}=\phi^{}_{101|[\mu,\nu]}=\phi^{\prime\prime\prime}_{76|[\nu,\mu]}
,\\
&& \label{basphi3} \hspace{-2em}i\partial^\nu
\phi^{(v)}_{1|\mu,\nu} =- \phi^{}_{017|\mu}.
\end{eqnarray}
From the equations at $\eta_1^+$  (in $Q|\chi_g\rangle_{(s)_3} =
0$) we may to  express the fields $\phi^{ (vi)}_{01|
\mu,[\nu,\rho]}$ in terms of only initial tensor
$\Phi_{\mu\nu,[\rho,\sigma]}$. To do this, we with using of the
following from the Eqs. (\ref{bas23ant})--(\ref{basphi2})
expressions for $\phi^{\prime\prime\prime }_{76| \mu,\rho},
\phi^{(v) }_{1|\nu,\rho}, \phi^{ }_{101| \nu,\rho}$
\begin{equation}
    \label{basphifin} \Phi^\rho{}_{\rho,[\mu,\nu]}
  =  \phi^{(v)}_{1|[\mu,\nu]}, \quad
\Phi^\rho{}_{[\mu,[\rho,\nu]]} \ =\
 \phi^{\prime\prime\prime}_{76|[\mu,\nu]}, \quad
\Phi^\rho{}_{[\mu,[\nu],\rho]} =
 \phi^{}_{101|[\mu,\nu]},
\end{equation}
 find
\begin{equation}\label{eqsxpress1}
  \phi^{ (vi)}_{01| \mu,[\nu,\rho]} =  - \textstyle {i}\partial_{\nu}
\Phi^\sigma{}_{[\mu,[\sigma,\rho]]}  - {i}\partial_{\mu}
\Phi^\sigma{}_{\sigma,[\nu,\rho]}  - {i}\partial_{\rho}
\Phi^\sigma{}_{[\mu,[\nu],\sigma]}  +2i
\partial^{\sigma} \Phi_{\mu\sigma,[\nu,\rho]}
.
\end{equation}
The set of the relations \ref{ident0f}, \ref{ident1f}
(\ref{basphifin}), (\ref{eqsxpress1}) together with one
independent Young symmetry condition from the Eqs.(\ref{bas23ant})
for nonvanishing components $\Phi_{\mu\sigma,[\nu,\rho]}$:
\begin{equation}\label{Young symmetry1}
   \Phi_{\{\mu\nu,[\rho\},\sigma]}=0,
\end{equation}
permit  to express all the rest components from
$|\chi_g\rangle_{(s)_3}$ in terms of only initial tensor
$\Phi_{\mu\sigma,[\nu,\rho]}$. This tensor due to the property
(\ref{Young symmetry1}) is doubly traceless.  Note, that
equivalently, we may work with the tensor
$\widehat{\Phi}_{\mu\nu,[\rho,\sigma]}$,
\begin{equation}\label{irrepYTensor}
    \widehat{\Phi}_{\mu\nu,[\rho,\sigma]} \equiv
\Phi_{\mu\nu,[\rho,\sigma]}-\frac{1}{2}\Phi_{[\rho\mu,[\nu,\sigma]]}-\frac{1}{2}\Phi_{\mu[\rho,[\mu,\sigma]]}
\end{equation}
which satisfies to the Eq.(\ref{Young symmetry1}) identically.

The corresponding field vector $|\chi_g\rangle_{(s)_3}$ reduces to
the form of $|\chi_{f}\rangle_{(s)_3}$, ($|\chi_{f}\rangle_{(s)_3}
= |\chi_{fb}\rangle_{(s)_3}+\eta_0|\chi_{0f}\rangle_{(s)_3}$),
\begin{eqnarray} \label{x-0fin}
|\chi_{f}\rangle_{(s)_3} & = &|\Phi\rangle_{(2,1,1)}+
\mathcal{P}_{1}^+\eta_1^+|\phi_{1|f}\rangle_{(0,1,1)}+\mathcal{P}_{2}^+\eta_1^+|\phi_{76|f}\rangle_{(1,0,1)}+
\mathcal{P}_{3}^+\eta_1^+|\phi_{101|f}\rangle_{(1,1,0)}\\
&& \hspace{-2em}  +\
\eta_0\Bigl[\mathcal{P}_{1}^+\Bigl\{|\phi^{}_{01|f}\rangle_{(1,1,1)}-
\eta_1^+\Bigl(\mathcal{P}_{2}^+|\phi^{}_{02|f}\rangle_{(0,0,1)}+
\mathcal{P}_{3}^+|\phi^{}_{017|f} \rangle_{(0,1,0)}\Bigr)
\Bigr\}\nonumber \\
&&\hspace{-2em} +\
\mathcal{P}_{2}^+\Bigl\{|\phi^{}_{075|f}\rangle_{(2,0,1)}+
\mathcal{P}_{3}^+
\eta_1^+|\phi^{}_{076|f}\rangle_{(1,0,0)}\Bigr\}+
\mathcal{P}_3^+|\phi^{}_{0100|f} \rangle_{(2,1,0)}
\Bigr],\nonumber
\end{eqnarray}
where all the vectors $|\phi^{}_{...|f} \rangle_{(s)_3}$ are
expressed in terms of only components of
$\Phi_{\mu\sigma,[\nu,\rho]}$. So, the vectors $|\phi^{}_{0n|f}
\rangle_{(s)_3}$ have the structure,
\begin{align}
 \label{0fieldfin1} &
\phi^{}_{0100|\mu[\rho,\nu]} = i
\partial^{\sigma} \Phi_{\{\mu\sigma,[\rho,\nu\}]}-
\frac{i}{2}\partial_{\{\nu} \Phi^\sigma{}_{\mu\},[\rho,\sigma]} -
\frac{i}{2}\partial_{\{\mu} \Phi^\sigma{}_{\sigma,[\rho,\nu\}]}
 +
\frac{i}{2}\partial_{\{\mu} \Phi^\sigma{}_{\rho,[\nu\},\sigma]},\\
  \label{0fieldfin12}
& \phi^{}_{076|\mu} = \frac{i}{2}\partial^\rho\Bigl(
\Phi^{\sigma}{}_{\rho,[\sigma,\mu]} -
\Phi^{\sigma}{}_{\sigma,[\rho,\mu]} \Bigr) , \quad
\phi^{}_{076|\mu}= \phi^{}_{02|\mu} =\ - \phi^{}_{017|\mu},\\
& \phi^{}_{0100|\mu[\rho,\nu]}= -
\phi^{}_{075|\mu[\rho,\nu]}.\nonumber
\end{align}
Corresponding bra-vector ${}_{(s)_3}\langle\chi_{f}|$
 from dual Fock space reads as,
\begin{eqnarray}\label{x-0fbra}
{}_{(s)_3}\langle\chi_{f}| &=& {}_{(2,1,1)}\langle\Phi| +
{}_{(0,1,1)}\langle\phi_{1|f}|\eta_1\mathcal{P}_{1}{}_{(1,0,1)}+
{}_{(1,0,1)}\langle\phi_{76|f}|\eta_1\mathcal{P}_{2}+{}_{(1,1,0)}\langle\phi_{101|f}|\eta_1
\mathcal{P}_{3}\\
&& \hspace{-2em} + \Bigl[\Bigl\{{}_{(1,1,1)}\langle\phi_{01|f}| -
\Bigl( {}_{(0,0,1)}\langle\phi_{02|f}|\mathcal{P}_{2} +
{}_{(0,1,0)}\langle\phi_{017|f}|\mathcal{P}_3\Bigr)
|\eta_{1}\Bigr\}\mathcal{P}_{1}
\nonumber\\
&& \hspace{-2em} + \Bigl\{{}_{(2,0,1)}\langle\phi_{075|f}|+
{}_{(1,0,0)}\langle\phi_{076}|\eta_{1}\mathcal{P}_3\Bigr\}\mathcal{P}_2
+
{}_{(2,1,0)}\langle\phi_{0100|f}|\mathcal{P}_3\Bigr]\eta_0,\nonumber
\end{eqnarray}
where, for instance, bra-vector  ${}_{(2,1,1)}\langle\Phi|$
 has the form,
\begin{eqnarray}
\label{brabasdecomp} && {}_{(2,1,1)}\langle\Phi| =
\frac{1}{2}\Phi_{\mu\nu,[\rho,\sigma]}\langle0|
a_1^{\mu}a_1^{\nu}a_2^{\rho} a_3^{\sigma}.
 \label{braxdecompf}
\end{eqnarray}

Then the relation (\ref{Scl3}) with account for  $K_{(2,1,1)}=1$
on Fock space $\mathcal{H}\bigotimes \mathcal{H}_{gh}$ gives the
following action for $ (s)_3=(2,1,1)$,
\begin{eqnarray}\label{S(2,1,1)}
{\cal S}_{(s)_3} =       {}_{(s)_3}\langle \chi_{fb}
|\hspace{-0.1em}\Bigl\{l_0 | \chi_{fb}\rangle_{(s)_3}-\Delta Q|
\chi_{0f}\rangle_{(s)_3}
 \hspace{-0.1em}\Bigr\}\hspace{-0.1em} -\hspace{-0.1em} {}_{(s)_3}\langle \chi_{0f} |\Bigl\{\hspace{-0.1em}
\Delta Q| \chi_{fb}\rangle_{(s)_3}-
\hspace{-0.1em}\sum_i\hspace{-0.1em}\eta_i^+ \eta_i
|\chi_{0f}\rangle_{(s)_3}\hspace{-0.1em}\Bigr\}\hspace{-0.1em}.
\end{eqnarray}
Explicitly Lagrangian looks in terms of the tensor
$\Phi_{\mu\nu,[\rho,\sigma]}$ and all the  auxiliary fields (which
 should be then expressed in terms of only $\Phi^{\mu\nu,[\rho,\sigma]}$) as\footnote{in order to provide reality of the
Lagrangian due to sign "-" in the commutation relations,
$[a^{\mu}_i,a^{+\nu}_j] = - \eta^{\mu\nu}\delta_{ij}$,  we should
later to connect with each of the Lorentz index the imaginary unit
for the vector, second, third, fourth and so on rank tensors by
the rule, $\phi^{}_{\mu_1,...,\mu_n} \to \imath^n
\hat{\phi}^{}_{\mu_1,...,\mu_n}$, so that
$\phi^{}_{\mu,\nu,\rho,\sigma} =
\hat{\phi}^{}_{\mu,\nu,\rho,\sigma}$, $\phi^{}_{\mu,\nu,\rho} =
-\imath\hat{\phi}^{}_{\mu,\nu,\rho}$, $\phi^{}_{\mu,\nu} =
-\hat{\phi}^{}_{\mu,\nu}$, $\phi^{}_{\mu} =
\imath\hat{\phi}^{}_{\mu}$
 and must change the corresponding signs
in all above formulas with components with writing the hat "$
\hat{\phantom{\phi}}$" over the fields},
\begin{eqnarray}\label{S(2,1,1)c}
{\cal L}_{(2,1, 1)}\hspace{-0.3em} &\hspace{-0.5em} = \hspace{-0.5em}& \frac{1}{2}\Phi^{\mu\nu,[\rho,\sigma]}
\Box \Phi_{\mu\nu,[\rho,\sigma]}-
\frac{1}{4}\hat{\phi}^{(v)}_{1}{}^{[\mu,\nu]}\Box\hat{\phi}^{(v)}_{1|[\mu,\nu]}+
\frac{1}{4}\Phi^{\mu\nu,[\rho,\sigma]}\Bigl\{
\partial_{\{\mu} \hat{\phi}^{ (vi)}_{01| \nu\},[\rho,\sigma]}+ 2\partial_{\rho} \hat{\phi}_{075| \mu[\nu,\sigma]}
\\ && + 2\partial_{\sigma} \hat{\phi}_{0100| \mu[\nu,\rho]}\Bigr\}
 - \frac{1}{2}\hat{\phi}^{(v)}_{1}{}^{[\mu,\nu]}
\Bigl\{ \partial_\mu\hat{\phi}^{}_{02|\nu} +
\partial_\nu\hat{\phi}^{}_{017|\mu}+\frac{1}{2}\partial^\rho \hat{\phi}^{(vi)}_{01|\rho,[\mu,\nu]}\Bigr\}\nonumber\\
&& -\frac{1}{4}\hat{\phi}^{(vi)}_{01}{}^{\mu,[\nu,\rho]} \Bigl\{
\partial_\mu\hat{\phi}^{(v)}_{1|[\nu,\rho]} +
\partial_\nu\hat{\phi}^{\prime\prime\prime}_{76|[\mu,\rho]}+
\partial_\rho\hat{\phi}^{}_{101|[\mu,\nu]}+
2\partial^\sigma \Phi_{\sigma\mu,[\nu,\rho]}+ \hat{\phi}^{(vi)}_{01|\mu,[\nu,\rho]}\Bigr\}\nonumber\\
&&
+\frac{1}{4}\Bigl\{\hat{\phi}^{\prime\prime\prime}_{76}{}^{[\mu,\nu]}
-\hat{\phi}_{101}{}^{[\mu,\nu]}\Bigr\}\partial^\rho
\hat{\phi}^{(vi)}_{01|\mu,[\rho,\nu]}
+\frac{1}{2}\Bigl\{\hat{\phi}^{\mu}_{02}-\hat{\phi}^{\mu}_{017}\Bigr\}\partial^\rho\hat{\phi}^{(v)}_{1|[\rho,\mu]}+\hat{\phi}^{\mu}_{02}\hat{\phi}_{02|\mu}+\hat{\phi}^{\mu}_{017}\hat{\phi}_{017|\mu}
,\nonumber\\
&& - \frac{1}{2}\hat{\phi}_{075}{}^{\mu[\nu,\sigma]}\Bigl\{
\hat{\phi}_{075|\mu[\nu,\sigma]}+\partial^\rho\Phi_{\mu\nu,[\rho,\sigma]}\Bigr\}
- \frac{1}{2}\hat{\phi}_{0100}{}^{\mu[\nu,\sigma]}\Bigl\{
\hat{\phi}_{0100|\mu[\nu,\sigma]}+\partial^\rho\Phi_{\mu\nu,[\sigma,\rho]}\Bigr\}
.\nonumber
\end{eqnarray}
Finally,
\begin{eqnarray}\label{S(2,1,1)fintot1}
{\cal L}_{(2,1, 1)} & = &\frac{1}{2}\Phi^{\mu\nu,[\rho,\sigma]}
\Bigl\{\Box \Phi_{\mu\nu,[\rho,\sigma]}+
\partial_{\{\mu} \bigl[\textstyle
\partial_{\rho} \Phi^\tau{}_{[\nu\},[\tau,\sigma]]}  +
{}\partial_{\nu\}} \Phi^\tau{}_{\tau,[\rho,\sigma]}  +
\partial_{\sigma} \Phi^\tau{}_{[\nu\},[\rho],\tau]}  \\
&&  -2
\partial^{\tau} \Phi_{\nu\}\tau,[\rho,\sigma]}\bigr] + 2\partial_{\rho} \bigl[\partial_{\{\mu}
\Phi^\tau{}_{\tau,[\sigma,\nu\}]}  +
\partial_{\{\nu} \Phi^\tau{}_{[\mu\},[\sigma],\tau]}  -2
\partial^{\tau} \Phi_{\{\mu\tau,[\sigma,\nu\}]}\bigr]\Bigr\} \nonumber\\
&&
 - \frac{1}{4}\Phi_\tau{}^{\tau,[\mu,\nu]}
\Bigl\{\Box\Phi^\sigma{}_{\sigma,[\mu,\nu]} -
2\partial_\nu\partial^\rho\Bigl[\Phi^\sigma{}_{\sigma,[\rho,\mu]}-\Phi^\sigma{}_{\rho,[\sigma,\mu]}\Bigr]
\Bigr\}\nonumber\\
&& -  2\Phi_\sigma{}^{[\mu,[\sigma,\nu]]}
\partial^{\rho}\partial^{\tau} \Phi_{\mu\tau,[\rho,\nu]}
- \Phi^\sigma{}_{\sigma,}{}^{[\mu,\nu]}
\partial^{\rho}\partial^{\tau} \Phi_{\rho\tau,[\mu,\nu]} + 2
 \Phi^{\mu\nu,[\rho,\sigma]}\partial_{\nu}
\partial^{\tau} \Phi_{\mu\tau,[\rho,\sigma]}\nonumber\\
&& +\frac{1}{2}\Phi_\sigma{}^{\mu}{}_,{}^{[\sigma,\nu]}
\partial_\mu\partial^\rho\Bigl\{
\Phi^\tau{}_{\rho,[\tau,\nu]}-\Phi^\tau{}_{\tau,[\rho,\nu]}\Bigr\}
 .\nonumber
\end{eqnarray}
The Lagrangian  (\ref{S(2,1,1)fintot1}) is invariant with accuracy
up to total derivative with respect to  the gauge transformations
\begin{align}
    & \delta \Phi_{\mu\nu,[\rho,\sigma]} = - \textstyle\frac{1}{2}
\partial_{\{\mu} \phi^{(vi)1}_{1|\nu\},[\rho,\sigma]}+
\frac{{1}}{2}
\partial_{\rho} \phi^{(vi)1}_{1|\{\mu,[\nu\},\sigma]} +
\frac{{1}}{2}
\partial_{\sigma} \phi^{(vi)1}_{1|\{\mu,[\hat{\rho}, \nu]\}}
,\label{gfinal}
\end{align}
where gauge transformations for the gauge parameters
$\phi^{(vi)1}_{1|\rho,[\mu,\nu]}$ now reduce to the final form
with omitting the "$\hat{}$" over the gauge parameters
\begin{equation}
\delta \phi^{(vi) 1}_{1|\mu,[\nu,\rho]} = 2
\partial_{[\nu}\phi^{
\prime\prime\prime2}_{1|\mu,\rho]}\texttt{ and } \delta\phi^{
\prime\prime\prime2}_{1|\mu,\rho} = -\partial_\rho \phi^{
3}_{1|\mu]}. \label{redgfinal}
\end{equation}

Thus, we have obtained the gauge-invariant Lagrangian
(\ref{S(2,1,1)fintot1}) in terms of only initial free massless
mixed-symmetric tensor field $\Phi_{\mu\nu,[\rho,\sigma]}$. The
resulting theory is the second-stage reducible gauge theory. The
formulae (\ref{S(2,1,1)fintot1})--(\ref{redgfinal}) present our
basic result in the Section~\ref{ex211}. The case of massive
tensor field of spin $(2,1,1)$ maybe explicitly derived from the
Eqs. (\ref{S(2,1,1)fintot1})--(\ref{redgfinal}) by the dimensional
reduction procedure.


\section{Conclusions}

In the given paper, we have constructed a gauge-invariant
Lagrangian description of free integer  HS fields belonging to an
irreducible representation of the Poincare group $ISO(1,d-1)$ with
the arbitrary Young tableaux having $k$ rows in the ``metric-like"
formulation. The results of this study are the general and
obtained on the base of universal method which is applied by the
unique way to both massive and massless bosonic HS fields with a
mixed symmetry in a Minkowski space of any dimension. One should
be noted the Lagrangians for the massive arbitrary  HS fields on a
flat background have  not been derived until now in ``metric-like"
and ``frame-like" formulations\footnote{In principle, the massive
theory can be obtained from massless by dimensional reduction.
However, we suggest that it will require the same amount of work
as one for independent formulation.}.

We start from an embedding of bosonic HS fields into vectors of an
auxiliary Fock space, we treat the fields as coordinates of
Fock-space vectors and reformulate the theory in such terms. We
realize the conditions that determine an irreducible Poincare-group
representation  with a given mass and generalized spin in terms of
differential operator constraints imposed on the Fock space vectors.
These  constraints generate a closed Lie algebra of HS symmetry,
which contains, with the exception of $k$ basis generators of its
Cartan subalgebra, a system of first- and second-class constraints.
Above algebra coincides modulo isometry group generators with its
Howe dual $sp(2k)$ symplectic algebra.

We demonstrate  that the construction  of a correct Lagrangian
description requires a deformation of the initial symmetry
algebra, in order to obtain from the system of mixed-class
constraints a converted system with the same number of first-class
constraints alone, whose structure provides the appearance of the
necessary number of auxiliary tensor fields with lower generalized
spins. We have shown that this purpose can be achieved with the
help of an additional Fock space, by constructing an additive
extension of a $sp(2k)$ symmetry subalgebra which consists of the
subsystem of second-class constraints alone and of the generators
of the Cartan subalgebra.

We have realized the Verma module construction \cite{Dixmier} in
order to obtain an auxiliary representation in Fock space for the
above algebra with second-class constraints. As a consequence, the
converted Lie algebra of HS symmetry has the same algebraic
relations as the initial algebra; however, these relations are
realized in an enlarged Fock space. The generators of the
converted Cartan subalgebra contain linearly $k$ auxiliary
independent number parameters, whose choice provides the vanishing
of these generators in the corresponding subspaces of the total
Hilbert space extended by the ghost operators in accordance with
the minimal BFV--BRST construction for the converted HS symmetry
algebra. Therefore, the above generators, enlarged by the ghost
contributions up to the ``particle number'' operators  in the
total Hilbert space, covariantly determine Hilbert subspaces in
each of which the converted  symmetry algebra consists of
 the first-class constraints alone, labeled by the values
of the above parameters, and constructed from the initial
irreducible Poincare-group relations.

It is shown that the Lagrangian description corresponding to the
BRST operator, which encodes the converted HS symmetry algebra,
yields a consistent Lagrangian dynamics for bosonic fields of any
generalized spin.  The resulting Lagrangian description, realized
concisely in terms of the total Fock space, presents a set of
generating relations for the action and the sequence of gauge
transformations for given bosonic HS fields with a sufficient set
of auxiliary fields, and  proves to be a reducible gauge theory
with a finite number of reducibility stages, increasing with the
value of number of rows in the Young tableaux.

It is proved the fact that the solutions of the Lagrangian
equations of motion (\ref{Q12}) after a partial gauge-fixing and
resolution of the part of the equations of motion,
 correspond to the BRST cohomology space with a vanishing ghost number,
 which is determined only by the relations that extract the fields of
 an irreducible Poincare-group representation with a given value of
 generalized spin. One should be noted the case of totally
 antisymmetric tensors developed in Ref.\cite{brst1} is contained
 in the general Lagrangian formulation for $s_1=s_2=...=s_k=1$,
 $k=[d/2]$.

We  demonstrated that the general procedure contains as the
particular case the Lagrangian formulation for the mixed-symmetry
bosonic tensors subject to Young tableaux with two rows, developed
earlier in \cite{BurdikPashnev}, \cite{BuchKrycRysTak}, and
derived in the first time the new unconstrained Lagrangian
formulation in (\ref{Q123})--(\ref{Scl3}) for the mixed-symmetry
HS fields with three groups of symmetric indices subject to Young
tableaux with three rows. We applied the latter algorithm to get
 new Lagrangian (\ref{S(2,1,1)fintot1}) and its reducible gauge symmetries (\ref{gfinal}), (\ref{redgfinal})
for spin $(2,1,1)$
massless field in terms of only initial tensor of the fourth rank.
Obtained result permits to immediatedly enlarge the found
Lagrangian formulation on to one for  HS tensor of spin
$(2,1,\ldots,1))$ subject to Young tableaux with $k$ rows.

There are many directions for extensions of the results obtained
in this paper. We point out some of them. First, development of
the analogous approach to fermionic HS fields with arbitrary Young
tableaux. Second, Lagrangian construction for bosonic and
fermionic fields with arbitrary index symmetry on AdS space.
Third, derivation of component Lagrangians for simple enough but
new cases. Fourth, developing the unconstrained formulation for
fields with arbitrary Young tableaux analogously to component
formulation with minimal number of auxiliary fields given in
\cite{quartmixbosemas} for totally symmetric fields which  as well
(as it have shown in \cite{quartmixbosemas}) can be derived from
the obtained  general Lagrangian formulation by means of partial
gauge fixing procedure. We are going to study these problems in
the forthcoming works.

\section*{Acknowledgements}
The authors are grateful to V.A. Krykhtin, P.M. Lavrov and Yu.M.
Zinoviev for discussions,  to G. Bonelli, D. Francia, A. Sagnotti,
W. Siegel, A. Waldron for correspondence and to referee for
suggestion to consider the nontrivial example. A.R. is thankful to
K. Alkalaev and V. Gershun for the comments on conversion
procedure, structure of constraints and to D. Francia for useful
comments of the results. I.L.B is grateful to CAPES for supporting
his visit to to Physics Department of Universidade Federal de Juiz
de Fora, Brazil where the final part of the work was done. The
authors are grateful to RFBR grant, project No. 12-02-000121 and
grant for LRSS, project No. 224.2012.2 for partial support. Work
of I.L.B was supported in part by  the RFBR-Ukraine grant, project
No. 11-02-93986. Work of A.R. was partially supported by the RFBR
grant, project No. 11-02-08283.

\appendix
\section*{Appendix}

\section{Additional parts construction for $sp(2k)$ algebra}\label{addalgebra}
\renewcommand{\theequation}{\Alph{section}.\arabic{equation}}
\setcounter{equation}{0}

In this appendix, we describe the method of auxiliary
representation  construction (known for mathematicians as Verma
module \cite{Dixmier}) for the symplectic algebra $sp(2k)$ with
second-class constraints $\{o'_a, {o'}^+_a \} = \{l'_{ij}, ,
t^{\prime ij}, l^{\prime +}_{ij}, t^{\prime +}_{ij}\}$ and Cartan
subalgebra elements $g_0^{\prime i}$ having in mind the
identification of $sp(2k)$ elements and ones of HS symmetry
algebra $\mathcal{A}(Y(k), R^{1,d-1})$ given by the
Eqs.(\ref{sp2nhssa}).

Following to Poincare--Birkhoff--Witt  theorem, we start to
construct Verma module, based on Cartan  decomposition of $sp(2k)$
($i\leq j$, $l<m$, $i,j, l, m = 1,...,k$)
\begin{equation}\label{Cartandecomp}
    sp(2k) =  \{l^{\prime +}_{ij},
t^{\prime+}_{lm}\} \oplus \{g_0^{\prime i}\} \oplus \{l^{\prime
}_{ij}, t'_{lm}\} \equiv \mathcal{E}^-_k\oplus H_k
\oplus\mathcal{E}^+_k.\footnotemark
\end{equation}
\footnotetext{we may consider $sp(2k)$ in Cartan-Weyl basis for
unified description, however without loss of generality the basis
elements and structure constants of the algebra under
consideration will be chosen as in the table~\ref{table in}.}

 Note, that in contrast to the case of
 totally-symmetric bosonic HS fields on $\mathbb{R}^{1,d-1}$ the negative root vectors
 from  $\mathcal{E}^-_k$ do not commute already for $k
 \geq 2$ (see,
Refs. \cite{BurdikPashnev}, \cite{BuchKrycRysTak}). However, we
consider highest weight representation of the symplectic algebra
$sp(2k)$ with highest weight vector $|0\rangle_V$, which must
annihilate by the positive roots $E^{\alpha_i}\in
\mathcal{E}^+_k$, and being by the proper one for the Cartan
elements $g_0^i$,
\begin{align}\label{hwrep}
& E^{\alpha_i}|0\rangle_V =0 && g_0^i |0\rangle_V =
h^i|0\rangle_V.
\end{align}
The general vector of Verma module $V(sp(2k))$ compactly written
as, $|\vec{N}\rangle_V$, has the form in terms of occupation
numbers,$
 |\vec{N}\rangle_V = \left|
{\vec{n}}_{ij},{\vec{p}}_{rs} \rangle_V \right.$,
\begin{equation}\label{VMk}
\left| {\vec{n}}_{ij},{\vec{p}}_{rs} \rangle_V \right. =  \left|
{{n}}_{11},...,{{n}}_{1k}, n_{22},...,{{n}}_{2k},...,{{n}}_{kk};
{p}_{12},\ldots, {p}_{1k}, {p}_{23},\ldots,
{p}_{2k},\ldots,p_{k-1k}\rangle_V \right.,
\end{equation}
where the coordinates $n_{ij}, p_{rs}$ mean the exponents of
corresponding negative root vector $E^{-\alpha_i} \in
\mathcal{E}_k^-$, determined as
\begin{equation}\label{VM}
   |\vec{N}\rangle_V \equiv  \prod_{i,j=1,i\leq j}^k\bigl(l^{\prime
+}_{ij}\bigr){}^{n_{ij}}\prod_{r=1}^{k-1}\Bigr[\prod_{s=r+1}^{k}
\bigl(t^{\prime +}_{rs}\bigr){}^{p_{rs}}\Bigl] |0\rangle_V,
\end{equation}
and non-negative integers $n_{ij}, p_{rs}$. It is easy to obtain
the result of the action of negative root vectors, i.e.
$(l^{'+}_{i'j'}, t^{'+}_{r's'})$ and Cartan generators,
$g_{0}^{\prime i} $ on $|\vec{N}\rangle_V$
\begin{eqnarray}
l^{\prime+}_{i'j'}|\vec{N}\rangle_V & = & \left|\vec{N} +
\delta_{i'j',ij}\rangle_V
 \right. \,,
 \label{l'+ij} \\
 t^{\prime+}_{r's'}  |\vec{N}\rangle_V & = & \left|{\vec{n}}_{ij},
{\vec{p}}_{rs} + \delta_{r's',rs} \rangle_V \right. -
\sum_{k'=1}^{r'-1}p_{k'r'}\left|{\vec{n}}_{ij},
{\vec{p}}_{rs} - \delta_{k'r',rs}+ \delta_{k's',rs} \rangle_V \right.  \\
  && - \sum_{k'=1}^{k}(1+\delta_{k'r'})n_{r'k'}\left|\vec{N}-
  \delta_{r'k',ij}  + \delta_{s'k',ij}\rangle_V
 \right.\footnotemark\,,
 \label{t'+lm}
 \nonumber \\
 g_{0{}i}^{\prime } |\vec{N}\rangle_V & = & \left( \sum_{l} (1+\delta_{il})n_{il}  - \sum_{s>i}p_{is}+\sum_{r<i}p_{ri} + h^i\right)
 \left|\vec{N}\rangle_V
 \right.\,.\label{g'0i}
\end{eqnarray}
\footnotetext{explicitly, for instance, the notation for the
vectors $\left|\vec{N} + \delta_{i'j',ij}\rangle_V\right.$ in the
Eq.(\ref{l'+ij}) means subject to definition (\ref{VMk})
increasing of only the coordinate $n_{ij}$ in the vector
$|\vec{N}\rangle_V$, for $i=i', j=j'$, on unit with unchanged
values of the rest ones, whereas the vector $\left|{\vec{n}}_{ij},
{\vec{p}}_{rs} - \delta_{k'r',rs}+ \delta_{k's',rs} \rangle_V
\right.$ implies increasing of  the coordinate $p_{rs}$, for
$r=k', s=s'$, on unit and decreasing on unit the coordinate
$p_{rs}$, for $r=k', s=r'$, with unchanged values of the rest
coordinates in $|\vec{N}\rangle_V$.}In deriving, two last
relations we have used the algebraic relations for $sp(2k)$ from
table~\ref{table in} and the formula for the product of operators
$A$, $B$, $n\geq 0$,
\begin{eqnarray}
\label{product} &&    AB^n = \sum^{n}_{k=0} \frac{n!}{k!(n-k)!}
B^{n-k}\mathrm{ad}^k_B{}A\,,   \   \mathrm{ad}^k_B{}A=
[[...[A,\stackrel{ k{\,} {\rm times}}{ \overbrace{B\},...\},B}\}},
\end{eqnarray}
Second, the Eq.(\ref{product}) permits to find both the
identities,
\begin{equation}\label{identllm}
l^{\prime }_{l'm'} \left|{\vec{0}}_{ij}, {\vec{p}}_{rs}\rangle_V
\right. = 0
\end{equation}
and the equation
 in acting of the positive root vectors $ t^{'}_{l'm'}$  on the vector $|\vec{0}_{ij},
\vec{p}_{rs}\rangle_V$ (due to non-commutativity of the negative
root vectors $t^{\prime +}_{rs}$ among each other) in the form,
\begin{eqnarray}
 \label{t'recurr}
  t^{\prime}_{l'm'}
|\vec{0}_{ij},\vec{p}_{rs}\rangle_V &=&
\left|C^{l'm'}_{\vec{p}_{rs}}\rangle_V
 \right.-
\sum_{n'=1}^{l'-1}p_{n'm'}\left| \vec{0}_{ij},
\vec{p}_{rs}-\delta_{n'm',rs}+\delta_{n'l',rs}\rangle_V
 \right. \nonumber\\
  && + \sum_{k'=l'+1}^{m'-1}p_{l'k'}\Bigr[\prod_{r'< l', s'>r'}\prod_{r'=l', m'> s'>r'} \bigl(t^{\prime
+}_{r's'}\bigr){}^{p_{r's'}-\delta_{l'k',r's'}}\Bigl]
t'_{k'm'}\nonumber\\
&& \times\prod_{q'= l', t'\geq m'}\prod_{q'> l', t'>q'}
\bigl(t^{\prime +}_{q't'}\bigr){}^{p_{q't'}} \left|0\rangle_V
 \right.,
\end{eqnarray}
where the vector $ \left|C^{l'm'}_{\vec{p}_{rs}}\rangle_V
 \right.$, $l'<m'$, is determined as follows,
\begin{eqnarray}\label{Clmin}
 \left|C^{l'm'}_{\vec{p}_{rs}}\rangle_V
 \right. &=& p_{l'm'}\Big(h^{l'}-h^{m'}-\sum_{k'=m'+1}^{k}(p_{l'k'}+p_{m'k'})+\sum_{k'=l'+1}^{m'-1}p_{k'm'}-p_{l'm'}+1\Big)\times
 \nonumber
\\
&& \times \left|
\vec{0}_{ij}, \vec{p}_{rs}-\delta_{l'm',rs}\rangle_V
 \right.  + \sum_{k'=m'+1}^{k}p_{l'k'}\Bigl\{\left| \vec{0}_{ij},
\vec{p}_{rs}-\delta_{l'k',rs}+\delta_{m'k',rs}\rangle_V
 \right. \nonumber\\
  && - \sum_{n'=l'+1}^{m'-1} p_{n'm'}\left|\vec{0}_{ij},
\vec{p}_{rs}-\delta_{l'k',rs}-\delta_{n'm',rs}+\delta_{n'k',rs}\rangle_V
 \right. \Bigr\} .
 \end{eqnarray}
The recurrent relation (\ref{t'recurr}) maybe easily resolved  so
the solution has the form, (for $k'_{-1}\equiv 1$)
\begin{eqnarray}
 \label{t'fin}
  t^{\prime}_{l'm'}
|\vec{0}_{ij},\vec{p}_{rs}\rangle_V
&=& \hspace{-1em} \sum_{p=0}^{m'-l'-1}\bigg\{\sum_{k'_1=l'+1}^{m'-1}\ldots
\sum_{k'_p=l'+p}^{m'-1}\prod_{j=1}^{p}p_{k'_{j-1}k'_{j}}\bigg(
 \left|C^{k'_{p}m'}_{\vec{p}_{rs}-\sum_{j=1}^{p}\delta_{k'_{j-1}k'_j,rs}}\rangle_V
 \right.
   \nonumber\\
 &&\hspace{-1em}-
\sum_{n'=k'_{p-1}}^{k'_p-1}\hspace{-0.25em}p_{n'm'}\left| \vec{0}_{ij},
\vec{p}_{rs}-\hspace{-0.1em}\sum_{j=1}^{p}\delta_{k'_{j-1}k'_j,rs}-\delta_{n'm',rs}+\delta_{n'k'_p,rs}\rangle_V
 \right.\hspace{-0.3em}\bigg)\hspace{-0.25em}\bigg\} ,    k'_{0}\equiv l'.
 \end{eqnarray}
Therefore the final result for the action of $t^{\prime}_{l'm'}$
on a vector $|\vec{N}\rangle_V$ maybe written as follows,
 \begin{eqnarray}
\label{t'lm}
  t^{\prime}_{l'm'}
|\vec{N}\rangle_V &=& -\sum_{k'=1}^{k}(1+\delta_{k'm'})n_{k'm'}
\hspace{-0.2em}\left|\vec{n}_{ij}-\delta_{k'm',ij}+\delta_{k'l',ij},
\vec{p}_{rs}\rangle_V
 \right. \nonumber\\
  && +   \sum_{p=0}^{m'-l'-1} \bigg\{\sum_{k'_1=l'+1}^{m'-1}\ldots \sum_{k'_p=l'+p}^{m'-1}\prod_{j=1}^{p}p_{k'_{j-1}k'_{j}}\bigg(
 \left|C^{k'_{p}m'}_{ \vec{n}_{ij}, \vec{p}_{rs}-\sum_{j=1}^{p}\delta_{k'_{j-1}k'_j,rs}}\rangle_V
 \right.\nonumber\\
  && -
\sum_{n'=k'_{p-1}}^{k'_p-1}p_{n'm'}\left| \vec{n}_{ij},
\vec{p}_{rs}-\sum_{j=1}^{p}\delta_{k'_{j-1}k'_j,rs}-\delta_{n'm',rs}+\delta_{n'k'_p,rs}\rangle_V
 \right.\bigg)\bigg\}.
 \end{eqnarray}
Now, it is easy to derive the rest  formulae for the positive root
vectors $l'_{l'm'}$, for $l' = m'$, account of the relations
(\ref{identllm})
     \begin{eqnarray}
  \label{l'll}
  l^{\prime}_{l'l'}
\hspace{-0.2em}|\vec{N}\rangle_V
\hspace{-0.25em}& \hspace{-0.25em}= \hspace{-0.25em}&
 \hspace{-0.25em} -\frac{1}{2}\sum_{k'=1}^{l'-1}{n}_{k'l'}\Biggl[
   \sum_{p=0}^{l'-k'-1}\hspace{-0.25em}
  \bigg\{\sum_{k'_1=k'+1}^{l'-1}\ldots \hspace{-0.25em}\sum_{k'_p=k'+p}^{l'-1}\prod_{j=1}^{p}p_{k'_{j-1}k'_{j}}
 \bigg(\left|C^{k'_{p}l'}_{ \vec{n}_{ij}-\delta_{k'l',ij}, \vec{p}_{rs}-\sum_{j=1}^{p}\delta_{k'_{j-1}k'_j,rs}}\rangle_V
 \right.
\nonumber\\
  && -
\sum_{n'=k'_{p-1}}^{k'_p-1}p_{n'l'}\left| \vec{N}-\delta_{k'l',ij}-\sum_{j=1}^{p}\delta_{k'_{j-1}k'_j,rs}-\delta_{n'l',rs}+\delta_{n'k'_p,rs}\rangle_V
 \right.
 \bigg)
\bigg\}
 \nonumber\\
   && -\sum_{n'=k'+1}^{k}(1+\delta_{n'l'})n_{n'l'}\hspace{-0.2em}
\left|
\vec{n}_{ij}-\delta_{k'l',ij}-\delta_{n'l',ij}+\delta_{k'n',ij},
\vec{p}_{rs}\rangle_V
 \right. \Biggr]
 \nonumber\\
 && + n_{l'l'}\left(n_{l'l'}-1 +  \sum_{k'> l'} n_{k'l'}  - \sum_{s>l'}p_{l's}+\sum_{r<l'}p_{rl'} + h^{l'}\right)\left|\vec{N} -
 \delta_{l'l',lm}\rangle_V
 \right.\nonumber\\
 && - \frac{1}{2}\sum_{k'=l'+1}^{k}{n}_{l'k'}\Biggl[
\left|{\vec{n}}_{ij}-\delta_{l'k',ij}, {\vec{p}}_{rs}
+ \delta_{l'k',rs} \rangle_V \right.\nonumber\\
&& -
\sum\limits_{n'=1}^{l'-1}p_{n'l'}\left|{\vec{n}}_{ij}-\delta_{l'k',ij},
{\vec{p}}_{rs} - \delta_{n'l',rs}+ \delta_{n'k',rs} \rangle_V \right.  \nonumber\\
  && - \sum_{n'=k'+1}^{k}(1+\delta_{n'l'})n_{n'l'}
\left|\vec{N}-\delta_{l'k',lm}-\delta_{l'n',lm}  +
\delta_{k'n',lm}\rangle_V
 \right.
 \Biggr]\nonumber\\
 && + \frac{1}{2}\sum_{k'=1,k'\neq l'}^{k}\frac{{n}_{l'k'}({n}_{l'k'}-1)}{2}
 \left|\vec{N} -2\delta_{l'k',lm} + \delta_{k'k',lm}\rangle_V
 \right.,
    \end{eqnarray}
and for $l' \ne m'$,
   \begin{eqnarray}
\label{l'lm} \hspace{-1em} l^{\prime }_{l'm'}
\bigl|\vec{N}\bigr\rangle_V &\hspace{-0.5em} =&
\frac{1}{4}\sum\limits_{k'=1}^{m'-1}(1+\delta_{k'l'}){n}_{k'l'}
\Biggl[
\sum_{n'=k'}^{k}(1+\delta_{n'm'})n_{n'm'}
\hspace{-0.2em}\left|\vec{n}_{ij}-\delta_{k'l',ij}-
\delta_{n'm',ij}+\delta_{k'n',ij}, \vec{p}_{rs}\rangle_V
 \right.
\nonumber\\
  && -
   \sum_{p=0}^{m'-k'-1} \bigg\{\sum_{k'_1=k'+1}^{m'-1}\ldots \sum_{k'_p=k'+p}^{m'-1}\prod_{j=1}^{p}p_{k'_{j-1}k'_{j}}\bigg(
 \left|C^{k'_{p}m'}_{ \vec{n}_{ij}-\delta_{k'l',ij}, \vec{p}_{rs}-\sum_{j=1}^{p}\delta_{k'_{j-1}k'_j,rs}}\rangle_V
 \right.
  \nonumber\\
  &&
  -
\sum_{n'=k'_{p-1}}^{k'_p-1}p_{n'm'}\left| \vec{n}_{ij}-\delta_{k'l',ij},
\vec{p}_{rs}-\sum_{j=1}^{p}\delta_{k'_{j-1}k'_j,rs}-\delta_{n'm',rs}+\delta_{n'k'_p,rs}\rangle_V
 \right. \bigg)\bigg\} \Biggr]
 \nonumber\\
&& - \frac{1}{4}\sum\limits_{k'=m'+1}^{k}{n}_{l'k'}
\Biggl[ \left|{\vec{n}}_{ij}-\delta_{l'k',ij},
{\vec{p}}_{rs} + \delta_{m'k',rs} \rangle_V \right. \nonumber\\
&&-
\sum\limits_{n'=1}^{m'-1}p_{n'm'}\left|{\vec{n}}_{ij}-\delta_{l'k',ij},
{\vec{p}}_{rs} - \delta_{n'm',rs}+ \delta_{n'k',rs} \rangle_V \right. \nonumber \\
   && - \sum_{n'=l'+1}^{k}(1+\delta_{n'm'})n_{m'n'}\left|
{\vec{n}}_{ij}-\delta_{l'k',ij}-\delta_{n'm',ij}  +
\delta_{k'n',ij}\rangle_V
 \right.  \Biggr]\nonumber\\
&&   + \frac{1}{4}n_{l'm'}\Bigl(n_{l'm'}-1 + \sum_{k'>
l'}(1+\delta_{k'm'}) n_{k'm'} + \sum_{k'>
m'}n_{l'k'} - \sum_{s>l'}p_{l's} \nonumber\\
&& - \sum_{s>m'}p_{m's}+\sum_{r<l'}p_{rl'} +\sum_{r<m'}p_{rm'} +
h^{l'}+
h^{m'}\Bigr)\bigl|\vec{n}_{ij}-\delta_{l'm',ij},\vec{p}_{rs}\bigr\rangle_V
\nonumber
\end{eqnarray}
\vspace{-3ex}
 \begin{eqnarray}
&& -\frac{1}{4}  \sum\limits_{k'=1}^{l'-1}{n}_{k'm'}
\Biggl[   \sum_{p=0}^{l'-k'-1}\bigg\{\sum_{k'_1=k'+1}^{l'-1}\ldots \sum_{k'_p=k'+p}^{l'-1}\prod_{j=1}^{p}p_{k'_{j-1}k'_{j}}
 \bigg(\left|C^{k'_{p}l'}_{ \vec{n}_{ij}-\delta_{k'm',ij}, \vec{p}_{rs}-\sum_{j=1}^{p}\delta_{k'_{j-1}k'_j,rs}}\rangle_V
 \right.
\nonumber\\
  && -
\sum_{n'=k'_{p-1}}^{k'_p-1}p_{n'l'}\left| \vec{n}_{ij}-\delta_{k'm',ij},
\vec{p}_{rs}-\sum_{j=1}^{p}\delta_{k'_{j-1}k'_j,rs}-\delta_{n'l',rs}+\delta_{n'k'_p,rs}\rangle_V
 \right.
 \bigg)
\bigg\}\nonumber\\
  && -\sum_{n'=k'+1}^{k}(1+\delta_{n'l'})n_{n'l'}
\hspace{-0.2em}\left|\vec{n}_{ij}-\delta_{k'm',ij}-
\delta_{n'l',ij}+\delta_{k'n',ij}, \vec{p}_{rs}\rangle_V
 \right. \Biggr]\nonumber\\
&& -
\frac{1}{4}\textstyle\sum\limits_{k'=l'+1}^{k}(1+\delta_{k'm'}){n}_{m'k'}
\Biggl[ \left|{\vec{n}}_{ij}-\delta_{m'k',ij}, {\vec{p}}_{rs} +
\delta_{l'k',rs} \rangle_V \right
. \nonumber\\
&&- \sum\limits_{n'=1}^{l'-1}p\
_{n'l'}\left|{\vec{n}}_{ij}-\delta_{m'k',ij}, {\vec{p}}_{rs} -
\delta_{n'l',rs}+ \delta_{n'k',rs} \rangle_V \right. \Biggr].
\end{eqnarray}

So, the formulae (\ref{l'+ij})-- (\ref{g'0i}), (\ref{t'lm}) --
(\ref{l'lm}) completely solve the problem of auxiliary
representation  (Verma module) construction for the symplectic
$sp(2k)$ algebra.

\subsection{note on additional parts construction for  massive HS fields}\label{addalgebram}}

To solve the same problem as above described in the
Appendix~\ref{addalgebra}, but for auxiliary representation
construction for  HS symmetry massive algebra
$\mathcal{A}(Y(k),\mathbb{R}^{1,d-1})$  we may to enlarge the
Cartan decomposition (\ref{Cartandecomp}) up to one for
$\mathcal{A}(Y(k),\mathbb{R}^{1,d-1})$.  Then we could make all
the same steps again with only the fact, that the Cartan
subalgebra would now contain the element $l_0'$ whereas the
highest weight vector $|0\rangle_V$ and basis vector
$|\vec{N}^m\rangle_V$ of $\mathcal{A}(Y(k),\mathbb{R}^{1,d-1})$ in
addition to definitions (\ref{hwrep})--(\ref{VM}) determines as
follows,
\begin{align}\label{masshwrep}
& l'_i|0\rangle_V =0 && l'_0 |0\rangle_V = m^2|0\rangle_V,\\
& |\vec{N}^m\rangle_V \sim
\prod_{i}^k\textstyle\bigl(\frac{l^{\prime
+}_{i}}{m_i}\bigr){}^{n_{i}}|\vec{N}\rangle_V,
\end{align}
for some parametes $m_i \in \mathbb{R}_+$ of dimension of mass, so
that central charge $m^2$ in the initial algebra
$\mathcal{A}(Y(k),\mathbb{R}^{1,d-1})$ will vanish in the
converted algebra $\mathcal{A}_c(Y(k),\mathbb{R}^{1,d-1})$ because
of the additive composition law
\begin{align}\label{vancentcharge}
    & m^2 \to M^2 = m^2+{m'}^2 =0, &&  l_0  \to L_0 = l_0 + l'_0 = l_0
+m^2,
\end{align}
for the central elements $m^2, {m'}^2$ and Casimir operators $l_0,
l'_0$ respectively of the original algebra of $o_I$ and algebra of
additional parts $o'_I$.

\section{Oscillator realization  of the
additional parts in a new Fock space} \label{oscrealsp2kdet}
\setcounter{equation}{0}

 Following general Burdik's result
of Refs. \cite{Burdik} and making use of the mapping between basis
of Verma module for $sp(2k)$ given by $|\vec{N}\rangle_V$
(\ref{VM}) and one in new Fock space $\mathcal{H}'$,
\begin{equation}\label{map}
    \left| \vec{n}_{ij}, \vec{p}_{rs}\rangle_V \right.
    \leftrightarrow \left| \vec{n}_{ij}, \vec{p}_{rs}\rangle
    \right.,\qquad
\left| \vec{n}_{ij}, \vec{p}_{rs}\rangle
    \right. = \prod_{i,j\geq i}^k\bigl(b^{+}_{ij}\bigr){}^{
 n_{ij}}\prod_{r,s,s>r}^k\bigl(d^{+}_{rs}\bigr){}^{p_{rs}}|0\rangle\,,
\end{equation}
where the vector $\left| \vec{n}_{ij},
\vec{p}_{rs}\rangle\right.$, first, has the same structure as the
vector $|\vec{N}\rangle_V$ in the Eq.(\ref{VMk}), for $n_{ij},
{p}_{rs} \in \mathbb{N}_0$ and, second, appears by the basis
vectors of a Fock space $\mathcal{H}'$ generated by new
 bosonic, $b^{+}_{ij}, d^+_{rs}, b_{ij},
d_{rs}$, $i,j,r,s =1,\ldots, k; i\leq j; r<s$, creation and
annihilation operators with the only nonvanishing commutation
relations
\begin{equation}\label{commrelationsf}
 [b_{i_1j_1}, b^+_{i_2j_2}] =
 \delta_{i_1i_2}\delta_{j_1j_2}\,, \   \qquad [d_{r_1s_1}\,,d^+_{r_2s_2}]
 =\delta_{r_1r_2}\delta_{s_1s_2}\,,
\end{equation}
we can represent the action of the elements $o'_I$ on
$|\vec{N}\rangle_V$ given by the Eqs. (\ref{l'+ij})--
(\ref{g'0i}), (\ref{t'lm}) -- (\ref{l'lm}) as polynomials in the
creation operators of the Fock space $\mathcal{H}'$. The only
requirement on the number of pairs of the above bosonic operators
that it must coincides with one for pairs of second-class
constraints, i.e. with the numbers of negative (or positive) root
vectors in Cartan decomposition of $sp(2k)$.

Finally, the oscillator realization of the elements $o'_I$
 may be uniquely presented as follows, for Cartan elements and negative
root vectors,
\begin{eqnarray}
g_0^{\prime i}& = &  \sum_{l\leq m}
 b_{lm}^+b_{lm}(\delta^{il}+\delta^{im}) + \sum_{r< s}d^+_{rs}d_{rs}(\delta^{is}-
 \delta^{ir}) +h^i
 \,,\label{g'0iFa} \\
  t^{\prime+}_{lm}   & = & d^+_{lm} - \sum_{n=1}^{l-1}d_{nl}d^+_{nm}
   - \sum_{n=1}^{k}(1+\delta_{nl})b^+_{nm}b_{ln}\,,
 \label{t'+lma}
 \\
   l^{\prime+}_{ij} & = & b_{ij}^+\,,
 \label{l'+ijFa}
\end{eqnarray}
 for the elements $l^{\prime }_{lm}$ of upper-triangular
subalgebra $\mathcal{E}^+_k$ separately, for $l=m$ and for $l<m$
\begin{eqnarray}
\label{l'lla} l^{\prime }_{ll} &=&  \frac{1}{4}\sum_{n=1,n\neq
l}^{k}b^+_{nn}{b}^2_{ln} +  \frac{1}{2}\sum_{n=1}^{l-1}\Bigl[
\sum_{n'=n+1}^{k}(1+\delta_{n'l})b^+_{nn'}b_{n'l}\\
  && -
\sum_{p=0}^{l-n-1} \Big(\sum_{k_1=n+1}^{l-1}\ldots \sum_{k_p=n+p}^{l-1}\Big\{
 C^{k_{p}l}(d^+,d)- \sum_{n'=k_{p-1}}^{k_p-1}d^+_{n'k_p}d_{n'l} \Big\}\prod_{j=1}^{p}d_{k_{j-1}k_{j}} \Big)\Bigr]{b}_{nl}
 \nonumber\\
 && + \left(\sum_{n= l}^k b^+_{nl}b_{nl}  - \sum_{s>l}d^+_{ls}d_{ls}+\sum_{r<l}d^+_{rl}
 d_{rl} + h^{l}\right)b_{ll}\nonumber\\
 && - \frac{1}{2}\sum_{n=l+1}^{k}\Bigl[d^+_{ln} -
\sum\limits_{n'=1}^{l-1}d^+_{n'n}d_{n'l} -
\sum_{n'=n+1}^{k}(1+\delta_{n'l})b^+_{n'n}b_{n'l}
 \Bigr]{b}_{ln}\nonumber ,
 \end{eqnarray}
\vspace{-3ex}
\begin{eqnarray} \label{l'lmbosea}
l^{\prime }_{lm}&=&
 -
\frac{1}{4}\sum\limits_{n=1}^{m-1}(1+\delta_{nl}) \Bigl[
-\sum_{n'=n}^{k}(1+\delta_{n'm})
 b^+_{n'n}b_{n'm} \\
 && +
\sum_{p=0}^{m-n-1}\Big(\sum_{k_1=n+1}^{m-1}\ldots \sum_{k_p=n+p}^{m-1}\Big\{
 C^{k_{p}m}(d^+,d)- \sum_{n'=k_{p-1}}^{k_p-1}d^+_{n'k_p}d_{n' m} \Big\}\prod_{j=1}^{p}d_{k_{j-1}k_{j}}\Big)\Bigr]b_{nl}
 \nonumber\\
&& - \frac{1}{4}\textstyle\sum\limits_{n=m+1}^{k} \Bigl[ d^+_{mn}-
\sum\limits_{n'=1}^{m-1} d^+_{n'n}d_{n'm} -
\sum\limits_{n'=l+1}^{k}(1+\delta_{n'm})b^+_{n'n}b_{mn'}  \Bigr]{b}_{ln}\nonumber\\
&& + \frac{1}{4}\Bigl(\sum_{n=m}^kb^+_{ln}b_{ln} + \sum_{n=
l+1}^k(1+\delta_{nm})b^+_{nm} b_{nm}  - \sum_{s>l}d^+_{ls}d_{ls} -
\sum_{s>m}d^+_{ms}d_{ms}\nonumber\\
&& +\sum_{r<l}d^+_{rl}d_{rl} +\sum_{r<m}d^+_{rm}d_{rm} + h^{l}+
h^{m}\Bigr)b_{lm} \nonumber\\
&& -\frac{1}{4}  \sum\limits_{n=1}^{l-1} \Bigl[
\sum_{p=0}^{l-n-1} \Big(\sum_{k_1=n+1}^{l-1}\ldots \sum_{k_p=n+p}^{l-1}\Big\{
 C^{k_{p}l}(d^+,d)- \sum_{n'=k_{p-1}}^{k_p-1}d^+_{n'k_p}d_{n'l} \Big\}\prod_{j=1}^{p}d_{k_{j-1}k_{j}} \Big)\nonumber\\
 && -\sum_{n'=n+1}^{k}(1+\delta_{n'l})b^+_{n'n}b_{n'l}
 \Bigr]{b}_{nm} - \frac{1}{4}\textstyle\sum\limits_{n=l+1}^{k}(1+\delta_{nm})
\Bigl[ d^+_{ln} - \sum\limits_{n'=1}^{l-1}d^+_{n'n}d_{n'l}
\Bigr]{b}_{mn}\nonumber,
 \end{eqnarray}
and for the "mixed-symmetry" elements $t^{\prime }_{lm}$,
  \begin{eqnarray}
t^{\prime }_{lm} &=& \sum_{p=0}^{m-l-1}\bigg[\sum_{k_1=l+1}^{m-1}\ldots \sum_{k_p=l+p}^{m-1}
 \Big\{C^{k_{p}m}(d^+,d)- \sum_{n'=k_{p-1}}^{k_p-1}d^+_{n'k_p}d_{n'm} \Big\}\prod_{j=1}^{p}d_{k_{j-1}k_{j}}\bigg]
 \label{t'lmFa}\\
  && -\sum_{n=1}^{k}(1+\delta_{nm})b^+_{nl}
b_{nm}
 \,, \qquad k_0\equiv l. \nonumber
\end{eqnarray}
where the  vector $\left|C^{lm}_{\vec{p}_{rs}}\rangle_V
 \right.$, $l<m$ given in (\ref{Clmin}), is transformed to
 the operator $C^{lm}(d,d^+)$ given by the Eq. (\ref{Clm}).

Thus, we have obtained the expressions of the additional parts
$o'_I(B,B^+)$ (\ref{g'0iF})--(\ref{t'lmF}) for the operator
algebra $sp(2k)$ given by the table~\ref{table in}.

Let us find an explicit expression for the operator $K'$  used in
the definition of the scalar product (\ref{newsprod})  and given
in an exact form in (\ref{explicit K}).

 One can show
by direct calculation that the following relation holds true:
\begin{equation}
{}_V\left\langle\vec{n}'_{lm}, \vec{p}'_{rs}\right.
\left|\vec{n}_{lm},\vec{p}_{rs}\rangle_V\right.\sim
\prod_{l=1}^k\delta^{\textstyle\sum_i(1+\delta_{il})n_{il}-\sum_{i>l}
p_{li}+\sum_{i<l}
p_{il}}_{\textstyle\sum_i(1+\delta_{il})n'_{il}-\sum_{i>l}
p'_{li}+\sum_{i<l} p'_{il}}.
\end{equation}
 For practical calculations for  low values of sets of $k$
numbers
\begin{equation} (\sum_i(1+\delta_{i1})n_{1i}-\sum_{i>1}
p_{1i}, \sum_i(1+\delta_{i2})n_{2i}-\sum_{i>2} p_{2i} + p_{12},...
, \sum_i(1+\delta_{ik})n_{ik} + \sum_{i<k} p_{ik}),
\end{equation}
with $p_{rt}, n_{ij}$ being the numbers of ``particles''
associated with $d^+_{rt}, b_{ij}^+$ for $i\leq j, r<t$ (where
$d^+_{rt}$ reduces the spin number $s_r$ by one unit and increases
the spin number $s_t$ by one unit simultaneously), the operator
$K'$ reads with use of normalization condition
${}_V\langle0|0\rangle_V = 1$
\begin{eqnarray}\label{Ka}
K' &=& |0\rangle\langle0| +
\sum_{r<s}(h^r-h^s)d^+_{rs}|0\rangle\langle0|d_{rs}
 +\sum_{i\le j}\Big(h^i(1+2\delta^{ij})+h^j\Big)
b_{ij}^+|0\rangle\langle 0|b_{ij}
 \end{eqnarray}
\vspace{-3ex}
\begin{eqnarray}
\phantom{K'} && + \frac{1}{2}\sum_{l<i}(h^i-h^l)\Big(
b^+_{ii}|0\rangle\langle0|b_{li}d_{li}+
b^+_{li}d^+_{li}0\rangle\langle0|b_{ii} \Big)\nonumber\\
&& +\frac{1}{4} \sum_{i<j}b_{ij}^+|0\rangle\langle0|\Bigl(
 \sum_{l<i}b_{lj}d_{li} (h^i-h^l) + (1+\delta^{li})\sum_{l < j}(h^j-h^l)b_{il}d_{lj}
 \Bigr)\nonumber\\
\phantom{K'}&& + \frac{1}{4}
\sum_{i<j}\Bigl(\sum_{l<i}b_{lj}^+d^+_{li}
|0\rangle\langle0|(h^i-h^l)
  + \sum_{l<j} b_{lj}^+d^+_{li}
|0\rangle\langle0|(1+\delta^{li})(h^j-h^l)
 \Bigr) b_{ij} + \ldots \,.
\nonumber
\end{eqnarray}
The above expression for the operator $K'$ may be used to
construct LF for HS fields with low value of spin.

Thus, we have found the auxiliary scalar representation of the
symplectic algebra $sp(2k)$ for the additional parts of the
constraints $o'_I$ in the new Fock space $\mathcal{H}'$.

\section{Reduction to the initial irreducible relations}\label{reductionC}
\setcounter{equation}{0}

Here, we consider preferably massive case, making then comments on
massless HS fields.  We show that the equations of motion
(\ref{Eq-0bm}), (\ref{Eq-1b})--(\ref{Eq-3b}) [or equivalently
(\ref{l0}), for $\tilde{l}_0 = l_0+m^2$, and (\ref{lilijt})] can
be obtained from the Lagrangian (\ref{Scl}) after gauge-fixing and
removing the auxiliary fields by using a part of the equations of
motion. Let us start with gauge-fixing.

\subsection{Gauge-fixing}

Let us consider the field $|\chi^l \rangle$, for $l=1,...,k(k+1)$,
at some fixed values of the spin $(n_1,\ldots, n_k)$. In this
section we will omit the subscripts associated with the
eigenvalues of the $\sigma_i$ operators (\ref{statem}). Then we
extract dependence of $Q$ (\ref{Q}) on zero-mode ghosts $\eta_0$
and $\mathcal{P}_0$
\begin{eqnarray}
Q &=& \eta_0L_0 + {\imath}\sum_m\eta_m^+\eta^m {\cal{}P}_0+ \Delta
Q
\\
\Delta Q &=& \eta_i^+L^i +\sum\limits_{l\leq m}\eta_{lm}^+L^{lm} +
\sum\limits_{l< m}\vartheta^+_{lm}T^{lm} -\sum\limits_{i<l<j}
(\vartheta^+_{lj}\vartheta^+_{i}{}^l -
\vartheta^+_{il}\vartheta^{+l}{}_{j})\lambda^{ij}
 \\\hspace{-0.4em}
&& {}\hspace{-0.4em}  -
\sum\limits_{l<n<m}\vartheta_{lm}^+\vartheta^{l}{}_n\lambda^{nm} +
\sum\limits_{n<l<m}\vartheta_{lm}^+\vartheta_{n}{}^m\lambda^{+nl}
- \sum_{n,l<m}(1+\delta_{ln})\vartheta_{lm}^+\eta^{l+}{}_{n}
\mathcal{P}^{mn} \nonumber
\end{eqnarray}
\vspace{-2ex}
\begin{eqnarray}&& +
\sum_{n,l<m}(1+\delta_{mn})\vartheta_{lm}^+\eta^{m}{}_{n}
\mathcal{P}^{+ln}+ \frac{1}{2}\sum\limits_{l<m,n\leq
m}\eta^+_{nm}\eta^{n}{}_l\lambda^{lm}
\nonumber\\
\hspace{-0.4em} && \hspace{-0.4em}  
 - \Bigl(\frac{1}{2}\sum\limits_{l\leq
m}(1+\delta_{lm})\eta^m\eta_{lm}^+ +
\sum\limits_{l<m}\vartheta_{lm} \eta^{+m}
+\sum\limits_{m<l}\vartheta^+_{ml} \eta^{+m} \Bigr)\mathcal{P}^l
+Herm.C.\nonumber
 \label{DQ}
\end{eqnarray}
and do the same for the fields and gauge parameters
\begin{equation}\label{decompeta0}
|\chi^l\rangle=|S^l\rangle +\eta_0|B^l\rangle.
\end{equation}
Then the equations of motion and gauge transformations
(\ref{dx0})--(\ref{dxs}) can be rewritten as follows
\begin{align}\label{regthr}
& \delta |S^{l - 1} \rangle = \Delta Q |S^l \rangle -
\sum_m\eta_m^+\eta^m |B^l \rangle && \delta |S^{-1} \rangle
\equiv0,
\\
& \delta |B^{l - 1} \rangle = L_0 |S^l \rangle - \Delta Q|B^l
\rangle && \delta |B^{-1} \rangle \equiv 0,
\end{align}
for $l=0,\ldots , k(k+1)$.

As the following step, we consider the lowest level gauge
transformation
\begin{eqnarray}\label{lowgtr}
\delta |S^{k(k+1)-1} \rangle = \Delta Q|S^{k(k+1)} \rangle,
&\qquad& \delta |B^{k(k+1)-1} \rangle = L_0 |S^{k(k+1)} \rangle,
\end{eqnarray}
where due to the ghost number restriction one has used that
$|B^{k(k+1)}\rangle\equiv 0$. Extracting explicitly dependence of
the gauge parameters and of the operator $\Delta{}Q$ (\ref{DQ}) on
$\eta_{11}$, $\mathcal{P}_{11}^+$ ghost coordinate and momentum
\begin{eqnarray}\label{chil}
|\chi^l\rangle=|\chi^l_0\rangle+P_{11}^+|\chi^l_{1}\rangle,
&\qquad& \Delta Q=\Delta
Q_{11}+\eta_{11}T_{11}^++U_{11}\mathcal{P}_{11}^+,
\end{eqnarray}
where the quantities $|\chi^l_0\rangle$, $|\chi^l_{1}\rangle$,
$T_{11}^+$, $U_{11}$, $\Delta Q_{11}$ do not depend on
$\eta_{11}$, $\mathcal{P}_{11}^+$ we obtain the gauge
transformation of $|S^{k(k+1)-1}_0\rangle$ [with the same
decomposition for $|S^{k(k+1)-1}\rangle$, $|S^{k(k+1)-1}\rangle  =
|S^{k(k+1)-1}_0\rangle$ $+
\mathcal{P}_{11}^+|S^{k(k+1)-1}_1\rangle$ as one for
$|\chi^l\rangle$ (\ref{chil})]
\begin{eqnarray}
\delta |S^{k(k+1)-1}_0 \rangle & = & T_{11}^+ |S^{k(k+1)}_1
\rangle. \label{dSk(k+1)}
\end{eqnarray}
Here we have used that $|S^{k(k+1)}_0\rangle\equiv0$ due to the
ghost number restriction. Since $T_{11}^+ = L_{11}^+ + O
(\mathcal{C})= b_{11}^++\ldots$, as it's follows from  the
structure of $\Delta Q$ in Eq. (\ref{DQ}), we can remove
dependence of $|S^{k(k+1)-1}_0\rangle$ on $b_{11}^+$ operator
using all the degrees of freedom of $|S^{k(k+1)}_1\rangle$.
Therefore, after the gauge fixing at the lowest level of the gauge
transformations we have conditions on $|S_0^{k(k+1)-1}\rangle$
\begin{eqnarray}
b_{11}|S^{k(k+1)-1}_0\rangle=0 &\Longleftrightarrow&
b_{11}\mathcal{P}_{11}^+|\chi^{k(k+1)-1}\rangle=0,
\label{gk(k+1)-1}
\end{eqnarray}
so that the theory became by the $k(k+1)-1$-reducible gauge
theory.

 Let us turn to the next level of the gauge
transformation. Extracting explicit dependence of the gauge
parameters and $\Delta{}Q$ on $\eta_{11}$, $\mathcal{P}_{11}^+$,
$\eta_{12}$, $\mathcal{P}_{12}^+$  and using similar arguments as
at the previous level of the gauge transformation one can show
that the gauge on $|\chi^{k(k+1)-2}\rangle$
\begin{eqnarray}\label{gk(k+1)-2}
b_{11}\mathcal{P}_{11}^+|\chi^{k(k+1)-2}\rangle=0, &\qquad&
b_{12}\mathcal{P}_{11}^+\mathcal{P}_{12}^+|\chi^{k(k+1)-2}\rangle=0.
\end{eqnarray}
can be imposed. To obtain these gauge conditions all degrees of
freedom of the gauge parameters $|\chi^{k(k+1)-1}\rangle$
restricted by the Eq.(\ref{gk(k+1)-1}) have to be used.

Applying an above described  procedure one can obtain step by
step, first, for $l=k(k+1)-3$,
\begin{eqnarray}\label{gk(k+1)-3}
b_{11}\mathcal{P}_{11}^+|\chi^{l}\rangle=0, \quad
b_{12}\mathcal{P}_{11}^+\mathcal{P}_{12}^+|\chi^{l}\rangle=0,
\quad b_{13}\prod_{i}^3\mathcal{P}_{1i}^+|\chi^{l}\rangle=0.
\end{eqnarray}
Then, for $l=k(k+1)-4$
\begin{eqnarray} \label{gk(k+1)-4}
\left(b_{11}\mathcal{P}_{11}^+,\
b_{12}\prod_i^2\mathcal{P}_{1i}^+,\
b_{13}\prod_{i}^3\mathcal{P}_{1i}^+,\
b_{14}\prod_{i}^4\mathcal{P}_{1i}^+\right)|\chi^l\rangle=0.
\end{eqnarray}
Defining the set of the operators used in
(\ref{gk(k+1)-1})--(\ref{gk(k+1)-4}) by the rule,
\begin{eqnarray} \label{Al}
[\mathcal{A}^l] = \Bigl(b_{11}\mathcal{P}_{11}^+,...,
b_{1k}\prod_i^k\mathcal{P}_{1i}^+,\ldots,
b_{k-1{}k}\prod_i^{k-1}\mathcal{P}_{ik}^+\hspace{-0.3em}\prod_{i,j=1,
i\leq j}^{k-1} \hspace{-0.3em}\mathcal{P}_{ij}^+,
b_{k{}k}\hspace{-0.3em}\prod_{i,j=1, i\leq j}^{k}\hspace{-0.3em}
\mathcal{P}_{ij}^+\Bigr),  l=1,... , \textstyle\frac{k(k+1)}{2},
\end{eqnarray}
(where, for instance  $(k+1)$-th component of the set
$[\mathcal{A}^l]$ is equal to $\mathcal{A}^{k+1} =
b_{22}\mathcal{P}_{22}^+\prod_i^k\mathcal{P}_{1i}^+$) we may
rewrite equivalently gauge conditions
(\ref{gk(k+1)-1})--(\ref{gk(k+1)-4}) and all subsequent ones which
are based on the decomposition of the gauge parameters in all the
ghost momenta $P_{ij}^+, i\leq j$, as follows,
\begin{eqnarray} \label{Apgauge}
[\mathcal{A}^l]|\chi^{k(k+1)-l}\rangle=0,\texttt{ for }l =
1,\ldots, \frac{ k(k+1)}{2}.
\end{eqnarray}
Next, we apply the same procedure as above but starting from
 the gauge parameter $|\chi^{\frac{k(k+1)}{2}-1}\rangle$ and extract
 from it, from
the operator $\Delta{}Q$ (\ref{DQ}) of the ghost coordinates and
momenta $\eta_{ij}$, $\mathcal{P}_{ij}^+$, $i\leq j$ and $\eta_1$,
$\mathcal{P}_{1}^+$. As a result, we have obtained the set of the
gauge conditions on the parameter
$|\chi^{\frac{k(k+1)}{2}-1}\rangle$,
\begin{equation}
\label{gk(k+1)2-1} \Bigl([\mathcal{A}^{\frac{1}{2}k(k+1)}],
b_1\mathcal{P}_1^+\prod_{i,j=1, i\leq j}^{k}
\mathcal{P}_{ij}^+\Bigr)|\chi^{\frac{k(k+1)}{2}-1}\rangle =0 .
\end{equation}
Continue the process with ghosts $\eta_1$, $\eta_2$
$\mathcal{P}_{1}^+$, $\mathcal{P}_{2}^+$ and so on extraction we
obtain the $k$ sets of the gauge conditions on the parameters
$|\chi^{\frac{k(k+1)}{2}-m}\rangle$,
$m=1,\ldots,k$
\begin{eqnarray} \label{gk(k-1)2}
&& \Bigl([\mathcal{A}^{\frac{1}{2}k(k+1)}],
b_1\mathcal{P}_1^+\prod_{i,j=1, i\leq j}^{k} \mathcal{P}_{ij}^+,
b_2\prod_m^2\mathcal{P}_m^+\prod_{i,j=1, i\leq j}^{k}
\mathcal{P}_{ij}^+ \Bigr)|\chi^{\frac{k(k+1)}{2}-2}\rangle =0 , \\
&& \hspace{3cm}\ldots \ldots \ldots \ldots \ldots \ldots\ldots \ldots \ldots\nonumber\\
 \label{gk(k-1)2f} &&
\Bigl([\mathcal{A}^{\frac{1}{2}k(k+1)}],
b_1\mathcal{P}_1^+\prod_{i,j=1, i\leq j}^{k} \mathcal{P}_{ij}^+,
 \ldots,
b_k\prod_m^k\mathcal{P}_m^+\prod_{i,j=1, i\leq j}^{k}
\mathcal{P}_{ij}^+ \Bigr)|\chi^{\frac{k(k-1)}{2}}\rangle =0.
\end{eqnarray}
At last, realizing the same algorithm  as above but initiating
from
  gauge parameter $|\chi^{\frac{k(k-1)}{2}-1}\rangle$ and extract
 from it, from
the operator $\Delta{}Q$ (\ref{DQ})  the ghost coordinates and
momenta $\eta_m$, $\mathcal{P}_{m}^+$, $\eta_{ij}$,
$\mathcal{P}_{ij}^+$, $i\leq j$ and $\vartheta_{ps}$,
$\lambda_{ps}^+$, for $p<s$, we have obtained the
$\frac{1}{2}k(k-3)$ sets of the gauge conditions on the parameter
$|\chi^{\frac{k(k-1)}{2}-m}\rangle$, for $m=1, \ldots,
\frac{1}{2}k(k-1)-1$
\begin{eqnarray} \label{gk(k-1)2-1}
\hspace{-1.5em}&\hspace{-0.5em}&
\hspace{-1.0em}\Bigl([\mathcal{B}^{\frac{1}{2}k(k+3)}],
d_{12}\lambda_{12}^+\prod_m^k\mathcal{P}_m^+\prod_{i,j=1, i\leq
j}^{k} \mathcal{P}_{ij}^+\Bigr)|\chi^{\frac{k(k-1)}{2}-1}\rangle =0 , \\
\hspace{-1.5em}&\hspace{-0.5em}&\hspace{-1.0em} \hspace{3cm}\ldots \ldots \ldots \ldots \ldots \ldots\ldots \ldots \ldots\nonumber\\
 \label{gk(k-1)2fin} \hspace{-1.5em}&\hspace{-0.5em}&
\hspace{-1.0em}\Bigl([\mathcal{B}^{\frac{1}{2}k(k+3)}],
d_{12}\lambda_{12}^+\prod_m^k\mathcal{P}_m^+\hspace{-0.3em}\prod_{i,j=1,
i\leq j}^{k}\hspace{-0.3em} \mathcal{P}_{ij}^+,
 \ldots,
d_{k-2{}k}\hspace{-0.3em}\prod_r^{k-2}\hspace{-0.1em}\lambda_{rk}^+\hspace{-0.5em}
\prod_{p,s=1, p<
s}^{k-1}\hspace{-0.5em}\lambda_{ps}^+\prod_m^k\mathcal{P}_m^+\hspace{-0.5em}\prod_{i,j=1,
i\leq j}^{k}\hspace{-0.5em} \mathcal{P}_{ij}^+
\Bigr)\hspace{-0.1em}|\chi^{1}\rangle
\hspace{-0.1em}=\hspace{-0.1em}0.
\end{eqnarray}
In the Eqs. (\ref{gk(k-1)2-1}), (\ref{gk(k-1)2fin}) the set of the
operators $[\mathcal{B}^{p}]$ is determined from the Eqs.
(\ref{gk(k-1)2f}) as
\begin{equation}
[\mathcal{B}^{\frac{1}{2}k(k+3)}] =
\Bigl([\mathcal{A}^{\frac{1}{2}k(k+1)}],
b_1\mathcal{P}_1^+\prod_{i,j=1, i\leq j}^{k} \mathcal{P}_{ij}^+,
 \ldots,
b_k\prod_m^k\mathcal{P}_m^+\prod_{i,j=1, i\leq j}^{k}
\mathcal{P}_{ij}^+ \Bigr)
\end{equation}
And finally we obtain gauge conditions on the field
$|\chi^0\rangle$
\begin{equation}
\Bigl([\mathcal{B}^{\frac{1}{2}k(k+3)}],
d_{12}\lambda_{12}^+\prod_m^k\mathcal{P}_m^+\prod_{i,j=1, i\leq
j}^{k} \mathcal{P}_{ij}^+,
 \ldots,
d_{k-1{}k}\prod_{p,s=1, p<
s}^{k}\lambda_{ps}^+\prod_m^k\mathcal{P}_m^+\prod_{i,j=1, i\leq
j}^{k} \mathcal{P}_{ij}^+ \Bigr)|\chi^{0}\rangle =0. \label{G1}
\end{equation}

Let us now turn to removing the auxiliary fields using the
equations of motion.

\subsection{Removing auxiliary fields by resolution of equations
of motion}

As the, first step, we decompose the fields $|S^0\rangle$ as
follows
\begin{align}
&|S^0 \rangle = |S^0_0 \rangle + \mathcal{P}_{11}^+ |S^0_{1}
\rangle, \hspace{-2em}
&&\hspace{-2em}|S^0_{(0)_{\frac{1}{2}k(k+1)}} \rangle =
|S^0_{(0)_{\frac{1}{2}k(k+1)}0} \rangle + \mathcal{P}_{1}^+
|S^0_{(0)_{\frac{1}{2}k(k+1)}1}\rangle,
\\
&|S^0_{0}\rangle = |S^0_{00} \rangle + \mathcal{P}_{12}^+
|S^0_{01} \rangle, \hspace{-2em} &&\hspace{-2em}
|S^0_{(0)_{\frac{1}{2}k(k+1)}0} \rangle \hspace{-0.1em}
=\hspace{-0.1em} |\hspace{-0.1em}S^0_{(0)_{\frac{1}{2}k(k+1)+2}}
\rangle \hspace{-0.1em}+\hspace{-0.1em} \mathcal{P}_{2}^+
|\hspace{-0.1em}S^0_{(0)_{\frac{1}{2}k(k+1)}01}\rangle,
\\
& \qquad\ldots , && \qquad\ldots ,
\nonumber \\
&|S^0_{(0)_k} \rangle = |S^0_{(0)_k0} \rangle + \mathcal{P}_{1k}^+
|S^0_{(0)_k1} \rangle,\hspace{-1em}
&&\hspace{-2em}|S^0_{(0)_{\frac{1}{2}k(k+3)}} \rangle =
|S^0_{(0)_{\frac{1}{2}k(k+3)}0} \rangle + \mathcal{P}_{k}^+
|S^0_{(0)_{\frac{1}{2}k(k+3)}1} \rangle,b
\\
&|S^0_{(0)_{k+1}} \rangle = |S^0_{(0)_{k+2}} \rangle +
\mathcal{P}_{22}^+ |S^0_{(0)_{k+1}1} \rangle, \hspace{-1em}
&&\hspace{-2em} |\hspace{-0.1em}S^0_{(0)_{\frac{1}{2}k(k+3)}0}
\rangle\hspace{-0.1em} =\hspace{-0.1em}
|\hspace{-0.1em}S^0_{(0)_{\frac{1}{2}k(k+3)}00} \rangle
\hspace{-0.1em}+ \hspace{-0.1em}\lambda_{12}^+
|\hspace{-0.1em}S^0_{(0)_{\frac{1}{2}k(k+3)}01} \rangle,
\\
&\qquad\ldots , \hspace{-2em} &&\hspace{-2em} \qquad \ldots ,
\nonumber\\
\hspace{-0.5em}&\hspace{-0.5em}|\hspace{-0.1em}S^0_{(0)_{\frac{1}{2}k(k+1)-1}}\hspace{-0.1em}
\rangle \hspace{-0.2em}=\hspace{-0.2em}
|\hspace{-0.1em}S^0_{(0)_{\frac{1}{2}k(k+1)}}
\hspace{-0.2em}\rangle \hspace{-0.1em}+
\hspace{-0.1em}\mathcal{P}_{kk}^+\hspace{-0.1em}
|\hspace{-0.1em}S^0_{(0)_{\frac{1}{2}k(k+1)-1}1}
\hspace{-0.2em}\rangle, \hspace{-0.5em}
&&\hspace{-0.1em}|\hspace{-0.1em}S^0_{(0)_{k(k+1)-1}}
\hspace{-0.2em}\rangle \hspace{-0.15em}=
\hspace{-0.15em}|\hspace{-0.1em}S^0_{(0)_{k(k+1)}}
\hspace{-0.1em}\rangle\hspace{-0.15em} +
\hspace{-0.2em}\lambda_{k-1{}k}^+\hspace{-0.1em}
|\hspace{-0.1em}S^0_{(0)_{k(k+1)-1}1} \hspace{-0.1em}\rangle
\end{align}
and make the same for the vector $|B^0\rangle$
\begin{eqnarray} \label{Bdecomp}
|B^0 \rangle &=&  \mathcal{P}_{11}^+ |B^0_{1} \rangle +
\ldots.+\mathcal{P}_{1k}^+ |B^0_{(0)_k1} \rangle +
\mathcal{P}_{22}^+ |B^0_{(0)_{k+1}1} \rangle + \ldots +
\mathcal{P}_{kk}^+ |B^0_{(0)_{\frac{1}{2}k(k+1)-1}1}
\rangle\\
&& + \mathcal{P}_{1}^+ |B^0_{(0)_{\frac{1}{2}k(k+1)}1}\rangle +
\ldots + \mathcal{P}_{k}^+ |B^0_{(0)_{\frac{1}{2}k(k+3)}1}
\rangle
\nonumber\\
&& + \lambda_{12}^+ |B^0_{(0)_{\frac{1}{2}k(k+3)}01} \rangle+
\ldots + \lambda_{k-1{}k}^+|B^0_{(0)_{k(k+1)-1}1} \rangle
 ,\nonumber
\end{eqnarray}
where the term independent of the ghost momenta is absent due to
the ghost number condition, $gh(|B^0 \rangle)=-1$ and spin value.
One should be noted that due to $gh(|S^0\rangle)=0$ and spin
value, vector $|S^0_{(0)_{k(k+1)}} \rangle$ do not depend on the
ghost coordinates and momenta as a consequence of the gauge
conditions (\ref{G1}), $|S^0_{(0)_{k(k+1)}}\rangle=|\Phi\rangle$,
with $|\Phi\rangle$ being the physical field (\ref{PhysState}).

After that, analogously to the fields we extract in $\Delta{}Q$
(\ref{DQ}) first dependence on $\eta_{11}$, $\mathcal{P}_{11}^+$,
 $\eta_{12}$, $\mathcal{P}_{12}^+$, \ldots ,
$\eta_{1k}$, $\mathcal{P}_{1k}^+$, next dependence on $\eta_{l}$,
$\mathcal{P}_{l}^+$, $l=1,\ldots , k$,  and on $\vartheta_{ps}$,
$\lambda_{ps}^{+}$,  $p<s$ respectively.

Substituting these $k(k+1)$ decompositions into the equation of
motion
\begin{eqnarray}
\tilde{l}_0 |S^0 \rangle - \Delta Q|B^0 \rangle=0
\end{eqnarray}
and using  gauge conditions (\ref{G1}) one can show that first
$|\hspace{-0.1em}B^0_{(0)_{k(k+1)-1}1}\rangle\hspace{-0.1em}=\hspace{-0.1em}0$,
then $\hspace{-0.1em}|B^0_{(0)_{k(k+1)-2}1}\hspace{-0.1em}
\rangle$ = $0$, and so on till $|B^0_{1}\rangle=0$ which means
that
\begin{eqnarray}
\tilde{l}_0|S^0\rangle=0, &\qquad& |B^0\rangle=0. \label{E1}
\end{eqnarray}

Analogously we consider the second equation of motion
\begin{eqnarray}
\Delta Q |S^0 \rangle=0, \label{E2}
\end{eqnarray}
where $|B^0\rangle=0$ has been used. After the same decomposition
we derive one after another that
\begin{eqnarray}
|S^0_{(0)_{k(k+1)-1}1}\rangle=|S^0_{(0)_{k(k+1)-2}1}\rangle=
\ldots = |S^0_{01}\rangle= |S^0_1\rangle=0. \label{E3}
\end{eqnarray}
Eqs. (\ref{E1}) and (\ref{E3}) mean that all the auxiliary fields
vanish and as a result we have
$|\chi^0\rangle_{(s)_k}=|\Phi\rangle$ and the equations of motion
(\ref{Eq-0bm}), (\ref{Eq-1b})--(\ref{Eq-3b}) hold true. Therefore
we proved that the space of BRST cohomologies of the operator $Q$
(\ref{Q}) with a vanishing ghost number is determined only by the
constraints corresponding to an irreducible Poincare-group
representation.

One should be noted, that for the massless case the above proof of
one-to-one correspondence of the Lagrangian equations of motion
(\ref{Q12}) to the Eqs. (\ref{Eq-0b})--(\ref{Eq-3b}) is changed
non-significantly because of the $k$ gauge fixing conditions
(\ref{gk(k+1)2-1})--(\ref{gk(k-1)2f}) do not hold and in the rest
of the Eqs.(\ref{gk(k+1)2-1})--(\ref{G1}) there are not the
operators $b_k \mathcal{P}^+_l$. However, we can  straightforward
show on the validity of the same conclusion as for the massive HS
fields for the Lagrangian formulation  of the massless HS fields
as well.

\section{Decomposition of field and gauge Fock space vectors  for spin $\mathbf{s}=(2,1,1)$ tensor}\label{example211}
\setcounter{equation}{0}

We consider here only the structure of corresponding Fock space
$\mathcal{H}_{tot}$ vectors $|\chi^0\rangle_{(2,1,1)},
|\chi^1\rangle_{(2,1,1)}$, $\ldots , |\chi^{12}\rangle_{(2,1,1)}$
in (\ref{dx0})--(\ref{dxs}) for the example of
Subsection~\ref{ex211} to be used for Lagrangian formulation for
massless tensor $\Phi_{\mu\nu, \rho,\sigma}$  in $d$-dimensional
flat space-time which is characterized by the Poincare group
irreducible
coniditions (\ref{Eq-0b})--(\ref{Eq-3b}) and Young tableux $\begin{array}{|c|c|}\hline
  \!\mu \!&\! \nu\!  \\
   \hline
    \! \rho\!   \\
  \cline{1-1}
    \! \sigma\!   \\
   \cline{1-1}
\end{array}\
$.

One should be noted, the maximal stage of reducibility $L_k=
k(k+1)-1$ for the tensor with $k$  group of symmetric indices is
reached only starting from the rectangular Young tableaux with
length of any row  equal to $(2k+1)$. Thus, for the case of $k=3$
rows a value $L_3 = 11$ for nonvanishing vector
$|\chi^{12}\rangle_{(s_1,s_2,s_3)}$ may appears for the Lagrangian
construction starting from tensor with spin $\mathbf{s}=(7,7,7)$.
In our case, of the tensor with spin $\mathbf{s}=(2,1,1)$, the
nonvanishing gauge vector with minimal ghost number $gh_{\min}$
appears for $gh_{\min} = - 5$. Indeed, it is easy to show that
Eqs. (\ref{nidecompos}), (\ref{ghnum}) have not admissible
solutions among the set  of non-negative integers $(n_l,n_{ij},
p_{rs}, n_{f{}0}, n_{f{}i}$, $n_{p{}j},n_{f{}lm}, n_{p{}no}$,
$n_{f{}rs}, n_{\lambda{}tu}, p_i)$, for $l,i,j,r,s,l,m,n,o, t,u
=1,2,3, \ i\leq j, r<s, l\leq m, n\leq o, t<u$, in (\ref{chi}) and
(\ref{PhysState}) for the vectors $|\chi^l\rangle_{(s)_3}$, $l >
5$.

Therefore, the gauge vector $|\chi^{5}\rangle_{(2,1,1)}$ from the
general expression (\ref{chi})  subject to the conditions
(\ref{nidecompos}), (\ref{ghnum}) for the respective values
$\mathbf{s}=(2,1,1)$ and $l=5$ has the representation (below, we
suppose for this example only that $(s)_3 \equiv (2,1,1)$)
\begin{eqnarray}
|\chi^{5}\rangle_{(s)_3} &=&
\mathcal{P}_1^+\mathcal{P}_2^+\mathcal{P}_{11}^+\lambda^+_{12}\lambda^+_{23}
|\phi^{5} \rangle_{(0)}, \quad  |\phi^{5} \rangle_{(0)} =
|0\rangle\phi^{5}(x) , \label{x-5}
\end{eqnarray}
where, we have used the notation, $(0) \equiv (0,0,0)$, for
$|\phi^{5} \rangle_{(0)}\equiv |\phi^{5} \rangle_{(0,0,0)}$.

 In turn,
for the reducible gauge parameter $|\chi^{4}\rangle_{(s)_3}$ of
the fourth level for $l=4$ in the Eq.(\ref{ghnum}) and the same
Eqs.(\ref{nidecompos}) for spin components distribution, we have
the decomposition in even powers of ghosts from $(11+1)$ summands
with the fourth order in momenta $\mathcal{P}_I$ in
$\eta_0$-independent terms,
\begin{eqnarray}\label{x-4}
|\chi^{4}\rangle_{(s)_3} &=&
\mathcal{P}_1^+\hspace{-0.2em}\Bigl[\mathcal{P}_2^+\hspace{-0.2em}\Bigl\{\mathcal{P}_{11}^+\hspace{-0.2em}\Bigl(\lambda^+_{13}|\phi^{4}_1
\rangle_{(0)}+ \lambda^+_{12}|\phi^{4}_2 \rangle_{(0,-1,1)} +
\lambda^+_{23}|\phi^{4}_3 \rangle_{(-1,1,0)}\hspace{-0.2em}\Bigr)
\hspace{-0.2em}+\mathcal{P}_{12}^+\lambda^+_{23}|\phi^{4}_4\rangle_{(0)}\\
&+&  \lambda^+_{12}\lambda^+_{23} |\phi^{4}_5
\rangle_{(2,0,0)}\hspace{-0.2em}\Bigr\}\hspace{-0.2em}+
\mathcal{P}_{3}^+\mathcal{P}_{11}^+\lambda^+_{12} |\phi^{4}_6
\rangle_{(0)}+\mathcal{P}_{11}^+\lambda^+_{12}\bigl(\lambda^+_{23}
|\phi^{4}_7 \rangle_{(0,1,0)}+\lambda^+_{13} |\phi^{4}_8
\rangle_{(1,0,0)}\bigr)
\nonumber\\
&+& \mathcal{P}_{12}^+\lambda^+_{12}\lambda^+_{23} |\phi^{4}_9
\rangle_{(1,0,0)}
\hspace{-0.2em}\Bigr]+\mathcal{P}_2^+\mathcal{P}_{11}^+\lambda^+_{12}\lambda^+_{23}
|\phi^{4}_{10} \rangle_{(1,0,0)}+
\mathcal{P}_{11}^+\mathcal{P}_{12}^+\lambda^+_{12}\lambda^+_{23}
|\phi^{4}_{11} \rangle_{(0)}
+\eta_0|\chi^{4}_0\rangle_{(s)_3}\nonumber
 \,,
\end{eqnarray}
where, first, the vector $|\chi^{4}_0\rangle_{(s)_3}$,
$|\phi^{4}_0\rangle_{(s)_3}=|0\rangle\phi^{4}_0(x)$, has the same
definition and properties as $|\chi^{5}\rangle_{(s)_3}$ in the Eq.
(\ref{x-5}) and, second, the decomposition of ghost-independent
vectors in powers of initial and auxiliary creation operators from
$\mathcal{H}\bigotimes \mathcal{H}'$ reads
\begin{eqnarray} \label{x-4decomp}
|\phi^{4}_n \rangle_{(0)} =|0\rangle \phi^{4}_{n} (x)  \,,
&\qquad& |\phi^{4}_2 \rangle_{(0,-1,1)}=d^+_{23}|0\rangle
\phi^{4}_{2} (x) \,,\\ |\phi^{4}_3 \rangle_{(-1,1,0)} =
d^+_{12}|0\rangle \phi^{4}_{3} (x)\,, &\qquad& |\phi^{4}_5
\rangle_{(2,0,0)} = a_1^{+\mu}a_1^{+\nu} |0\rangle
\phi^{4}_{5|\mu\nu} (x)+ b_{11}^+|0\rangle
\phi^{4}_{5} (x)\,,\label{x-4decompi}\\
|\phi^{4}_p \rangle_{(1,0,0)}  =  a_1^{+\mu} |0\rangle
\phi^{4}_{p|\mu} (x) \,, &\qquad& |\phi^{4}_7 \rangle_{(0,1,0)} =
a_2^{+\mu} |0\rangle \phi^{4}_{7|\mu} (x) + a_1^{+\mu}d^+_{12}
|0\rangle \phi^{\prime 4}_{7|\mu} (x)\,, \label{x-4decompf}
\end{eqnarray}
for $ n=0,1,4,6,11$  and  $p=8,9,10$.

For the reducible gauge parameter of the third level
$|\chi^{3}\rangle_{(s)_3}$ for $l=3$ in the Eq.(\ref{ghnum}) and
for the same spin value $(2,1,1)$ in the Eqs.(\ref{nidecompos}) we
get from the general expression (\ref{chi}) the decomposition  in
odd powers of ghosts from $(45+11)$ summands starting from the
third order in $\mathcal{P}_I$,
\begin{eqnarray}\label{x-3}
|\chi^{3}\rangle_{(s)_3}& =&
\mathcal{P}_{1}^+\biggl[\mathcal{P}_{2}^+\Bigl\{\mathcal{P}_{3}^+|\phi^{3}_{1}
\rangle_{(1,0,0)}+\mathcal{P}_{11}^+\Bigl(|\phi^{3}_{2}
\rangle_{(-1,0,1)}+\vartheta^+_{23}\lambda^+_{12}|\phi^{3}_3
\rangle_{(0)} + \vartheta^+_{12}\lambda^+_{23}|\phi^{3}_4
\rangle_{(0)}\Bigr)
\\
&& \hspace{-3em} +  \mathcal{P}_{12}^+|\phi^{3}_{5}
\rangle_{(0,-1,1)}+\mathcal{P}_{13}^+|\phi^{3}_{6}
\rangle_{(0)}+\lambda^+_{12} |\phi^{3}_7
\rangle_{(2,-1,1)}+\lambda^+_{12}\lambda^+_{23}
\Bigl(\eta_{1}^+|\phi^{3}_8\rangle_{(1,0,0)_3} +\eta_{11}^+
|\phi^{3}_9\rangle_{(0)}\Bigr)\nonumber
\\
&& \hspace{-3em} +\lambda^+_{13} |\phi^{3}_{10}
\rangle_{(2,0,0)}+\lambda^+_{23} |\phi^{3}_{11}
\rangle_{(1,1,0)}\Bigr\}+
\mathcal{P}_{3}^+\Bigl\{\mathcal{P}_{12}^+|\phi^{3}_{12}
\rangle_{(0)}+ \lambda^+_{12} |\phi^{3}_{13} \rangle_{(2,0,0)}
+\mathcal{P}_{11}^+ |\phi^{3}_{14} \rangle_{(-1,1,0)}
\Bigr\}\nonumber
\\
&& \hspace{-3em} +\mathcal{P}_{11}^+\Bigl\{\lambda^+_{12}
|\phi^{3}_{15} \rangle_{(0,0,1)}+
\eta_1^+\lambda_{12}^+\lambda^+_{13}|\phi^{3}_{16} \rangle_{(0)}+
\lambda^+_{12}\lambda^+_{23}\Bigl(
\eta_1^+|\phi^{3}_{17}\rangle_{(-1,1,0)_3}+\eta_{2}^+
|\phi^{3}_{18}\rangle_{(0)}\nonumber\\
&& \hspace{-3em}
+\vartheta_{12}^+|\phi^{3}_{19}\rangle_{(1,0,0)_3}\Bigr)
+\lambda^+_{13} |\phi^{3}_{20} \rangle_{(0,1,0)}+ \lambda^+_{23}
|\phi^{3}_{21}
\rangle_{(-1,2,0)}\Bigr\}\nonumber\\
&& \hspace{-3em} + \mathcal{P}_{12}^+\Bigl\{\lambda^+_{23}
|\phi^{3}_{22} \rangle_{(0,1,0)}+\lambda^+_{12} |\phi^{3}_{23}
\rangle_{(1,-1,1)}+\lambda^+_{13} |\phi^{3}_{24}
\rangle_{(1,0,0)}+\eta_1^+\lambda^+_{12}\lambda^+_{23}
|\phi^{3}_{25}\rangle_{(0)} \Bigr\} \nonumber\\
&& \hspace{-3em} + \mathcal{P}_{13}^+\lambda^+_{12} |\phi^{3}_{26}
\rangle_{(1,0,0)}+ \mathcal{P}_{22}^+\lambda_{23}^+|\phi^{3}_{27}
\rangle_{(1,0,0)}+ \lambda^+_{12}\Bigl\{\lambda^+_{23}
|\phi^{3}_{28} \rangle_{(2,1,0)}+\lambda^+_{13} |\phi^{3}_{29}
\rangle_{(3,0,0)}\Bigr\} \biggr]
\nonumber\\
&& \hspace{-3em} +
\mathcal{P}_2^+\biggl[\mathcal{P}_{3}^+\mathcal{P}_{11}^+|\phi^{3}_{30}
\rangle_{(0)} + \mathcal{P}_{11}^+\Bigl\{\lambda^+_{23}
|\phi^{3}_{31} \rangle_{(0,1,0)}+\lambda^+_{13} |\phi^{3}_{32}
\rangle_{(1,0,0)}+\lambda^+_{12} |\phi^{3}_{33}
\rangle_{(1,-1,1)}\nonumber\\
&& \hspace{-3em} + \eta_1^+\lambda^+_{12}\lambda^+_{23}
|\phi^{3}_{34}
\rangle_{(0)}\Bigr\}+\mathcal{P}_{12}^+\lambda^+_{23}|\phi^{3}_{35}
\rangle_{(1,0,0)} + \lambda^+_{12}\lambda^+_{23} |\phi^{3}_{36}
\rangle_{(3,0,0)}\biggr] +
\mathcal{P}_3^+\mathcal{P}_{11}^+\lambda^+_{12} |\phi^{3}_{37}
\rangle_{(1,0,0)}
\nonumber\\
&& \hspace{-3em} +
\mathcal{P}_{11}^+\biggl[\mathcal{P}_{12}^+\Bigl\{\lambda^+_{12}
|\phi^{3}_{38} \rangle_{(0,-1,1)}+\lambda^+_{13} |\phi^{3}_{39}
\rangle_{(0)}+\lambda^+_{23} |\phi^{3}_{40}
\rangle_{(-1,1,0)}\Bigr\}+\mathcal{P}_{13}^+\lambda^+_{12}
|\phi^{3}_{41}
\rangle_{(0)}\nonumber\\
&& \hspace{-3em} + \mathcal{P}_{22}^+\lambda_{23}^+|\phi^{3}_{42}
\rangle_{(0)}+\lambda_{12}^+\lambda^+_{13} |\phi^{3}_{43}
\rangle_{(2,0,0)} + \lambda^+_{12}\lambda^+_{23} |\phi^{3}_{44}
\rangle_{(1,1,0)}\biggr]+
\mathcal{P}_{12}^+\lambda^+_{12}\lambda^+_{23} |\phi^{3}_{45}
\rangle_{(2,0,0)}\nonumber\\
 && \hspace{-2em} +\eta_0|\chi^{3}_0\rangle_{(s)_3}
\,,\nonumber
\end{eqnarray}
where,  the vector $|\chi^{3}_0\rangle_{(s)_3}$ has the same
definition and properties as $|\chi^{4}\rangle_{(s)_3}$ in the
Eqs. (\ref{x-4}), (\ref{x-4decomp})--(\ref{x-4decompi}) and the
decomposition of ghost-independent vectors from
$\mathcal{H}\bigotimes \mathcal{H}'$  looks like,
\begin{align}\label{x-3decomp}
&|\phi^{3}_l \rangle_{(1,0,0)}  =  a_1^{+\mu} |0\rangle
\phi^{3}_{l|\mu}   \,, && |\phi^{3}_r \rangle_{(0,1,0)} =
a_2^{+\mu} |0\rangle \phi^{3}_{r|\mu}  + a_1^{+\mu}d^+_{12}
|0\rangle \phi^{\prime 3}_{r|\mu}  \,,\\
&|\phi^{3}_n \rangle_{(0)} =|0\rangle \phi^{3}_{n}  \,, &&
|\phi^{3}_p \rangle_{(2,0,0)} = a_1^{+\mu}a_1^{+\nu} |0\rangle
\phi^{3}_{p|\mu\nu}+ b_{11}^+|0\rangle
\phi^{3}_{p} \,,\\
& |\phi^{3}_t \rangle_{(0,-1,1)}=d^+_{23}|0\rangle \phi^{3}_{t}
\,, &&|\phi^{3}_d \rangle_{(-1,1,0)} = d^+_{12}|0\rangle
\phi^{3}_{d}
 \,,
\\
& |\phi^{3}_a \rangle_{(1,-1,1)}=a_1^{+\mu}d^+_{23}|0\rangle
\phi^{3}_{a|\mu}  \,, && |\phi^{3}_{21} \rangle_{(-1,2,0)} =
d^+_{12}\bigl(a_2^{+\mu} |0\rangle \phi^{3}_{21|\mu}  +
a_1^{+\mu}d^+_{12} |0\rangle \phi^{\prime3}_{21|\mu} \bigr)\,,
\\
& |\phi^{3}_2 \rangle_{(-1,0,1)}=d^+_{13}|0\rangle
\phi^{3}_{2}+d^+_{12}d^+_{23}|0\rangle \phi^{\prime 3}_{2} \,, &&
|\phi^{3}_7 \rangle_{(2,-1,1)} =
d^+_{23}\bigl(a_1^{+\mu}a_1^{+\nu} |0\rangle \phi^{3}_{7|\mu\nu} +
b_{11}^+|0\rangle \phi^{3}_{7} \bigr)\,,
\end{align}
\vspace{-3.5ex}
\begin{eqnarray}
|\phi^{3}_{15} \rangle_{(0,0,1)} & = & a_3^{+\mu}|0\rangle \phi^{
3}_{15|\mu} +a_2^{+\mu}d^+_{23} |0\rangle \phi^{\prime 3}_{15|\mu}
+ a_1^{+\mu}d^+_{13} |0\rangle \phi^{\prime\prime
3}_{15|\mu}+a_1^{+\mu}d^+_{12}d^+_{23} |0\rangle
\phi^{\prime\prime\prime 3}_{15|\mu}  \,, \\
 |\phi^{3}_{28}
\rangle_{(2,1,0)}&=& a_1^{+\mu}a_1^{+\nu}\bigl(a_2^{+\rho}
|0\rangle \phi^{3}_{28|\mu\nu,\rho}+a_1^{+\rho}d_{12}^+|0\rangle
\phi^{3}_{28|\mu\nu\rho}\bigr) + b_{11}^+a_2^{+\mu}|0\rangle
\phi^{\prime\prime3}_{28| \mu}\\
 && +a_1^{+\mu}\bigl(b_{11}^+d_{12}^+
|0\rangle \phi^{3}_{28|\mu} + b_{12}^+|0\rangle
\phi^{\prime3}_{28|\mu}\bigr) \,,\nonumber\\
|\phi^{3}_y \rangle_{(1,1,0)}&=&a_1^{+\mu}\bigl(a_2^{+\nu}
|0\rangle \phi^{3}_{y|\mu,\nu}+a_1^{+\nu}d_{12}^+|0\rangle
\phi^{3}_{y|\mu\nu}\bigr)  +b_{11}^+d_{12}^+ |0\rangle
\phi^{3}_{y} + b_{12}^+|0\rangle
\phi^{\prime3}_{y}\,,\\
|\phi^{3}_b \rangle_{(3,0,0)} & = &
a_1^{+\mu}\bigl(a_1^{+\nu}a_1^{+\rho} |0\rangle
\phi^{3}_{b|\mu\nu\rho}+b_{11}^+|0\rangle
\phi^{3}_{b|\mu}\bigr),\label{x-3decompf}
\end{eqnarray}
for $l=1,8,19, 24,26,27,32,35,37$, $p=10,13,43,45$, $
n=3,4,6,9,12,16,18,25,30,34,39$, $41$, $42$, $r=20,22,31$, $t =5,38$, $a
= 23,33$, $d =  14,17,40$, $y= 11,44$ and $b=29,36$.

Then, for the reducible gauge parameter of the second level
$|\chi^{2}\rangle_{(s)_3}$ for the  value $l=2$ in the
Eq.(\ref{ghnum}) and for the spin value $(2,1,1)$ in the
Eqs.(\ref{nidecompos}) we have decomposition  in even powers of
ghosts from $(102+45)$ summands starting from the second order in
$\mathcal{P}_I$,
\begin{eqnarray}\label{x-2}
|\chi^{2}\rangle_{(s)_3} &=&
\mathcal{P}_{1}^+\biggl[\mathcal{P}_{2}^+\Bigl\{|\phi^{2}_{1}\rangle_{(1,0,1)}+
\eta_1^+\mathcal{P}_{3}^+|\phi^{2}_{2}
\rangle_{(0)}+\mathcal{P}_{11}^+\Bigl(\vartheta^+_{13}|\phi^{2}_{3}
\rangle_{(0)}+\vartheta^+_{23}|\phi^{2}_4 \rangle_{(-1,1,0)}
\\
&& \hspace{-3em} + \vartheta^+_{12}|\phi^{2}_5
\rangle_{(0,-1,1)}\hspace{-0.1em}\Bigr)\hspace{-0.1em} +
\vartheta^+_{23}\lambda^+_{12}|\phi^{2}_6 \rangle_{(2,0,0)}+
\vartheta^+_{12}\lambda^+_{23}|\phi^{2}_7 \rangle_{(2,0,0)}
+\vartheta^+_{23}\mathcal{P}_{12}^+|\phi^{2}_{8}
\rangle_{(0)}+\hspace{-0.1em}\lambda^+_{13}\hspace{-0.1em}
\Bigl(\hspace{-0.1em}\eta_{1}^+|\phi^{2}_9\rangle_{(1,0,0)}
\nonumber
\\
&& \hspace{-3em}  +\eta_{11}^+ |\phi^{2}_{10}\rangle_{(0)}\Bigr)
+\lambda^+_{12}
\hspace{-0.1em}\Bigl(\hspace{-0.1em}\eta_{1}^+|\phi^{2}_{11}\rangle_{(1,-1,1)}
+\eta_{11}^+
|\phi^{2}_{12}\rangle_{(0,-1,1)}\hspace{-0.1em}\Bigr)\hspace{-0.1em}
+ \lambda^+_{23}
\hspace{-0.1em}\Bigl(\hspace{-0.1em}\eta_{1}^+|\phi^{2}_{13}\rangle_{(0,1,0)}+\eta_{2}^+
|\phi^{2}_{102}\rangle_{(1,0,0)} \nonumber
\\
&& \hspace{-3em} +\eta_{11}^+ |\phi^{2}_{14}\rangle_{(-1,1,0)}
+\eta_{12}^+
|\phi^{2}_{15}\rangle_{(0)}\hspace{-0.1em}\Bigr)\hspace{-0.1em}\Bigr\}\hspace{-0.1em}+
\mathcal{P}_{3}^+\Bigl\{|\phi^{2}_{16}
\rangle_{(1,1,0)}+\mathcal{P}_{11}^+\vartheta^+_{12}|\phi^{2}_{17}
\rangle_{(0)}+\lambda^+_{12}[\eta_{1}^+{|\phi^{2}_{18}
\rangle_{(1,0,0)}}\nonumber
\\
&& \hspace{-3em} +\eta_{11}^+|\phi^{2}_{19} \rangle_{(0)}\Bigr\} +
\mathcal{P}_{11}^+\Bigl\{|\phi^{2}_{20}
\rangle_{(-1,1,1)}+\eta_3^+\lambda^+_{12}|\phi^{2}_{21}
\rangle_{(0)}+\eta_2^+\bigl(\lambda^+_{12}{|\phi^{2}_{22}
\rangle_{(0,-1,1)}}+\lambda^+_{13}|\phi^{2}_{23} \rangle_{(0)}
\nonumber
\\
&& \hspace{-3em} +\lambda^+_{23}{|\phi^{2}_{24}
\rangle_{(-1,1,0)}}\bigr) + \eta_1^+
\bigl(\lambda^+_{12}{|\phi^{2}_{25}
\rangle_{(-1,0,1)}}+\lambda^+_{13}{|\phi^{2}_{26}
\rangle_{(-1,1,0)}}+\lambda^+_{23}{|\phi^{2}_{27}
\rangle_{(-2,2,0)}}\bigr) \nonumber
\\
&& \hspace{-3em} +\vartheta^+_{23}\lambda^+_{12}|\phi^{2}_{28}
\rangle_{(0,1,0)} + \vartheta^+_{12}\lambda^+_{13}|\phi^{2}_{29}
\rangle_{(1,0,0)}+\vartheta^+_{13}\lambda^+_{12}|\phi^{2}_{30}
\rangle_{(1,0,0)}+\vartheta^+_{12}\lambda^+_{23}|\phi^{2}_{31}
\rangle_{(0,1,0)}  \nonumber
\\
&& \hspace{-3em}
+\vartheta^+_{12}\lambda^+_{12}\bigl[{|\phi^{2}_{32}
\rangle_{(1,-1,1)}} +  \eta_1^+\lambda^+_{23}{|\phi^{2}_{33}
\rangle_{(0)}}\bigr]\Bigr\}+\mathcal{P}_{12}^+\Bigl\{|\phi^{2}_{34}
\rangle_{(0,0,1)}+ \eta_1^+\bigl(\lambda^+_{12}|\phi^{2}_{35}
\rangle_{(0,-1,1)}\nonumber
\\
&& \hspace{-3em} +\lambda^+_{13}|\phi^{2}_{36} \rangle_{(0)}
+\lambda^+_{23}|\phi^{2}_{37} \rangle_{(-1,1,0)}\bigr) +
\eta_2^+\lambda^+_{23}|\phi^{2}_{38}
\rangle_{(0)}+\vartheta^+_{23}\lambda^+_{12} |\phi^{2}_{39}
\rangle_{(1,0,0)}+\vartheta^+_{12}\lambda^+_{23} |\phi^{2}_{40}
\rangle_{(1,0,0)}\Bigr\}\nonumber\\
&& \hspace{-3em} + \mathcal{P}_{22}^+\Bigl\{|\phi^{2}_{41}
\rangle_{(1,-1,1)}+
 \eta_1^+\lambda^+_{23}|\phi^{2}_{42} \rangle_{(0)}\Bigr\}+
\mathcal{P}_{13}^+\Bigl\{|\phi^{2}_{43} \rangle_{(0,1,0)}+
\eta_1^+\lambda^+_{12}|\phi^{2}_{44}
\rangle_{(0)}\Bigr\}+ \mathcal{P}_{23}^+|\phi^{2}_{45}\rangle_{(1,0,0)}\nonumber\\
&& \hspace{-3em} + \lambda^+_{12} |\phi^{2}_{46}
\rangle_{(2,0,1)}+\lambda^+_{12}\lambda^+_{23}
\Bigl(\eta_{1}^+|\phi^{2}_{47}\rangle_{(1,1,0)} +\eta_{11}^+
|\phi^{2}_{48}\rangle_{(0,1,0)}+\eta_{12}^+
|\phi^{2}_{49}\rangle_{(1,0,0)}+ \vartheta_{12}^+
{|\phi^{2}_{50}\rangle_{(3,0,0)}} \nonumber
\\
&& \hspace{-3em} + \eta_{2}^+ {|\phi^{2}_{51}\rangle_{(2,0,0)}}
\hspace{-0.15em}\Bigr)\hspace{-0.15em}+
\lambda^+_{13}\hspace{-0.1em}\Bigl\{\hspace{-0.1em} |\phi^{2}_{52}
\rangle_{(2,1,0)}\hspace{-0.1em}+
\eta_1^+\lambda^+_{12}|\phi^{2}_{53}
\rangle_{(2,0,0)}\hspace{-0.1em}+
\eta_{11}^+\lambda^+_{12}|\phi^{2}_{54}
\rangle_{(1,0,0)}\hspace{-0.15em}\Bigr\}\hspace{-0.15em}+
\lambda^+_{23} |\phi^{2}_{55} \rangle_{(1,2,0)}
\hspace{-0.15em}\biggr]
\nonumber\\
&& \hspace{-3.5em} +\mathcal{P}_2^+
\biggl[\mathcal{P}_{3}^+|\phi^{2}_{56} \rangle_{(2,0,0)} +
\mathcal{P}_{11}^+\Bigl\{|\phi^{2}_{57}
\rangle_{(0,0,1)}+\eta_2^+\lambda^+_{23} |\phi^{2}_{58}
\rangle_{(0)}+\eta_1^+\lambda^+_{13} |\phi^{2}_{59}
\rangle_{(0)}\nonumber\\
&& \hspace{-3em} +\eta_1^+\lambda^+_{12} |\phi^{2}_{60}
\rangle_{(0,-1,1)}+\vartheta^+_{23}\lambda^+_{12} |\phi^{2}_{61}
\rangle_{(1,0,0)}+\vartheta^+_{12}\lambda^+_{23} |\phi^{2}_{62}
\rangle_{(1,0,0)}+\eta_1^+\lambda^+_{23} |\phi^{2}_{63}
\rangle_{(-1,1,0)}\Bigr\}\nonumber\\
&& \hspace{-3em} + \mathcal{P}_{12}^+\Bigl\{|\phi^{2}_{64}
\rangle_{(1,-1,1)}+\eta_1^+\lambda^+_{23}|\phi^{2}_{65}
\rangle_{(0)} \Bigr\} + \lambda^+_{13}|\phi^{2}_{66}
\rangle_{(3,0,0)}+ \lambda^+_{23}|\phi^{2}_{67} \rangle_{(2,1,0)}
\nonumber\\
&& \hspace{-3em} + \lambda^+_{12}\Bigl\{|\phi^{2}_{68}
\rangle_{(3,-1,1)} + \eta_1^+\lambda^+_{23} |\phi^{2}_{69}
\rangle_{(2,0,0)}+\eta_{11}^+\lambda^+_{23} |\phi^{2}_{70}
\rangle_{(1,0,0)}\Bigr\}+ \mathcal{P}_{13}^+|\phi^{2}_{71}
\rangle_{(1,0,0)}\biggr]
\nonumber
\end{eqnarray}
\vspace{-3.5ex}
\begin{eqnarray}
&& \hspace{-3.5em} +  \mathcal{P}_3^+\Bigl[\mathcal{P}_{11}^+
\Bigl\{|\phi^{2}_{72} \rangle_{(0,1,0)} +
\eta_1^+\lambda^+_{12}|\phi^{2}_{73} \rangle_{(0)}\Bigr\}+
\mathcal{P}_{12}^+|\phi^{2}_{74} \rangle_{(1,0,0)} +\lambda^+_{12}
|\phi^{2}_{75}
\rangle_{(3,0,0)}\Bigr]\nonumber\\
&& \hspace{-3.5em} +
\mathcal{P}_{11}^+\biggl[\mathcal{P}_{12}^+\Bigl\{|\phi^{2}_{76}
\rangle_{(-1,0,1)}+ \vartheta^+_{23}\lambda^+_{12} |\phi^{2}_{77}
\rangle_{(0)}+\vartheta^+_{12}\lambda^+_{23} |\phi^{2}_{78}
\rangle_{(0)}\Bigr\}+\mathcal{P}_{13}^+ |\phi^{2}_{79}
\rangle_{(-1,1,0)} \nonumber\\
&& \hspace{-3em} +\mathcal{P}_{22}^+|\phi^{2}_{80}
\rangle_{(0,-1,1)}+\mathcal{P}_{23}^+|\phi^{2}_{81}
\rangle_{(0)}+\lambda_{12}^+\Bigl\{|\phi^{2}_{82}
\rangle_{(1,0,1)}+ \eta_1^+\lambda^+_{13} |\phi^{2}_{83}
\rangle_{(1,0,0)}+\eta_{11}^+\lambda^+_{13} |\phi^{2}_{84}
\rangle_{(0)}\nonumber\\
&& \hspace{-3em} + \eta_1^+\lambda^+_{23} |\phi^{2}_{85}
\rangle_{(0,1,0)}+\eta_2^+\lambda^+_{23} |\phi^{2}_{101}
\rangle_{(1,0,0)}+\eta_{11}^+\lambda^+_{23} |\phi^{2}_{86}
\rangle_{(-1,1,0)}+\eta_{12}^+\lambda^+_{23} |\phi^{2}_{87}
\rangle_{(0)} \nonumber\\
&& \hspace{-3em} + \vartheta_{12}^+\lambda^+_{23} {|\phi^{2}_{88}
\rangle_{(2,0,0)}}\Bigr\}+ \lambda^+_{13}|\phi^{2}_{89}
\rangle_{(1,1,0)}+\lambda^+_{23} |\phi^{2}_{90}
\rangle_{(0,2,0)}\biggr]
\nonumber\\
&& \hspace{-3.5em}
+\mathcal{P}_{12}^+\Bigl[\mathcal{P}_{13}^+|\phi^{2}_{91}
\rangle_{(0)}+\lambda^+_{12}\Bigl\{|\phi^{2}_{92}
\rangle_{(2,-1,1)}+\eta_1^+\lambda_{23}^+|\phi^{2}_{93}
\rangle_{(1,0,0)}+\eta_{11}^+\lambda_{23}^+|\phi^{2}_{94}
\rangle_{(0)}\Bigr\} \nonumber\\
&& \hspace{-3em} + \lambda^+_{13}|\phi^{2}_{95} \rangle_{(2,0,0)}
+ \lambda^+_{23}|\phi^{2}_{96} \rangle_{(1,1,0)}\Bigr]+
\mathcal{P}_{22}^+\lambda^+_{23} |\phi^{2}_{97}
\rangle_{(2,0,0)}+\mathcal{P}_{13}^+\lambda^+_{12} |\phi^{2}_{98}
\rangle_{(2,0,0)}\nonumber\\
&& \hspace{-3.5em} + \lambda^+_{12}
\Bigl[\lambda^+_{13}|\phi^{2}_{99}
\rangle_{(4,0,0)}+\lambda^+_{23}|\phi^{2}_{100}
\rangle_{(3,1,0)}\Bigr]\ +\
\eta_0|\chi^{2}_0\rangle_{(s)_3}\nonumber
 \,,
\end{eqnarray}
where,  the vector $|\chi^{2}_0\rangle_{(s)_3}$ has the same
definition and properties as $|\chi^{3}\rangle_{(s)_3}$  in the
Eqs. (\ref{x-3}), (\ref{x-3decomp})--(\ref{x-3decompf}) without
$\eta_0$-dependent terms and the decomposition of
ghost-independent vectors in $\mathcal{H}\bigotimes \mathcal{H}'$
with different values of spins than in
(\ref{x-3decomp})--(\ref{x-3decompf}) writes as follows,
\begin{eqnarray}\label{x-2decomp}
&& |\phi^{2}_l \rangle_{(1,0,1)} =
a_1^{+\mu}\hspace{-0.15em}\Bigl(\hspace{-0.15em}
a_1^{+\nu}\bigl\{d^+_{13} |0\rangle
\phi^{2}_{l|\mu\nu}+\hspace{-0.1em}d^+_{12}d^+_{23} |0\rangle
\phi^{\prime2}_{l|\mu\nu}\bigr\}+
\hspace{-0.1em}a_2^{+\nu}d^+_{23} |0\rangle
\phi^{\prime\prime2}_{l|\mu,\nu}+\hspace{-0.1em}a_3^{+\nu}
|0\rangle \phi^{\prime\prime\prime2}_{l{}\mu,\nu}
\hspace{-0.15em}\Bigr)\\
&&\quad +b_{11}^+\bigl\{d^+_{13}|0\rangle
\phi^{2}_{l}+d^+_{12}d^+_{23} |0\rangle
\phi^{\prime2}_{l}\bigr\}+b_{12}^+d^+_{23} |0\rangle
\phi^{\prime\prime
2}_{l}+b_{13}^+|0\rangle\phi^{\prime\prime\prime2}_{l}\,,\nonumber
\\ \label{-111}
&& |\phi^{2}_m \rangle_{(-1,1,1)} =
a_1^{+\mu}d^+_{12}\bigr\{d^+_{13} |0\rangle
\phi^{2}_{m|\mu}+d^+_{12}d^+_{23} |0\rangle
\phi^{\prime2}_{m|\mu}\bigr\}+ a_2^{+\mu}\bigr\{d^+_{13} |0\rangle
\phi^{\prime\prime2}_{m|\mu}\\
&&\quad +d^+_{12}d^+_{23} |0\rangle
\phi^{\prime\prime\prime2}_{m|\mu}\bigr\}+a_3^{+\mu}d^+_{12}
|0\rangle \phi^{(iv)2}_{m|\mu} \,,
\nonumber\\
\label{2p} && |\phi^{2}_p \rangle_{(2,0,1)} =
a_1^{+\mu}a_1^{+\nu}\Bigl(a_3^{+\rho} |0\rangle
\phi^{2}_{p|\mu\nu,\rho}+a_1^{+\rho}\bigr\{d^+_{13} |0\rangle
|0\rangle \phi^{\prime2}_{p|\mu\nu\rho}+ d^+_{12}d^+_{23}
|0\rangle \phi^{\prime\prime2}_{p|\mu\nu\rho}\bigr\}\\
&&\quad + a_2^{+\rho} d^+_{23}|0\rangle
\phi^{\prime\prime\prime2}_{p|\mu\nu,\rho}\Bigr) +
b_{11}^+\Bigl(a_3^{+\mu} |0\rangle \phi^{2}_{p|\mu}
+a_2^{+\mu}d_{23}^+ |0\rangle
\phi^{\prime2}_{p|\mu}+a_1^{+\mu}d_{13}^+ |0\rangle
\phi^{\prime\prime2}_{p|\mu} \nonumber\\
&& \quad+a_1^{+\mu}d_{12}^+ d_{23}^+ |0\rangle
\phi^{\prime\prime\prime2}_{p|\mu}\Bigr)+ a_1^{+\mu}\Bigl(b_{13}^+
|0\rangle \phi^{(iv)2}_{p|\mu}+b_{12}^+
d_{23}^+|0\rangle \phi^{(v)2}_{p|\mu}\Bigr), \nonumber\\
&&\label{120r} |\phi^{2}_r \rangle_{(1,2,0)} =
a_1^{+\mu}\bigl(a_1^{+\nu}a_1^{+\rho}(d^+_{12})^2 |0\rangle
\phi^{2}_{r|\mu\nu\rho}+a_1^{+\nu}a_2^{+\rho}d^+_{12} |0\rangle
\phi^{\prime2}_{r|\mu\nu,\rho}+ a_2^{+\nu}a_2^{+\rho} |0\rangle
\phi^{\prime\prime2}_{r|\mu,\nu\rho}\\
&& \quad +b_{22}^+|0\rangle \phi^{(iv)2}_{r|\mu}
+b_{11}^+(d^+_{12})^2 |0\rangle \phi^{2}_{r|\mu}+b_{12}^+d^+_{12}
|0\rangle \phi^{\prime
2}_{r|\mu}\bigr)+a_2^{+\mu}\bigl(b_{11}^+d^+_{12} |0\rangle
\phi^{\prime \prime2}_{r|\mu}+b_{12}^+ |0\rangle
\phi^{\prime\prime\prime 2}_{r|\mu}\bigr), \nonumber\\
&& \label{310n}|\phi^{2}_{100} \rangle_{(3,1,0)} =
a_1^{+\mu}a_1^{+\nu}\Bigl(a_1^{+\rho}a_1^{+\sigma}
d_{12}^+|0\rangle
\phi^{2}_{100|\mu\nu\rho\sigma}+a_1^{+\rho}a_2^{+\sigma} |0\rangle
\phi^{\prime 2}_{100|\mu\nu\rho,\sigma} +b_{12}^+|0\rangle
\phi^{\prime 2}_{100|\mu\nu}
\\
&&\quad + b_{11}^+d_{12}^+|0\rangle \phi^{2}_{100|\mu\nu}\Bigr) +
a_1^{+\mu}a_2^{+\nu}b_{11}^+ |0\rangle \phi^{ 2}_{100|\mu,\nu}
+b_{11}^+\bigl(b_{11}^+d_{12}^+|0\rangle \phi^{2}_{100}+
b_{12}^+|0\rangle \phi^{\prime 2}_{100}\bigr), \nonumber\\
&& |\phi^{2}_{99} \rangle_{(4,0,0)} =
a_1^{+\mu}a_1^{+\nu}\bigl(a_1^{+\rho}a_1^{+\sigma} |0\rangle
\phi^{2}_{{99}|\mu\nu\rho\sigma}+b^+_{11} |0\rangle
\phi^{2}_{{99}|\mu\nu}\bigr) + (b_{11}^+)^2 |0\rangle
\phi^{2}_{{99}},\label{400n} \\
&& |\phi^{3}_t \rangle_{(3,-1,1)}  =
a_1^{+\mu}d_{23}^+\bigl(a_1^{+\nu}a_1^{+\rho} |0\rangle
\phi^{3}_{t|\mu\nu\rho}+b_{11}^+|0\rangle
\phi^{3}_{t|\mu}\bigr), \\
&& |\phi^{2}_u \rangle_{(0,2,0)} =
a_1^{+\mu}\bigl(a_1^{+\nu}(d^+_{12})^2 |0\rangle
\phi^{2}_{u|\mu\nu}+a_2^{+\nu}d^+_{12} |0\rangle
\phi^{\prime2}_{u|\mu,\nu}\bigr)+ a_2^{+\mu}a_2^{+\nu} |0\rangle
\phi^{\prime\prime2}_{u|\mu\nu}\\
&& \quad +b_{11}^+(d^+_{12})^2 |0\rangle
\phi^{2}_{u{}}+b_{12}^+d^+_{12} |0\rangle \phi^{\prime
2}_{u{}}+b_{22}^+ |0\rangle \phi^{\prime\prime 2}_{u{}},\nonumber\\
&& |\phi^{2}_{27} \rangle_{(-2,2,0)} =(d^+_{12})^2 |0\rangle
|\phi^{2}_{27} \label{x-2decompf}
\end{eqnarray}
for $l = 1, 82, m=20 , p=46,   r=55,  u=90, t = 68$.

Analogously, we decompose the  gauge parameter
$|\chi^{1}\rangle_{(s)_3}$ for values $l=1$ and $(s)_3 = (2,1,1)$
in the Eq.(\ref{ghnum}) and Eqs.(\ref{nidecompos}) respectively,
in odd powers of ghosts from $(164+102)$ summands starting from
the first order in momenta $\mathcal{P}_I$,
\begin{eqnarray}\label{x-1}
|\chi^{1}\rangle_{(s)_3} &=&
\mathcal{P}_{1}^+\biggl[|\phi^{1}_{1}\rangle_{(1,1,1)}+
\mathcal{P}_{2}^+\Bigl\{\eta_1^+\bigl(|\phi^{1}_{2}\rangle_{(0,0,1)}+\eta_2^+
\lambda_{23}^+|\phi^{1}_{3}\rangle_{(0)}+\vartheta_{12}^+
\lambda_{23}^+|\phi^{1}_{4}\rangle_{(1,0,0)}
\\
&& \hspace{-3.5em} +\vartheta_{23}^+
\lambda_{12}^+|\phi^{1}_{5}\rangle_{(1,0,0)}\bigr) +
\eta_2^+|\phi^{1}_{6}\rangle_{(1,-1,1)}
+\eta_3^+|\phi^{1}_{7}\rangle_{(1,0,0)}+
\eta_{11}^+\bigl(|\phi^{1}_{8}\rangle_{(-1,0,1)}+
\vartheta_{12}^+\lambda_{23}^+|\phi^{1}_{9}\rangle_{(0)}\nonumber\\
&& \hspace{-3.5em}+
\vartheta_{23}^+\lambda_{12}^+|\phi^{1}_{10}\rangle_{(0)}\bigr)+
\eta_{13}^+|\phi^{1}_{11}\rangle_{(0)} +
\eta_{12}^+|\phi^{1}_{12}\rangle_{(0,-1,1)}+
\vartheta_{12}^+|\phi^{1}_{13}\rangle_{(2,-1,1)}+\vartheta_{13}^+
|\phi^{1}_{14}\rangle_{(2,0,0)}\nonumber\\
&& \hspace{-3.5em} +
\vartheta_{23}^+\bigl(|\phi^{1}_{15}\rangle_{(1,1,0)}+
\mathcal{P}_{11}^+\vartheta_{12}^+|\phi^{1}_{16}\rangle_{(0)}\bigr)\Bigr\}+
\mathcal{P}_{3}^+\Bigl\{\eta_1^+|\phi^{1}_{17} \rangle_{(0,1,0)} +
\eta_2^+|\phi^{1}_{18} \rangle_{(1,0,0)}+
\eta_{11}^+|\phi^{1}_{19} \rangle_{(-1,1,0)}
\nonumber\\
&& \hspace{-3.5em} + \eta_{12}^+|\phi^{1}_{20} \rangle_{(0)}+
\vartheta_{12}^+|\phi^{1}_{21} \rangle_{(2,0,0)}\Bigr\} +
\mathcal{P}_{11}^+\Bigl\{\eta_1^+\bigl(|\phi^{1}_{22}\rangle_{(-2,1,1)}+
\vartheta_{12}^+
\lambda_{23}^+|\phi^{1}_{23}\rangle_{(-1,1,0)}\nonumber\\
&& \hspace{-3.5em} +\vartheta_{23}^+
\lambda_{12}^+|\phi^{1}_{24}\rangle_{(-1,1,0)}+\vartheta_{12}^+
\lambda_{13}^+|\phi^{1}_{25}\rangle_{(0)} +  \vartheta_{13}^+
\lambda_{12}^+|\phi^{1}_{26}\rangle_{(0)}+\vartheta_{12}^+
\lambda_{12}^+|\phi^{1}_{27}\rangle_{(0,-1,1)}\bigr)  \nonumber\\
&& \hspace{-3.5em} +\eta_2^+
\bigl(|\phi^{1}_{28}\rangle_{(-1,0,1)} + \vartheta_{12}^+
\lambda_{23}^+|\phi^{1}_{29}\rangle_{(0)}+\vartheta_{23}^+
\lambda_{12}^+|\phi^{1}_{30}\rangle_{(0)}\bigr) +
\eta_3^+|\phi^{1}_{31}\rangle_{(-1,1,0)}
+\vartheta^+_{13}|\phi^{1}_{32} \rangle_{(0,1,0)}
\nonumber\\
&& \hspace{-3.5em} + \vartheta^+_{12}\bigl(|\phi^{1}_{33}
\rangle_{(0,0,1)}+\vartheta^+_{23}\lambda_{12}^+ |\phi^{1}_{34}
\rangle_{(1,0,0)}\bigr)+ \vartheta^+_{23}|\phi^{1}_{35}
\rangle_{(-1,2,0)}\Bigr\}+
\mathcal{P}_{12}^+\Bigl\{\eta^+_1\bigl(|\phi^{1}_{36}
\rangle_{(-1,0,1)} \nonumber\\
&& \hspace{-3.5em} + \vartheta_{12}^+\lambda^+_{23}|\phi^{1}_{37}
\rangle_{(0)}\bigr)+\eta^+_2|\phi^{1}_{38} \rangle_{(0,-1,1)}
+\eta^+_3|\phi^{1}_{39} \rangle_{(0)}+\vartheta^+_{12}
|\phi^{1}_{40} \rangle_{(1,-1,1)}+ \vartheta^+_{13} |\phi^{1}_{41}
\rangle_{(1,0,0)}
\nonumber\\
&& \hspace{-3.5em}+ \vartheta^+_{23}\bigl(|\phi^{1}_{42}
\rangle_{(0,1,0)}+\eta_1^+\lambda^+_{12}|\phi^{1}_{43}
\rangle_{(0)}\bigr)\Bigr\} +
\mathcal{P}_{22}^+\Bigl\{\eta_1^+|\phi^{1}_{44}
\rangle_{(0,-1,1)}+ \vartheta_{23}^+|\phi^{1}_{45}
\rangle_{(1,0,0)} \Bigr\} \nonumber\\
&& \hspace{-3.5em} +
\mathcal{P}_{13}^+\Bigl\{\eta_1^+|\phi^{1}_{46}
\rangle_{(-1,1,0)}+ \eta_2^+|\phi^{1}_{47} \rangle_{(0)}+
\vartheta_{12}^+|\phi^{1}_{164} \rangle_{(1,0,0)} \Bigr\} +
\eta_1^+\mathcal{P}_{23}^+|\phi^{1}_{48}
\rangle_{(0)}+\lambda^+_{13}
\Bigl\{\eta_{1}^+\bigl(|\phi^{1}_{49}\rangle_{(1,1,0)}
 \nonumber
\end{eqnarray}
\begin{eqnarray}
&& \hspace{-3.5em} +
\eta_{11}^+\lambda^+_{12}|\phi^{1}_{50}\rangle_{(0)}\bigr)
+\eta_{2}^+|\phi^{1}_{51}\rangle_{(2,0,0)} +\eta_{11}^+
|\phi^{1}_{52}\rangle_{(0,1,0)}
+\eta_{12}^+|\phi^{1}_{53}\rangle_{(1,0,0)} \hspace{-0.1em}+
\vartheta_{12}^+|\phi^{1}_{54}\rangle_{(3,0,0)}\Bigr\}\nonumber\\
&& \hspace{-3.5em} \hspace{-0.15em}+ \lambda^+_{12}
\hspace{-0.15em}\Bigl\{\hspace{-0.15em}\eta_{1}^+\hspace{-0.1em}\bigl(\hspace{-0.1em}|\phi^{1}_{55}
\hspace{-0.1em}\rangle_{\hspace{-0.1em}(1,0,1)}\hspace{-0.1em}+
\eta_{11}^+\hspace{-0.1em}\lambda_{23}^+
|\phi^{1}_{56}\hspace{-0.1em}\rangle_{\hspace{-0.1em}(-1,1,0)}\hspace{-0.1em}+
\eta_{2}^+\hspace{-0.1em}\lambda^+_{23}|\phi^{1}_{57}\hspace{-0.1em}\rangle_{\hspace{-0.1em}(1,0,0)}\hspace{-0.15em}
+
\eta_{12}^+\hspace{-0.1em}\lambda^+_{23}|\phi^{1}_{58}\hspace{-0.1em}\rangle_{\hspace{-0.1em}(0)}\hspace{-0.15em}+
\vartheta_{12}^+\hspace{-0.1em}\lambda_{23}^+|\phi^{1}_{59}\hspace{-0.1em}\rangle_{\hspace{-0.1em}(2,0,0)}
\hspace{-0.15em}\bigr)
\nonumber\\
&& \hspace{-3.5em}
+\eta_{2}^+\bigl(|\phi^{1}_{60}\rangle_{(2,-1,1)}+
\eta_{11}^+\lambda_{23}^+ |\phi^{1}_{61}\rangle_{(0)}\bigr)
+\eta_3^+|\phi^{1}_{62}\rangle_{(2,0,0)}
+\eta_{11}^+\bigl(|\phi^{1}_{63}\rangle_{(0,0,1)} +
\vartheta_{12}^+\lambda^+_{23}
|\phi^{1}_{64}\rangle_{(1,0,0)}\bigr) \nonumber\\
&& \hspace{-3.5em}  +\eta_{12}^+|\phi^{1}_{65}
\rangle_{(1,-1,1)}+\vartheta_{12}^+
|\phi^{1}_{66}\rangle_{(3,-1,1)}+\vartheta_{13}^+|\phi^{1}_{67}\rangle_{(3,0,0)}
 +\vartheta^+_{23}|\phi^{1}_{68} \rangle_{(2,1,0)}\Bigr\} + \lambda^+_{23}
\Bigl\{\eta_{1}^+|\phi^{1}_{69}\rangle_{(0,2,0)}
 \nonumber
\\
&& \hspace{-3.5em} +\eta_2^+|\phi^{1}_{70}
\rangle_{(1,1,0)}+\eta_{11}^+
|\phi^{1}_{71}\rangle_{(-1,2,0)}+\eta_{12}^+
|\phi^{1}_{72}\rangle_{(0,1,0)} +\eta_{22}^+
|\phi^{1}_{73}\rangle_{(1,0,0)}+\vartheta^+_{12}|\phi^{1}_{74}
\rangle_{(2,1,0)}\Bigr\}\biggr]\nonumber
\\
&& \hspace{-4.0em} +
\mathcal{P}_{2}^+\biggl[|\phi^{2}_{75}\rangle_{(2,0,1)}+
\mathcal{P}_{3}^+\Bigl\{ \eta_1^+|\phi^{1}_{76}\rangle_{(1,0,0)}+
\eta_{11}^+|\phi^{1}_{77}\rangle_{(0)}\Bigr\}+
\mathcal{P}_{11}^+\Bigl\{\eta_1^+\bigl(|\phi^{1}_{78}\rangle_{(-1,0,1)}\nonumber\\
&& \hspace{-3.5em} +  \vartheta_{12}^+\lambda^+_{23}
|\phi^{1}_{79}\rangle_{(0)}+\vartheta_{23}^+\lambda^+_{12}
|\phi^{1}_{80}\rangle_{(0)}\bigr)+\eta_2^+|\phi^{1}_{81}\rangle_{(0,-1,1)}+\eta_3^+|\phi^{1}_{82}\rangle_{(0)}
+\vartheta^+_{12}|\phi^{1}_{83} \rangle_{(1,-1,1)}\nonumber
\\
&& \hspace{-3.5em} +  \vartheta^+_{13}|\phi^{1}_{84}
\rangle_{(1,0,0)} + \vartheta^+_{23}|\phi^{1}_{85}
\rangle_{(0,1,0)}\Bigr\}+
\mathcal{P}_{12}^+\Bigl\{\eta^+_1|\phi^{1}_{86}
\rangle_{(0,-1,1)}+  \vartheta^+_{23}|\phi^{1}_{87}
\rangle_{(1,0,0)}\Bigr\}
\nonumber\\
&& \hspace{-3.5em} +  \mathcal{P}_{13}^+\eta_1^+|\phi^{1}_{88}
\rangle_{(0)}+ \lambda^+_{13}
\Bigl\{\eta_{1}^+|\phi^{1}_{89}\rangle_{(2,0,0)}+\eta_{11}^+
|\phi^{1}_{90}\rangle_{(1,0,0)}\Bigr\}+ \lambda^+_{12}
\Bigl\{\eta_{1}^+\bigl(|\phi^{1}_{91}\rangle_{(2,-1,1)}\nonumber
\\
&& \hspace{-3.5em} +\eta_{11}^+\lambda^+_{23}
|\phi^{1}_{92}\rangle_{(0)}\bigr)
+\eta_{11}^+|\phi^{1}_{93}\rangle_{(1,-1,1)}+
\vartheta^+_{23}|\phi^{1}_{94} \rangle_{(3,0,0)}\Bigr\} +
\lambda^+_{23}
\Bigl\{\eta_{1}^+|\phi^{1}_{95}\rangle_{(1,1,0)} \nonumber\\
&& \hspace{-3.5em} +\eta_2^+|\phi^{1}_{96}
\rangle_{(2,0,0)}+\eta_{11}^+
|\phi^{1}_{97}\rangle_{(0,1,0)}+\eta_{12}^+
|\phi^{1}_{98}\rangle_{(1,0,0)}+\vartheta^+_{12}|\phi^{1}_{99}
\rangle_{(3,0,0)}\Bigr\}\biggr]\nonumber
\\
&& \hspace{-4.0em} +
 \mathcal{P}_3^+\Bigl[|\phi^{1}_{100}
\rangle_{(2,1,0)}+  \mathcal{P}_{11}^+
\Bigl\{\eta_1^+|\phi^{1}_{101} \rangle_{(-1,1,0)} +
\eta_2^+|\phi^{1}_{102} \rangle_{(0)}
+\vartheta_{12}^+|\phi^{1}_{103}
\rangle_{(1,0,0)}\Bigr\}\nonumber\\
&& \hspace{-3.5em} + \mathcal{P}_{12}^+\eta_1^+|\phi^{1}_{104}
\rangle_{(0)} +\lambda^+_{12} \Bigl\{\eta_1^+|\phi^{1}_{105}
\rangle_{(2,0,0)}+\eta_{11}^+|\phi^{1}_{106}
\rangle_{(1,0,0)}\Bigr\}\Bigr]\nonumber\\
&& \hspace{-4.0em} +  \mathcal{P}_{11}^+\biggl[|\phi^{1}_{107}
\rangle_{(0,1,1)}+ \mathcal{P}_{12}^+ \bigl\{\vartheta^+_{12}
|\phi^{1}_{108} \rangle_{(0,-1,1)}+\vartheta^+_{13}
|\phi^{1}_{109} \rangle_{(0)}+\vartheta^+_{23} |\phi^{1}_{110}
\rangle_{(-1,1,0)}\bigr\}\nonumber\\
&& \hspace{-3.5em} + \mathcal{P}_{22}^+ \vartheta^+_{23}
|\phi^{1}_{111} \rangle_{(0)}+\mathcal{P}_{13}^+
\vartheta^+_{12}|\phi^{1}_{112} \rangle_{(0)} +
\lambda_{12}^+\Bigl\{\eta_1^+\bigl(|\phi^{1}_{113}
\rangle_{(0,0,1)}+ \eta_2^+\lambda^+_{23} |\phi^{1}_{114}
\rangle_{(0)} \nonumber\\
&& \hspace{-3.5em} +\vartheta_{12}^+\lambda^+_{23} |\phi^{1}_{115}
\rangle_{(1,0,0)}\bigr)+\eta_2^+|\phi^{1}_{116}
\rangle_{(1,-1,1)}+  \eta_3^+ |\phi^{1}_{117} \rangle_{(1,0,0)} +
\eta_{11}^+\bigl(|\phi^{1}_{118} \rangle_{(-1,0,1)}+\eta_{13}^+
|\phi^{1}_{121} \rangle_{(0)}\nonumber\\
&& \hspace{-3.5em}  + \vartheta^+_{12}\lambda^+_{23}
|\phi^{1}_{119} \rangle_{(0)}\bigr) +\eta_{12}^+ |\phi^{1}_{120}
\rangle_{(0,-1,1)}
 +  \vartheta^+_{12}|\phi^{1}_{122} \rangle_{(2,-1,1)} +
\vartheta^+_{13}|\phi^{1}_{123} \rangle_{(2,0,0)}+
\vartheta^+_{23}|\phi^{1}_{124} \rangle_{(1,1,0)}\Bigr\}
\nonumber
\end{eqnarray}
\begin{eqnarray}
&& \hspace{-3.5em} +
 \lambda^+_{13}\Bigl\{\eta_1^+|\phi^{1}_{125} \rangle_{(0,1,0)}
+ \eta_2^+|\phi^{1}_{126}
\rangle_{(1,0,0)}+\eta_{11}^+|\phi^{1}_{127}
\rangle_{(-1,1,0)}+\eta_{12}^+|\phi^{1}_{128}
\rangle_{(0)}+\vartheta^+_{12} |\phi^{1}_{129}
\rangle_{(2,0,0)}\Bigr\} \nonumber\\
 && \hspace{-3.5em} +
 \lambda^+_{23}\Bigl\{\eta_1^+|\phi^{1}_{130} \rangle_{(-1,2,0)}
+ \eta_2^+|\phi^{1}_{131}
\rangle_{(0,1,0)}+\eta_{11}^+|\phi^{1}_{132}
\rangle_{(-2,2,0)}+\eta_{12}^+|\phi^{1}_{133}
\rangle_{(-1,1,0)}\nonumber\\
&& \hspace{-3.5em} +  \eta_{22}^+|\phi^{1}_{134}
\rangle_{(0)}+\vartheta^+_{12} |\phi^{1}_{135}
\rangle_{(1,1,0)}\Bigr\}\biggr]\nonumber\\
&& \hspace{-4.0em} +
 \mathcal{P}_{12}^+\Bigl[|\phi^{1}_{136}
\rangle_{(1,0,1)}+\lambda^+_{12}\Bigl\{\eta_1^+|\phi^{1}_{137}
\rangle_{(1,-1,1)}+ \eta_{11}^+|\phi^{1}_{138} \rangle_{(0,-1,1)}+
\vartheta_{23}^+|\phi^{1}_{139}
\rangle_{(2,0,0)} \Bigr\} \nonumber\\
&& \hspace{-3.5em} +  \lambda^+_{13}\Bigl\{\eta_1^+|\phi^{1}_{140}
\rangle_{(1,0,0)} +\eta_{11}^+|\phi^{1}_{141}
\rangle_{(0)}\Bigr\}+ \lambda^+_{23}\Bigl\{\eta_1^+|\phi^{1}_{142}
\rangle_{(0,1,0)}+\eta_2^+|\phi^{1}_{143} \rangle_{(1,0,0)}+
\eta_{11}^+|\phi^{1}_{144}
\rangle_{(-1,1,0)}\nonumber\\
&& \hspace{-3.5em} +  \eta_{12}^+|\phi^{1}_{145} \rangle_{(0)}+
\vartheta_{12}^+|\phi^{1}_{146} \rangle_{(2,0,0)} \Bigr\}\Bigr]+
\mathcal{P}_{22}^+\Bigl[|\phi^{1}_{147}
\rangle_{(2,-1,1)}+\lambda^+_{23}\Bigl\{ \eta_1^+|\phi^{1}_{148}
\rangle_{(1,0,0)}+\eta_{11}^+|\phi^{1}_{149}
\rangle_{(0)}\Bigr\}\Bigr]
\nonumber\\
&& \hspace{-3.5em} +  \mathcal{P}_{13}^+\Bigl[|\phi^{1}_{150}
\rangle_{(1,1,0)}+\lambda^+_{12}\Bigl\{ \eta_1^+|\phi^{1}_{151}
\rangle_{(1,0,0)}+\eta_{11}^+|\phi^{1}_{152}
\rangle_{(0)}\Bigr\}\Bigr] + \ \mathcal{P}_{23}^+|\phi^{1}_{153}
\rangle_{(2,0,0)}+  \lambda^+_{12}\Bigl[|\phi^{1}_{154}
\rangle_{(3,0,1)}
\nonumber\\
&& \hspace{-3.5em} +\lambda^+_{13}\Bigl\{ \eta_1^+|\phi^{1}_{155}
\rangle_{(3,0,0)}+\eta_{11}^+|\phi^{1}_{156}
\rangle_{(2,0,0)}\Bigr\}+ \lambda^+_{23}\Bigl\{
\eta_1^+|\phi^{1}_{157} \rangle_{(2,1,0)} +
\eta_2^+|\phi^{1}_{158}
\rangle_{(3,0,0)}\nonumber\\
&& \hspace{-3.5em}+\eta_{11}^+|\phi^{1}_{159}
\rangle_{(1,1,0)}+\eta_{12}^+|\phi^{1}_{160}
\rangle_{(2,0,0)}+\vartheta_{12}^+|\phi^{1}_{161}
\rangle_{(4,0,0)}\Bigr\}\Bigr] +  \lambda^+_{13}|\phi^{1}_{162}
\rangle_{(3,1,0)} \nonumber\\
&& \hspace{-3.5em} +\lambda^+_{23}|\phi^{1}_{163}
\rangle_{(2,2,0)} +\eta_0|\chi^{1}_0\rangle_{(s)_3} \nonumber
\end{eqnarray}
where, as an usual the vector $|\chi^{1}_0\rangle_{(s)_3}$ has the
same definition and properties as the $\eta_0$-in\-de\-pen\-dent part of
$|\chi^{2}\rangle_{(s)_3}$ in the Eqs. (\ref{x-2}),
(\ref{x-2decomp})--(\ref{x-2decompf}) and the decomposition of
ghost-independent vectors in $\mathcal{H}\bigotimes \mathcal{H}'$
with only different values of spins than in
(\ref{x-3decomp})--(\ref{x-3decompf}),
(\ref{x-2decomp})--(\ref{x-2decompf}) writes as follows, for
$l=107$, $m=1$,   $r=163$, $t=154$,
\begin{eqnarray}
\label{x-1decomp} && |\phi^{1}_l \rangle_{(0,1,1)} =
a_1^{+\mu}\Bigl(a_3^{+\nu}d_{12}^+ |0\rangle \phi^{1}_{l|\mu,\nu}
+ a_1^{+\nu}d^+_{12}\bigl\{d^+_{13}|0\rangle
\phi^{\prime1}_{l|\mu\nu}+d^+_{12}d^+_{23}|0\rangle
\phi^{\prime\prime 1}_{l|\mu\nu}\bigr\}\\
&&\quad + a_2^{+\nu}\bigl\{d^+_{13}|0\rangle
\phi^{\prime\prime\prime 1}_{l|\mu,\nu}+d^+_{12}d^+_{23}|0\rangle
\phi^{(iv) 1}_{l|\mu,\nu}\bigr\}\Bigr)+a_2^{+\mu}\Bigl(a_3^{+\nu}
|0\rangle \phi^{(v)1}_{l|\mu,\nu} +
 a_2^{+\nu}d^+_{23}|0\rangle
\phi^{(6)1}_{l|\mu\nu}\Bigr)\nonumber\\
&&\quad + b_{11}^+d^+_{12} \bigl\{d^+_{13}|0\rangle
\phi^{1}_{l{}}+d^+_{12}d^+_{23}|0\rangle \phi^{\prime
1}_{l{}}\bigr\}+ b_{12}^+\bigl\{d^+_{13}|0\rangle
\phi^{\prime\prime1}_{l{}}+d^+_{12}d^+_{23}|0\rangle
\phi^{\prime\prime\prime
1}_{l{}}\bigr\}+b_{13}^+d^+_{12}\phi^{(iv) 1}_{l{}}\nonumber\\
&&\quad + b_{22}^+d^+_{23}
\phi^{(v)1}_{l}+b_{23}^+\phi^{(6)1}_{l}\,,
\nonumber\\
&& |\phi^{1}_m \rangle_{(1,1,1)} =
a_1^{+\mu}\hspace{-0.1em}\Bigl(\hspace{-0.1em}a_1^{+\nu}\hspace{-0.1em}
a_1^{+\rho} \hspace{-0.1em}d^+_{12}\bigl\{\hspace{-0.1em}d^+_{13}
|0\rangle \phi^{1}_{m|\mu\nu\rho}\hspace{-0.1em}+ d^+_{12}d^+_{23}
|0\rangle \phi^{\prime
1}_{m|\mu\nu\rho}\hspace{-0.1em}\bigr\}\hspace{-0.1em}+a_1^{+\nu}\hspace{-0.1em}a_2^{+\rho}\bigl\{\hspace{-0.1em}
d^+_{13} |0\rangle \phi^{\prime\prime
1}_{\hspace{-0.1em}m|\mu\nu,\rho}\label{111}\\
&& \quad +d^+_{12}\hspace{-0.1em}d^+_{23} |0\rangle
\phi^{\prime\prime\prime
1}_{m|\mu\nu,\rho}\hspace{-0.1em}\bigr\}+a_1^{+\nu}\hspace{-0.1em}a_3^{+\rho}
d^+_{12} |0\rangle
\phi^{(iv)1}_{m|\mu\nu,\rho}\hspace{-0.1em}+a_2^{+\nu}\hspace{-0.1em}a_2^{+\rho}
\hspace{-0.1em}d^+_{23} |0\rangle \phi^{(v)1}_{m|\mu,\nu\rho}
\hspace{-0.1em}+a_2^{+\nu}\hspace{-0.1em}a_3^{+\rho}  |0\rangle \phi^{(vi)1}_{\hspace{-0.1em}m|\mu,\nu,\rho}\nonumber \\
&& \quad + b_{11}^+ \hspace{-0.1em}d^+_{12}
\hspace{-0.1em}\bigl\{d^+_{13} |0\rangle \phi^{1}_{m|\mu}+
d^+_{12}\hspace{-0.1em}d^+_{23} |0\rangle \phi^{\prime
1}_{m|\mu}\hspace{-0.1em}\bigr\}+ b_{12}^+
\bigl\{\hspace{-0.1em}d^+_{13} |0\rangle \phi^{\prime\prime
1}_{m|\mu}+ d^+_{12}\hspace{-0.1em}d^+_{23} |0\rangle
\phi^{\prime\prime\prime 1}_{m|\mu}\hspace{-0.1em}\bigr\}+
b_{22}^+ \hspace{-0.1em}d^+_{23}
\phi^{(iv)1}_{m|\mu} \nonumber \\
&& \quad + b_{23}^+ |0\rangle \phi^{(v)1}_{m|\mu}\Bigr)+
b_{11}^+\Bigl( a_2^{+\mu}\bigl\{d^+_{13} |0\rangle \phi^{(6)
1}_{m|\mu}+ d^+_{12}d^+_{23} |0\rangle \phi^{(7) 1}_{m|\mu}\bigr\}
+a_3^{+\mu}d^+_{12} |0\rangle \phi^{(8)
1}_{m|\mu}\Bigr)\nonumber\\
&&\quad +b_{12}^+\Bigl( a_2^{+\mu}d^+_{23} |0\rangle \phi^{(9)
1}_{m|\mu} +a_3^{+\mu}|0\rangle \phi^{(10) 1}_{m|\mu}\Bigr) +
b_{13}^+\Bigl(a_2^{+\mu}|0\rangle \phi^{(11) 1}_{m|\mu}+a_1^{+\mu}d_{12}^+|0\rangle \phi^{(12) 1}_{m|\mu}\Bigr)\,, \nonumber\\
&& |\phi^{1}_r \rangle_{(2,2,0)}
=a_1^{+\mu}\hspace{-0.1em}a_1^{+\nu}\hspace{-0.15em}\Bigl(\hspace{-0.15em}a_1^{+\rho}\hspace{-0.1em}a_1^{+\sigma}
(d_{12}^+)^2|0\rangle
\phi^{1}_{\hspace{-0.1em}r|\mu\nu\rho\sigma}\hspace{-0.1em}+a_1^{+\rho}\hspace{-0.1em}a_2^{+\sigma}
d_{12}^+|0\rangle
\phi^{\prime1}_{\hspace{-0.1em}r|\mu\nu\rho,\sigma} \\
&&\quad +a_2^{+\rho}a_2^{+\sigma} |0\rangle
\phi^{\prime\prime1}_{r|\mu\nu,\rho\sigma} +
b_{11}^+(d_{12}^+)^2|0\rangle \phi^{1}_{r|\mu\nu} +
b_{12}^+d_{12}^+|0\rangle \phi^{\prime 1}_{r|\mu\nu} +
b_{22}^+|0\rangle \phi^{\prime\prime 1}_{r|\mu\nu} \Bigr)  \nonumber \\
&& \quad + b_{11}^+ \Bigl(a_1^{+\mu}a_2^{+\nu} d_{12}^+|0\rangle
\phi^{1}_{r|\mu,\nu} + a_2^{+\mu}a_2^{+\nu} |0\rangle
\phi^{\prime\prime\prime 1}_{r|\mu\nu}+
 b_{11}^+(d_{12}^+)^2|0\rangle
\phi^{1}_{r}+ b_{12}^+d_{12}^+|0\rangle \phi^{\prime 1}_{r}\nonumber \\
&& \quad + b_{22}^+|0\rangle \phi^{\prime\prime
1}_{r}\Bigr)+b_{12}^+\Bigl( a_1^{+\mu}a_2^{+\nu} |0\rangle
\phi^{\prime 1}_{r|\mu,\nu}+ b_{12}^+|0\rangle
\phi^{\prime\prime\prime 1}_{r}\Bigr), \nonumber
\\
&& \hspace{-2em}\label{301}|\phi^{1}_t \rangle_{(3,0,1)}  =
a_1^{+\mu}a_1^{+\nu}a_1^{+\rho}\Bigl(a_1^{+\sigma}\bigl\{d^+_{13}
|0\rangle \phi^{ 1}_{t|\mu\nu\rho\sigma}+ d^+_{12}d^+_{23}
|0\rangle \phi^{\prime 1}_{t|\mu\nu\rho\sigma}\bigr\}
+a_2^{+\sigma}d^+_{23} |0\rangle \phi^{\prime\prime
1}_{t|\mu\nu\rho,\sigma}\\
&&\quad +a_3^{+\sigma} |0\rangle \phi^{\prime\prime\prime
1}_{t|\mu\nu\rho,\sigma}\Bigr)+
a_1^{+\mu}a_1^{+\nu}\Bigl(b_{11}^+\bigl\{d^+_{13} |0\rangle \phi^{
1}_{t|\mu\nu}+ d^+_{12}d^+_{23} |0\rangle \phi^{\prime
1}_{t|\mu\nu}\bigr\} +b_{12}^{+}d^+_{23} |0\rangle
\phi^{\prime\prime 1}_{t|\mu\nu}\nonumber
\end{eqnarray}
\begin{eqnarray}
&& \quad +b_{13}^+|0\rangle \phi^{\prime\prime\prime
1}_{t|\mu\nu}\Bigr) + a_1^{+\mu}b_{11}^+\Bigl(a_2^{+\nu}d^+_{23}
|0\rangle \phi^{ 1}_{t|\mu,\nu}+ a_3^{+\nu} |0\rangle \phi^{\prime
1}_{t|\mu,\nu}\Bigr)+ b_{11}^+\bigl(b_{11}^+\bigl\{d^+_{13}
|0\rangle \phi^{
1}_{t{}}\nonumber\\
&& \quad+ d^+_{12}d^+_{23} |0\rangle \phi^{\prime 1}_{t{}}\bigr\}
+b_{12}^{+}d^+_{23} |0\rangle \phi^{\prime\prime
1}_{t{}}+b_{13}^+|0\rangle \phi^{\prime\prime\prime 1}_{t{}}\bigr)
. \nonumber
\end{eqnarray}
At last, the conditions (\ref{ghnum}) and (\ref{nidecompos})
applied for $l=0$ and $(s)_3 = (2,1,1)$ permit one to  decompose
the field vector $|\chi\rangle_{(s)_3}$ to be derived from general
Eq.(\ref{chi}), in even powers of ghosts from $(190+164)$ summands
starting from the ghost-independent vector
$|\Psi\rangle_{(2,1,1)}$,
\begin{eqnarray}\label{x-0}
|\chi\rangle_{(s)_3} &=&|\Psi\rangle_{(2,1,1)}+
\mathcal{P}_{1}^+\biggl[\eta_1^+|\phi_{1}\rangle_{(0,1,1)}+
\eta_2^+|\phi_{2}\rangle_{(1,0,1)}+\eta_3^+|\phi_{3}\rangle_{(1,1,0)}+\eta_{11}^+|\phi_{4}\rangle_{(-1,1,1)}
\\
&&\hspace{-3em}+\eta_{12}^+|\phi_{5}\rangle_{(0,0,1)}
+\eta_{13}^+|\phi_{6}\rangle_{(0,1,0)}+\eta_{22}^+|\phi_{7}\rangle_{(1,-1,1)}+\eta_{23}^+|\phi_{8}\rangle_{(1,0,0)}
+\vartheta_{12}^+|\phi_{9}\rangle_{(2,0,1)}\nonumber\\
&&\hspace{-3em}+\vartheta_{13}^+|\phi_{10}\rangle_{(2,1,0)}
+\vartheta_{23}^+|\phi_{11}\rangle_{(1,2,0)}+\mathcal{P}_{2}^+\Bigl\{\eta_1^+\Bigl(\eta_2^+|\phi_{12}\rangle_{(0,-1,1)}+
\eta_3^+|\phi_{13}\rangle_{(0)}+\vartheta_{12}^+|\phi_{14}\rangle_{(1,-1,1)}\nonumber\\
&&\hspace{-3em} +\vartheta_{13}^+|\phi_{15}\rangle_{(1,0,0)}+
\vartheta_{23}^+|\phi_{16}\rangle_{(0,1,0)}\Bigr)
+\eta_2^+\vartheta_{23}^+|\phi_{17}\rangle_{(1,0,0)}
+\eta_{11}^+\Bigl(\vartheta_{12}^+|\phi_{18}\rangle_{(0,-1,1)}\nonumber\\
&&\hspace{-3em}+ \vartheta_{13}^+|\phi_{19}\rangle_{(0)}+
\vartheta_{23}^+|\phi_{20}\rangle_{(-1,1,0)}\Bigr) +
\eta_{12}^+\vartheta_{23}^+|\phi_{21}\rangle_{(0)}+
\vartheta_{12}^+\vartheta_{23}^+|\phi_{22}\rangle_{(2,0,0)}
\Bigr\}\nonumber\\
&&\hspace{-3em}+
\mathcal{P}_{3}^+\Bigl\{\eta_1^+\Bigl(\eta_2^+|\phi_{23}\rangle_{(0)}
+\vartheta_{12}^+|\phi_{24}\rangle_{(1,0,0)}\Bigr)
+\eta_{11}^+\vartheta_{12}^+|\phi_{25}\rangle_{(0)} \Bigr\}+
\mathcal{P}_{11}^+\Bigl\{\eta_1^+\bigl(
\vartheta_{12}^+\bigl[|\phi_{26}\rangle_{(-1,0,1)}\nonumber\\
&&\hspace{-3em}+\vartheta_{23}^+\lambda_{12}^+|\phi_{27}\rangle_{(0)}\bigr]+\vartheta_{13}^+
|\phi_{28}\rangle_{(-1,1,0)}+
\vartheta_{23}^+|\phi_{29}\rangle_{(-2,2,0)}\bigr)+ \eta_2^+\bigl(
\vartheta_{12}^+|\phi_{30}\rangle_{(0,-1,1)}
\nonumber\\
&&\hspace{-3em}+\vartheta_{13}^+|\phi_{31}\rangle_{(0)}
+\vartheta_{23}^+|\phi_{32}\rangle_{(-1,1,0)}\bigr) +\eta_3^+
\vartheta_{12}^+|\phi_{33}\rangle_{(0)} +
\vartheta_{12}^+\bigl(\vartheta_{13}^+|\phi_{34}\rangle_{(1,0,0)}+
\vartheta_{23}^+|\phi_{35}\rangle_{(0,1,0)}\bigr)\Bigr\}\nonumber
\\
&&\hspace{-3em}+\mathcal{P}_{12}^+\Bigl\{\eta_1^+\bigl(
\vartheta_{12}^+|\phi_{36}\rangle_{(0,-1,1)}+\vartheta_{13}^+
|\phi_{37}\rangle_{(0)}+
\vartheta_{23}^+|\phi_{38}\rangle_{(-1,1,0)}\bigr) +\eta_2^+ \vartheta_{23}^+|\phi_{39}\rangle_{(0)}\nonumber\\
&&\hspace{-3em}+
\vartheta_{12}^+\vartheta_{23}^+|\phi_{40}\rangle_{(1,0,0)}
\Bigr\}+\mathcal{P}_{13}^+\eta_1^+
\vartheta_{12}^+|\phi_{41}\rangle_{(0)}+\mathcal{P}_{22}^+\eta_1^+
\vartheta_{23}^+|\phi_{42}\rangle_{(0)} +
\lambda_{12}^+\Bigl\{\eta_1^+\Bigl(
\eta_2^+|\phi_{43}\rangle_{(1,-1,1)}\nonumber\\
&&\hspace{-3em}+\eta_3^+|\phi_{44}\rangle_{(1,0,0)}+
\eta_{11}^+\bigl[|\phi_{45}\rangle_{(-1,0,1)}+
\vartheta_{12}^+\lambda_{23}^+|\phi_{46}\rangle_{(0)}\bigr]
+\eta_{12}^+|\phi_{47}\rangle_{(0,-1,1)}+
\eta_{13}^+|\phi_{48}\rangle_{(0)}
\nonumber\\
&&\hspace{-3em} + \vartheta_{12}^+ |\phi_{49}\rangle_{(2,-1,1)}+
\vartheta_{13}^+|\phi_{50}\rangle_{(2,0,0)}+
\vartheta_{23}^+|\phi_{51}\rangle_{(1,1,0)}\Bigr)+
\eta_2^+\Bigl(\eta^+_{11}|\phi_{52}\rangle_{(0,-1,1)} +
\vartheta_{23}^+|\phi_{53}\rangle_{(2,0,0)}\Bigr)\nonumber\\
&&\hspace{-3em} +\eta_3^+\eta^+_{11}|\phi_{54}\rangle_{(0)} +
\eta_{11}^+\Bigl(\vartheta_{12}^+|\phi_{55}\rangle_{(1,-1,1)}+
\vartheta_{13}^+|\phi_{56}\rangle_{(1,0,0)}+
\vartheta_{23}^+|\phi_{57}\rangle_{(0,1,0)}\Bigr)\nonumber\\
&&\hspace{-3em} +
\eta_{12}^+\vartheta_{23}^+|\phi_{58}\rangle_{(1,0,0)}+
\vartheta_{12}^+\vartheta_{23}^+|\phi_{59}\rangle_{(3,0,0)}
\Bigr\}\ + \
\lambda_{13}^+\Bigl\{\eta_1^+\Bigl(\eta_2^+|\phi_{60}\rangle_{(1,0,0)}+
\eta_{11}^+|\phi_{61}\rangle_{(-1,1,0)}\nonumber\\
&&\hspace{-3em} +\eta_{12}^+|\phi_{62}\rangle_{(0)}
+\vartheta_{12}^+|\phi_{63}\rangle_{(2,0,0)}\Bigr)+
\eta_2^+\eta^+_{11}|\phi_{64}\rangle_{(0)}  +
\eta_{11}^+\vartheta_{12}^+|\phi_{65}\rangle_{(1,0,0)}
\Bigr\}\nonumber\\
&&\hspace{-3em}+
\lambda_{23}^+\hspace{-0.1em}\Bigl\{\hspace{-0.1em}\eta_1^+\hspace{-0.1em}\Bigl(\hspace{-0.1em}
\eta_2^+|\phi_{66}\rangle_{(0,1,0)}+
\eta_{11}^+|\phi_{67}\rangle_{(-2,2,0)}
+\eta_{12}^+|\phi_{68}\rangle_{(-1,1,0)}+\eta_{22}^+|\phi_{69}\rangle_{(0)}
+\vartheta_{12}^+|\phi_{70}\rangle_{(1,1,0)}\hspace{-0.1em}\Bigr)\nonumber\\
&&\hspace{-3em}+
\eta_2^+\hspace{-0.1em}\bigl(\hspace{-0.1em}\eta^+_{11}|\phi_{71}\rangle_{(-1,1,0)}
\hspace{-0.1em}+ \eta^+_{12}|\phi_{72}\rangle_{(0)}\hspace{-0.1em}
+\vartheta_{12}^+|\phi_{73}\rangle_{(2,0,0)}
\hspace{-0.1em}\bigr)\hspace{-0.1em} +
\eta_{11}^+\vartheta_{12}^+|\phi_{74}\rangle_{(0,1,0)}\hspace{-0.1em}
+ \eta_{12}^+\vartheta_{12}^+|\phi_{75}\rangle_{(1,0,0)}\hspace{-0.1em}\Bigr\}\hspace{-0.1em}\biggr]\nonumber\\
&&\hspace{-3.5em}+
\mathcal{P}_{2}^+\biggl[\eta_1^+|\phi_{76}\rangle_{(1,0,1)}+
\eta_2^+|\phi_{77}\rangle_{(2,-1,1)}+\eta_3^+|\phi_{78}\rangle_{(2,0,0)}
+\eta_{11}^+|\phi_{79}\rangle_{(0,0,1)}+\eta_{12}^+|\phi_{80}\rangle_{(1,-1,1)}
\nonumber\\
&&\hspace{-3em}+
\eta_{13}^+|\phi_{81}\rangle_{(1,0,0)}+\vartheta_{12}^+|\phi_{82}\rangle_{(3,-1,1)}+
\vartheta_{13}^+|\phi_{83}\rangle_{(3,0,0)}
+\vartheta_{23}^+|\phi_{84}\rangle_{(2,1,0)}\nonumber\\
&&\hspace{-3em}+
\mathcal{P}_{11}^+\Bigl\{\eta_1^+\Bigl(\vartheta_{12}^+|\phi_{85}\rangle_{(0,-1,1)}
+\vartheta_{13}^+|\phi_{86}\rangle_{(0)}+\vartheta_{23}^+|\phi_{87}\rangle_{(-1,1,0)}\Bigr)
+\eta_2^+\vartheta_{23}^+|\phi_{88}\rangle_{(0)}
 \nonumber\\
&&\hspace{-3em}+
\vartheta_{12}^+\vartheta_{23}^+|\phi_{89}\rangle_{(1,0,0)}
\Bigr\} +
\mathcal{P}_{12}^+\eta_1^+\vartheta_{23}^+|\phi_{90}\rangle_{(0)}
+\lambda_{12}^+\Bigl\{\eta_1^+\Bigl(
\eta_{11}^+|\phi_{91}\rangle_{(0,-1,1)}+\vartheta_{23}^+|\phi_{92}\rangle_{(2,0,0)}\Bigr)\nonumber
\\
&&\hspace{-3em}+
\eta_{11}^+\vartheta_{23}^+|\phi_{93}\rangle_{(1,0,0)}
\Bigr\}+\lambda_{13}^+\eta_1^+ \eta_{11}^+|\phi_{94}\rangle_{(0)}+
\lambda_{23}^+\Bigl\{\eta_1^+\Bigl(\eta_2^+|\phi_{95}\rangle_{(1,0,0)}+
\eta_{11}^+|\phi_{96}\rangle_{(-1,1,0)} \nonumber\\
\phantom{|\chi\rangle_{(s)_3}}&&\hspace{-3em}+
\eta_{12}^+|\phi_{97}\rangle_{(0)}+\vartheta_{12}^+
|\phi_{98}\rangle_{(2,0,0)}\Bigr)+ \eta_2^+\eta^+_{11}|\phi_{99}
\rangle_{(0)} +
\eta_{11}^+\vartheta_{12}^+|\phi_{100}\rangle_{(1,0,0)}
\Bigr\}\biggr]\nonumber
\end{eqnarray}
\begin{eqnarray}
&&\hspace{-3.5em}+
\mathcal{P}_{3}^+\biggl[\eta_1^+|\phi_{101}\rangle_{(1,1,0)}+
\eta_2^+|\phi_{102}\rangle_{(2,0,0)}
+\eta_{11}^+|\phi_{103}\rangle_{(0,1,0)}+\eta_{12}^+|\phi_{104}
\rangle_{(1,0,0)}
+\vartheta_{12}^+|\phi_{105}\rangle_{(3,0,0)}\nonumber\\
&&\hspace{-3em}+
\mathcal{P}_{11}^+\eta_1^+\vartheta_{12}^+|\phi_{106}\rangle_{(0)}
+\lambda_{12}^+\eta_1^+\eta_{11}^+|\phi_{107}\rangle_{(0)}\biggr]\nonumber\\
&&\hspace{-3.5em}+
\mathcal{P}_{11}^+\biggl[\eta_1^+|\phi_{108}\rangle_{(-1,1,1)}+
\eta_2^+|\phi_{109}\rangle_{(0,0,1)}+
\eta_3^+|\phi_{110}\rangle_{(0,1,0)}
+\eta_{11}^+|\phi_{111}\rangle_{(-2,1,1)}+
\eta_{12}^+|\phi_{112}\rangle_{(-1,0,1)}
\nonumber\\
&&\hspace{-3em} +  \eta_{13}^+|\phi_{190}\rangle_{(-1,1,0)}+
\eta_{22}^+|\phi_{113}\rangle_{(0,-1,1)}+
\eta_{23}^+|\phi_{114}\rangle_{(0)}+\vartheta_{12}^+|\phi_{115}
\rangle_{(1,0,1)}+\vartheta_{13}^+|\phi_{116}\rangle_{(1,1,0)}
\nonumber\\
&&\hspace{-3em} +\vartheta_{23}^+|\phi_{117}\rangle_{(0,2,0)} +
\mathcal{P}_{12}^+
\vartheta_{12}^+\vartheta_{23}^+|\phi_{118}\rangle_{(0)} +
\lambda_{12}^+\Bigl\{\eta_1^+\Bigl(\eta_2^+|\phi_{119}\rangle_{(0,-1,1)}
+\eta_3^+|\phi_{120}\rangle_{(0)}\nonumber\\
&&\hspace{-3em} + \vartheta_{12}^+|\phi_{121}\rangle_{(1,-1,1)}+
\vartheta_{13}^+|\phi_{122}\rangle_{(1,0,0)}+
\vartheta_{23}^+|\phi_{123}\rangle_{(0,1,0)}\Bigr) + \eta_2^+
\vartheta_{23}^+|\phi_{124}\rangle_{(1,0,0)}
\nonumber\\
&&\hspace{-3em}
+\eta_{11}^+\hspace{-0.1em}\Bigl(\hspace{-0.1em}\vartheta_{12}^+|\phi_{125}\rangle_{(0,-1,1)}+
\vartheta_{13}^+|\phi_{126}\rangle_{(0)}\hspace{-0.1em}+
\vartheta_{23}^+|\phi_{127}\rangle_{(-1,1,0)}\hspace{-0.15em}\Bigr)\hspace{-0.1em}+
\eta_{12}^+\hspace{-0.1em}\vartheta_{23}^+|\phi_{128}\rangle_{(0)}\hspace{-0.15em}
+
\vartheta_{12}^+\hspace{-0.1em}\vartheta_{23}^+|\phi_{129}\rangle_{(2,0,0)}
\hspace{-0.15em}\Bigr\} \nonumber\\
&& \hspace{-3em} +
\lambda_{13}^+\Bigl\{\eta_1^+\bigl(\eta_2^+|\phi_{130}\rangle_{(0)}+
\vartheta_{12}^+|\phi_{131}\rangle_{(1,0,0)}\bigr)+
\eta_{11}^+\vartheta_{12}^+|\phi_{132}\rangle_{(0)} \Bigr\} +
\lambda_{23}^+\Bigl\{\eta_1^+\bigl(\eta_2^+|\phi_{133}\rangle_{(-1,1,0)}
\nonumber\\
&& \hspace{-3em}+
\vartheta_{12}^+|\phi_{134}\rangle_{(0,1,0)}\bigr)+
\eta_2^+\vartheta_{12}^+|\phi_{135}\rangle_{(1,0,0)}  +
\eta_{11}^+\vartheta_{12}^+|\phi_{136}\rangle_{(-1,1,0)} +
\eta_{12}^+\vartheta_{12}^+|\phi_{137}\rangle_{(0)}\Bigr\}
\biggr]\nonumber\\
&&\hspace{-3.5em} +
\mathcal{P}_{12}^+\biggl[\eta_1^+|\phi_{138}\rangle_{(0,0,1)}+
\eta_2^+|\phi_{139}\rangle_{(1,-1,1)}+\eta_3^+|\phi_{140}\rangle_{(1,0,0)}
+\eta_{11}^+|\phi_{141}\rangle_{(-1,0,1)}+
\eta_{12}^+|\phi_{142}\rangle_{(0,-1,1)}
\nonumber\\
&&\hspace{-3em} + \eta_{13}^+|\phi_{143}\rangle_{(0)}+
\vartheta_{12}^+|\phi_{144}\rangle_{(2,-1,1)}+
\vartheta_{13}^+|\phi_{145}\rangle_{(2,0,0)}
+\vartheta_{23}^+|\phi_{146}\rangle_{(1,1,0)}+
\lambda_{12}^+\Bigl\{\eta_1^+
\vartheta_{23}^+|\phi_{147}\rangle_{(1,0,0)}\nonumber
\\
&&\hspace{-3em} +
\eta_{11}^+\vartheta_{23}^+|\phi_{148}\rangle_{(0)} \Bigr\}+
\lambda_{23}^+\Bigl\{\eta_1^+\Bigl(\eta_2^+|\phi_{149}\rangle_{(0)}+
\vartheta_{12}^+ |\phi_{150}\rangle_{(1,0,0)}\Bigr)+
\eta_{11}^+\vartheta_{12}^+|\phi_{151}\rangle_{(0)}
\Bigr\}\biggr]\nonumber\\
&&\hspace{-3.5em} +
\mathcal{P}_{22}^+\Bigl[\eta_1^+|\phi_{152}\rangle_{(1,-1,1)}+
\eta_{11}^+|\phi_{153}\rangle_{(0,-1,1)} + \vartheta_{23}^+|\phi_{154}\rangle_{(2,0,0)}\Bigr]\nonumber\\
&&\hspace{-3.5em} +
\mathcal{P}_{13}^+\Bigl[\eta_1^+|\phi_{155}\rangle_{(0,1,0)}+\eta_2^+|\phi_{156}\rangle_{(1,0,0)}+\eta_{11}^+|\phi_{157}\rangle_{(-1,1,0)}
+\eta_{12}^+|\phi_{158}\rangle_{(0)}+
\vartheta_{12}^+|\phi_{159}\rangle_{(2,0,0)}\Bigr]\nonumber\\
&& \hspace{-3.5em} +
\mathcal{P}_{23}^+\Bigl[\eta_1^+|\phi_{160}\rangle_{(1,0,0)}+
\eta_{11}^+|\phi_{161}\rangle_{(0)} \Bigr]\nonumber\\
&& \hspace{-3.5em} +
\lambda_{12}^+\biggl[\eta_1^+|\phi_{162}\rangle_{(2,0,1)}+
\eta_2^+|\phi_{163}\rangle_{(3,-1,1)}+\eta_3^+|\phi_{164}
\rangle_{(3,0,0)}
+\eta_{11}^+|\phi_{165}\rangle_{(1,0,1)}+\eta_{12}^+|\phi_{166}
\rangle_{(2,-1,1)}
\nonumber\\
&&\hspace{-3em} +
\eta_{13}^+|\phi_{167}\rangle_{(2,0,0)}+
\vartheta_{12}^+|\phi_{168}\rangle_{(4,-1,1)}+
\vartheta_{13}^+|\phi_{169}\rangle_{(4,0,0)}
+\vartheta_{23}^+|\phi_{170}\rangle_{(3,1,0)}\nonumber\\
&&\hspace{-3em} +
\lambda_{13}^+\eta_1^+\eta_{11}^+|\phi_{171}\rangle_{(1,0,0)} +
\lambda_{23}^+\Bigl\{\eta_1^+\bigl(\eta_{2}^+|\phi_{172}\rangle_{(2,0,0)}+
\eta_{11}^+|\phi_{173}\rangle_{(0,1,0)}+
\eta_{12}^+|\phi_{174}\rangle_{(1,0,0)}\nonumber\\
&&\hspace{-3em} +
\vartheta_{12}^+|\phi_{175}\rangle_{(3,0,0)}\bigr) + \eta_2^+
\eta_{11}^+|\phi_{176}\rangle_{(1,0,0)}+\eta_{11}^+
\bigl(\eta_{12}^+|\phi_{177}\rangle_{(0)} +
\vartheta_{12}^+|\phi_{178}\rangle_{(2,0,0)}\bigr)\Bigr\}\biggr]\nonumber\\
&&\hspace{-3.5em} +
\lambda_{13}^+\Bigl[\eta_1^+|\phi_{179}\rangle_{(2,1,0)}+
\eta_2^+|\phi_{180}\rangle_{(3,0,0)}
+\eta_{11}^+|\phi_{181}\rangle_{(1,1,0)}+
\eta_{12}^+|\phi_{182}\rangle_{(2,0,0)} +
\vartheta_{12}^+|\phi_{183}\rangle_{(4,0,0)}
\Bigr]\nonumber\\
&&\hspace{-3.5em}+
\lambda_{23}^+\Bigl[\eta_1^+|\phi_{184}\rangle_{(1,2,0)}+
\eta_2^+|\phi_{185}\rangle_{(2,1,0)}
+\eta_{11}^+|\phi_{186}\rangle_{(0,2,0)}+\eta_{12}^+|\phi_{187}
\rangle_{(1,1,0)} +\eta_{22}^+|\phi_{188}\rangle_{(2,0,0)}
\nonumber\\
&&\hspace{-3em} + \vartheta_{12}^+|\phi_{189}\rangle_{(3,1,0)}
\Bigr]+\eta_0|\chi_0\rangle_{(s)_3}\,,\nonumber
\end{eqnarray}
where,
 the vector $|\chi_0\rangle_{(s)_3}$ has the same
definition and properties as $\eta_0$-independent part of
$|\chi^{1}\rangle_{(s)_3}$ in the Eqs. (\ref{x-1}),
(\ref{x-1decomp})--(\ref{301}) and the decomposition of
ghost-independent vectors in $\mathcal{H}\bigotimes \mathcal{H}'$
with only different values of spins than in
(\ref{x-3decomp})--(\ref{x-3decompf}),
(\ref{x-2decomp})--(\ref{x-2decompf}),
(\ref{x-1decomp})--(\ref{301}) has the form,
\begin{eqnarray}
\label{basdecomp} && \hspace{-0.5em}|\Psi \rangle_{(2,1,1)} =
a_1^{+\mu}\Bigl(a_1^{+\nu}a_1^{+\rho}\Bigl[a_1^{+\sigma}
d^+_{12}\bigl\{d^+_{13} |0\rangle \phi_{\mu\nu\rho\sigma}+
d^+_{12}d^+_{23} |0\rangle \phi^{\prime
}_{\mu\nu\rho\sigma}\bigr\}+a_2^{+\sigma} \bigl\{ d^+_{13}
|0\rangle \phi^{\prime\prime
}_{\mu\nu\rho,\sigma}\\
&& \quad +d^+_{12}d^+_{23} |0\rangle \phi^{\prime\prime\prime
}_{\mu\nu\rho,\sigma}\bigr\}+a_3^{+\sigma} d^+_{12} |0\rangle
\phi^{(iv)}_{\mu\nu\rho,\sigma}\Bigr]+a_1^{+\nu}a_2^{+\rho}a_2^{+\sigma}
d^+_{23} |0\rangle \phi^{(v)}_{\mu\nu,\rho\sigma}
\nonumber \\
&& \quad +a_1^{+\nu}a_2^{+\rho} a_3^{+\sigma} |0\rangle
\Phi_{\mu\nu,\rho,\sigma}+ a_1^{+\nu}b_{11}^{+}
d^+_{12}\bigl\{d^+_{13} |0\rangle \phi_{\mu\nu}+ d^+_{12}d^+_{23}
|0\rangle \phi^{\prime }_{\mu\nu}\bigr\}+a_1^{+\nu}b_{12}^{+}
\bigl\{ d^+_{13} |0\rangle \phi^{\prime\prime
}_{\mu\nu}\nonumber\\
&& \quad +d^+_{12}d^+_{23} |0\rangle \phi^{\prime\prime\prime
}_{\mu\nu}\bigr\}+a_1^{+\nu}b_{13}^{+} d^+_{12} |0\rangle
\phi^{(iv)}_{\mu\nu}+a_1^{+\nu}b_{22}^{+} d^+_{23} |0\rangle
\phi^{(v)}_{\mu\nu} +a_1^{+\nu}b_{23}^{+} |0\rangle
\phi^{(6)}_{\mu\nu} \nonumber\\
&& \quad +a_2^{+\nu}b_{11}^{+} \bigl\{ d^+_{13} |0\rangle
\phi^{(7) }_{\mu,\nu} + d^+_{12}d^+_{23} |0\rangle \phi^{(8)
}_{\mu,\nu}\bigr\}+a_3^{+\nu}b_{11}^{+} d^+_{12} |0\rangle
\phi^{(9)}_{\mu,\nu}+a_2^{+\nu}b_{12}^{+} d^+_{23} |0\rangle
\phi^{(10)}_{\mu,\nu}  \nonumber\\
&&\quad +a_2^{+\nu}b_{13}^{+} |0\rangle \phi^{(11)}_{\mu,\nu}
+a_3^{+\nu}b_{12}^{+} |0\rangle \phi^{(12)}_{\mu,\nu}\Bigr)+
b_{11}^+a_2^{+\mu}\Bigl(a_2^{+\nu}d_{23}^+|0\rangle
\phi^{(13)}_{\mu\nu} + a_3^{+\nu}|0\rangle
\phi^{(14)}_{\mu,\nu}\Bigr) \nonumber\\
&&\quad + b_{11}^+ \Bigl(b_{11}^+d^+_{12} \bigl\{d^+_{13}
|0\rangle \phi_{}+ d^+_{12}d^+_{23} |0\rangle \phi^{\prime
}\bigr\} + b_{12}^+ \bigl\{d^+_{13} |0\rangle \phi^{\prime\prime
}+ d^+_{12}d^+_{23} |0\rangle \phi^{\prime\prime\prime
}\bigr\}+b_{13}^+ d^+_{12}|0\rangle \phi^{(iv)}\nonumber\\
&&\quad + b_{22}^+ d^+_{23}|0\rangle \phi^{(v)}  + b_{23}^+
|0\rangle \phi^{(vi)}\Bigr) + b_{12}^+\bigl( b_{12}^+d^+_{23}
|0\rangle \phi^{(7)}+b_{13}^+ |0\rangle
\phi^{(8)}\bigr)\nonumber\\
&& \label{3-11}|\phi_m \rangle_{(3,-1,1)} =
a_1^{+\mu}d^+_{23}\bigl(a_1^{+\nu}a_1^{+\rho} |0\rangle
\phi_{m|\mu\nu\rho}+b^+_{11} |0\rangle \phi^{1}_{m|\mu}\bigr)\,,\ \\
&& |\phi_p \rangle_{(4,-1,1)} =
a_1^{+\mu}a_1^{+\nu}d^+_{23}\bigl(a_1^{+\rho}a_1^{+\sigma}
|0\rangle \phi_{p|\mu\nu\rho\sigma}+b^+_{11} |0\rangle
\phi_{p|\mu\nu}\bigr) + (b_{11}^+)^2d^+_{23} |0\rangle \phi_{p},\
 \label{xdecompf}
\end{eqnarray}
for $m =  163, p=168,$ and with initial tensor field
$\Phi_{\mu\nu,\rho,\sigma}$ describing massless particle with spin
$(2,1,1)$.

\end{document}